\documentclass[twocolumn,traditabstract]{aa}
\usepackage{caption}
\usepackage{titlesec}
\titlespacing{\section}{0pt}{*3}{*2}  
\titleformat{\subparagraph}[runin]{\normalfont\normalsize\bfseries}{\thesubparagraph}{4em}{}
\usepackage{xspace}
\usepackage{xcolor} 
\definecolor{model2}{HTML}{1F77B4}
\definecolor{model3}{HTML}{FF7F0E}
\definecolor{model4}{HTML}{2B5F4A}
\definecolor{model5}{HTML}{8C564B}
\definecolor{model6}{HTML}{D62728}
\definecolor{model7}{HTML}{EC0ABF}
\usepackage[switch]{lineno} 
\usepackage{amssymb}
\usepackage{hyperref}

\usepackage{placeins}
\usepackage{float}
\usepackage{pgfplots}
\usepackage{mathptmx}
\newcommand{\kpc}{\mathrm{kpc}}
\usepackage{multirow}
\usepackage{tablefootnote}
\usepackage{natbib}
\bibpunct{(}{)}{;}{a}{}{,}
\usepackage[]{graphicx}
\usepackage{booktabs}
\usepackage{longtable}
\usepackage{threeparttable} 
\usepackage{multirow}
\usepackage{titlesec}
\usepackage{tikz}
\usetikzlibrary{shapes.geometric, arrows}
\usepackage{enumitem}
\tikzstyle{startstop} = [rectangle, rounded corners, minimum width=3cm, minimum height=1cm, text centered, draw=black]
\tikzstyle{process} = [rectangle, minimum width=3cm, minimum height=1cm, text centered, draw=black]
\tikzstyle{chart} = [rectangle, minimum width=3cm, minimum height=1cm, text centered, draw=black, dashed]
\tikzstyle{decision} = [diamond, minimum width=3cm, minimum height=1cm, text centered, draw=black]
\tikzstyle{arrow} = [thick,->,>=stealth]
\usepackage{bm}

\DeclareRobustCommand{\uvec}[1]{{
  \ifcsname uvec#1\endcsname
     \csname uvec#1\endcsname
   \else
    \bm{\hat{\mathbf{#1}}}
   \fi
}}
\usepackage{cool}
\usepackage{subcaption}
\usepackage{amsmath}
\usepackage{lscape}
\usepackage{txfonts}
\usepackage{hyperref}
\hypersetup{colorlinks=true,citecolor=blue}

\usepackage{csquotes}

\usepackage{multirow}
\usepackage{tabularx}
\usepackage{siunitx}

\usepackage{orcidlink}
\usepackage{comment}
\usepackage{verbatim}
\usepackage{url}
\usepackage{graphicx}
\usepackage{subfig}

\usepackage{xcolor}    
\usepackage{soul}      
\usepackage{array}
\usepackage{xspace}

\usepackage{fontawesome}

\newcommand{\Rein}{\theta_{\rm E}}
\newcommand{\rs}{r_{\rm s}}

\usepackage{titlesec}
\makeatletter
\providecommand{\sf@counterlist}{}
\makeatother

\begin{document}

\defcitealias{2018Collier}{C18}
\defcitealias{2022Poci}{P22}
\title{The Stellar IMF and Dark Matter Halo of ESO0286: Constraints from Strong Lensing and Dynamics}

\author{Han~Wang\inst{1,2}\orcidlink{0000-0002-1293-5503}, Jens~Thomas\inst{3}, Mathias~Lipka\inst{3}\orcidlink{0000-0002-0730-0351}, Sherry H.~Suyu\inst{2,1}\orcidlink{0000-0001-5568-6052}, Aymeric~Galan\inst{4,1,2}\orcidlink{0000-0003-2547-9815}, Stefano~de~Nicola\inst{3} and 
Tian~Li\inst{5}}

\institute{
Max-Planck-Institut f\"ur Astrophysik, Karl-Schwarzschild-Str. 1, 85748 Garching, Germany \\
{\tt e-mail: wanghan@mpa-garching.mpg.de}
\and
Technical University of Munich, TUM School of Natural Sciences, Physics Department, James-Franck Str. 1, 85748 Garching, Germany
\and
Max-Planck-Institut für extraterrestrische Physik, Gießenbachstr 1, 85748 Garching, Germany
\and
Département d'Astronomie, Université de Genève, Chemin Pegasi 51, CH-1290 Versoix, Switzerland
\and
Institute of Cosmology and Gravitation, University of Portsmouth, Burnaby Rd, Portsmouth PO1 3FX, UK
}

\titlerunning{ESO0286 revealed by lensing and dynamics}

\authorrunning{H.~Wang et al.} \date{Received / Accepted}

\abstract{The internal mass structure of elliptical galaxies offers critical insights into galaxy formation, yet disentangling stellar mass from dark matter (DM) and determining the stellar initial mass function (IMF) remains challenging. We present a detailed analysis of ESO0286-G022 ($z=0.0312$), a rare nearby strong-lens system with a fast-rotating elliptical galaxy, combining high-resolution Hubble Space Telescope (HST) imaging with VLT/MUSE integral-field stellar kinematics. We construct axisymmetric and triaxial Schwarzschild orbit-superposition models to reconstruct its intrinsic shape and mass distribution. We find that despite being a fast rotator, ESO0286 exhibits clear kinematic signatures of intrinsic triaxiality, characterized by rotation along both the major and minor axes, making it only the second such confirmed case. Models based on kinematics alone yield significantly larger scatter in the predicted total mass at large radii. 
By incorporating the mass enclosed within the Einstein radius from strong lensing as a complementary constraint to the kinematic data, we anchor the total mass at a radius where the constraints from the IFU data alone are only weak. This significantly reduces the uncertainty on the outer mass profile and orbital structure: only models with strong radial anisotropy beyond the IFU FoV are compatible with the lensing constraints. In the inner regions, we robustly constrain an upper limit for the stellar mass around $r \sim 0.7$~kpc, ruling out an IMF more bottom-heavy than Kroupa. The data allow for a gentle gradient towards a slightly more bottom-heavy central IMF. This result aligns with recent dynamical studies of local well resolved massive early-type galaxies but contrasts with heavier IMFs reported for lenses detected at $z > 0.1$. Our work demonstrates the power of combining lensing and dynamical modeling to resolve the detailed inner structure of massive galaxies.
}


\maketitle
\section{Introduction}
\label{sec:intro}
The inner structure of elliptical galaxies plays a crucial role in both studies of galaxy evolution and cosmology. Compared to spiral galaxies, which have ongoing star formation, significant gas content, and complex structures such as bulges, discs, spiral arms and bars, elliptical galaxies are relatively simple in their composition. In most cases, the gas content in elliptical galaxies can be safely neglected \citep{2005Mamon, 2007Diehl, 2013Grillo}, leaving essentially three main components: stars, DM, and a central supermassive black hole. The black hole dominates only in the very central region, while at larger radii, the mass is primarily contributed by stars and DM. If these two components can be disentangled, we can gain valuable insights into their intrinsic properties. For example, the stellar initial mass function (IMF) characterizes the distribution of stellar masses at the time of star formation, while the DM component provides insights into its intrinsic properties, such as the core-cusp problem \citep{1994Flores, 2010deBlok, 2018Genina}, whether DM universally follows the Navarro-Frenk-White (NFW) \citep{Navarro1996, Navarro1997}. profile predicted by cosmological simulations, and how it evolves over time under the influence of baryonic feedback. Interestingly, even if the density profiles of stars and DM are not individually isothermal, elliptical galaxies often exhibit an approximately isothermal total mass distribution at the population level, a phenomenon commonly referred to as the “baryon-halo conspiracy” \citep{2001Gerhard,2003Koopmans,2004Treu,2006Humphrey,2007Thomas,2010Auger,2014Dutton,2016Cappellari,2021Shajib}.
 
Moreover, the majority of the lensing galaxies are elliptical galaxies. A detailed understanding of their inner mass distribution helps to break the mass-sheet degeneracy in strong lensing analyses, thereby improving the accuracy of cosmological measurements, such as the Hubble constant derived from time-delay lenses \citep[e.g.,][]{Holicow_wong, TDCOSMO1,TDCOSMO4,ShajibRXJ1131,2025TDCOSMO, 2020Akin, TDCOSMO13, 2025GLaD}. For these reasons, elliptical galaxies provide an exceptional laboratory for probing both the physics of galaxy formation and fundamental cosmological parameters.

The key challenge is to disentangle the DM and baryonic components with precision. Unlike DM, which we cannot observe directly and can only infer through its reaction to the gravitational force, baryonic matter produces measurable signals in the form of stellar spectra and photometry. Thus, our ability to improve DM mass models relies primarily on refining measurements of the baryonic component. In this context, accurately determining the IMF, especially its low-mass end, is essential. Faint dwarf stars, despite significantly contributing to the total stellar mass budget, are challenging to detect due to their faintness and slow evolution over cosmic time. If underestimated, their contribution may be misattributed to DM. Consequently, a strong degeneracy arises between the DM and IMF measurements \citep{2011Thomas,2015Newman, 2016Leier, 2025Sonnenfeld}.
 
A direct approach for inferring the IMF relies on single stellar population (SSP) synthesis libraries, though this method faces several significant limitations. As discussed in \citet{2009Conroy, 2018Dries, 2024Navarro}, fitting unresolved stellar populations with SSP models, which assume a single formation burst and uniform metallicity, can bias IMF slope estimates, particularly for galaxies with extended or complex star formation histories. In addition, these models place only weak constraints on the high-mass end of the IMF, since the most massive stars evolve quickly into compact remnants and leave little direct imprint on the observed spectrum.

An independent approach combines a basic SSP model for a reference IMF with either dynamical or lensing information. From the mismatch between the dynamical or lensing mass and the reference stellar mass-to-light ratio, the actual IMF can be inferred indirectly \citep{2010Treu, 2010Auger, 2011Thomas, 2013Barnabe, 2013Cap, 2015Spiniello}. By relying on the gravitational potential rather than solely on stellar spectra, this method provides an independent way to test for potential variations in the IMF. However, it only provides an upper limit on the IMF normalization (as long as the contribution of DM to the potential is unknown).
Dynamics and lensing constrain essentially the same quantity, i.e., the gravitational potential of the galaxy, but in complementary ways and at different scales. Dynamical modeling provides the strongest constraints in regions where kinematic data are available and reconstructs the mass in three dimensions.  Strong lensing is most sensitive at the radii where lensed images appear, such as the Einstein radius. It constrains the projected mass along the line of sight. By combining strong lensing and dynamical constraints, the mass model becomes more robust. Each method serves as a powerful complementary constraint for the other, restricting the allowed parameter space.

The ideal targets for studying the IMF and detailed mass structure of galaxies are nearby lensing galaxies, as they provide strong constraints from both dynamics and lensing. Such targets help minimize observational systematics caused by noise or resolution limits, particularly for key quantities in the inner regions of elliptical galaxies e.g., the stellar mass-to-light ratio within $r \lesssim 1~\rm kpc$, as suggested by \citet{2018Collett} and \citet{ 2024Mehrgan}. Typical lensing galaxies are located at $z \sim 0.5$, which largely limits the constraining power of kinematics. In these galaxies, the DM enclosed within the Einstein radius is non-negligible. Strong lensing, however, is primarily sensitive to the Einstein radius, which makes the separation between stellar and DM components less straightforward.

To date, only five low-redshift ($z<0.05$) lensing galaxies have been identified. The first, ESO325-G004 at $z = 0.0345$, was discovered using the Advanced Camera for Surveys on the Hubble Space Telescope \citep{2005Smith}. This galaxy is a slow rotator and exhibits a pronounced boxiness in its photometric structure. Three additional systems, including ESO0286-G022 (hereafter ESO0286 for simplicity), were identified in the SINFONI Nearby Elliptical Lens Locator Survey (SNELLS) \citep{SINFONI2015}. Among these, ESO0286 stands out as a fast rotator uniquely suited for detailed lensing and dynamical studies of its inner mass distribution. The other two SNELLS systems are observationally less ideal: one produces only a single lensed image, and the other lies very close to a massive satellite galaxy, complicating its analysis. The most recently discovered low-redshift lens, NGC 6505 \citep{2025Riordan}, was identified in the Euclid survey at $z = 0.042$ and is also classified as a slow rotator.

In this paper, we present a detailed study of the nearby lensing galaxy ESO0286 at redshift $z_{\rm lens} = 0.0312$, with a velocity dispersion of $\sigma_{\rm vel} = 310~\mathrm{km~s^{-1}}$ and a maximum rotation velocity of $V_{\rm rot} = 210~\mathrm{km~s^{-1}}$. These kinematic properties classify ESO0286 as a fast rotator, making it a valuable complementary case to the slow rotators that dominate the currently known sample of nearby lensing galaxies. We improve upon the lensing measurements of ESO0286 compared to \citet[][hereafter \citetalias{2018Collier}]{2018Collier} and explicitly incorporate the lensing constraints into our dynamical modeling. Moreover, we adopt a more flexible mass model to better capture the galaxy's internal mass distribution. Specifically, we differentiate our approach from the prior study by \citet[][hereafter \citetalias{2022Poci}]{2022Poci} in several key ways. Inspired by \citet{2024Mehrgan}, we advance triaxial Schwarzschild modeling by introducing a fully triaxial DM halo and a flexible stellar mass-to-light ratio gradient. Most importantly, we combine these advanced dynamical models with strong lensing data to obtain highly robust constraints on the total mass-to-light ratio. This enables us to dynamically measure the stellar IMF, directly compare our findings with the independent SSP-based estimates of \citetalias{2022Poci}, and demonstrate how the complementary nature of lensing and kinematics is essential for accurately constraining the mass distribution of ESO0286.

This paper is organized as follows. In Sect.~\ref{section:Data}, we detail the photometric and kinematic observations of ESO0286 that underpin our analysis. Sect~\ref{Sect:Strong lensing modeling} presents the strong lensing analysis, focusing on the modeling of the extended lensed images. In Sect.~\ref{sect:Dynamical modeling}, we describe the Schwarzschild dynamical modeling technique, while Sect.~\ref{sect:Model configuration and selection} outlines our specific modeling strategy and the bootstrap analysis used to quantify the uncertainties in the recovered mass distributions. We discuss our results in Sect.~\ref{sect:Results}, specifically examining the implications for the IMF and DM halo properties and demonstrating how combining dynamics with lensing significantly improves the model constraints. Finally, we summarize our findings and provide an outlook on future studies in Sect.~\ref{sect:Summary and outlook}. Throughout this paper, we adopt a standard cosmological model with $H_0 = 70.4~\mathrm{km~s^{-1}~Mpc^{-1}}$, $\Omega_{\rm m} = 0.27$, and $\Omega_\Lambda = 0.73$. At the lens redshift of $z = 0.0312$, this corresponds to a physical scale of $0.621~\mathrm{kpc}$ per arcsecond.


\section{Data}
\label{section:Data}
ESO0286 was discovered as part of the SNELLS survey (PI: R. J. Smith), which was designed to identify strong galaxy-galaxy lenses at very low redshifts using the Spectrograph for Integral Field Observations in the Near Infrared (SINFONI) at the European Southern Observatory Very Large Telescope (ESO VLT).


Within this program, ESO0286 was identified as the first newly discovered lens system. The lens is a compact early-type galaxy with an effective radius of $\approx 2$ kpc and ellipticity of $\approx 0.4$, around which two lensed images are observed. Follow-up near-IR spectroscopy detected H$\alpha$ and associated emission lines from the background source at $z_{\rm source} = 0.926$ \citep[see][]{SINFONI2015}, providing independent confirmation of the lensing interpretation. We describe the HST imaging of ESO0286 we use for lens modeling in Sect.~\ref{subsection:HST imaging} and the spectroscopy we use for the dynamical modeling in Sect.~\ref{subsection:Spectroscopy from MUSE}.

\subsection{HST imaging}
\label{subsection:HST imaging}
 As part of GO Cycle 23 (PI: R.~J.~Smith), HST imaging was obtained using the Wide Field Camera 3 (WFC3) UVIS channel. For ESO0286, this includes three dithered exposures in F814W totaling 1050\,s and three dithered exposures in F336W totaling 4413\,s. The F336W observations were chosen to lie shortward of the 4000\,\AA\ break in the lens galaxy while avoiding Ly$\alpha$ absorption from the background source.  

\begin{figure*}
    \centering
    \includegraphics[width=0.51\linewidth]{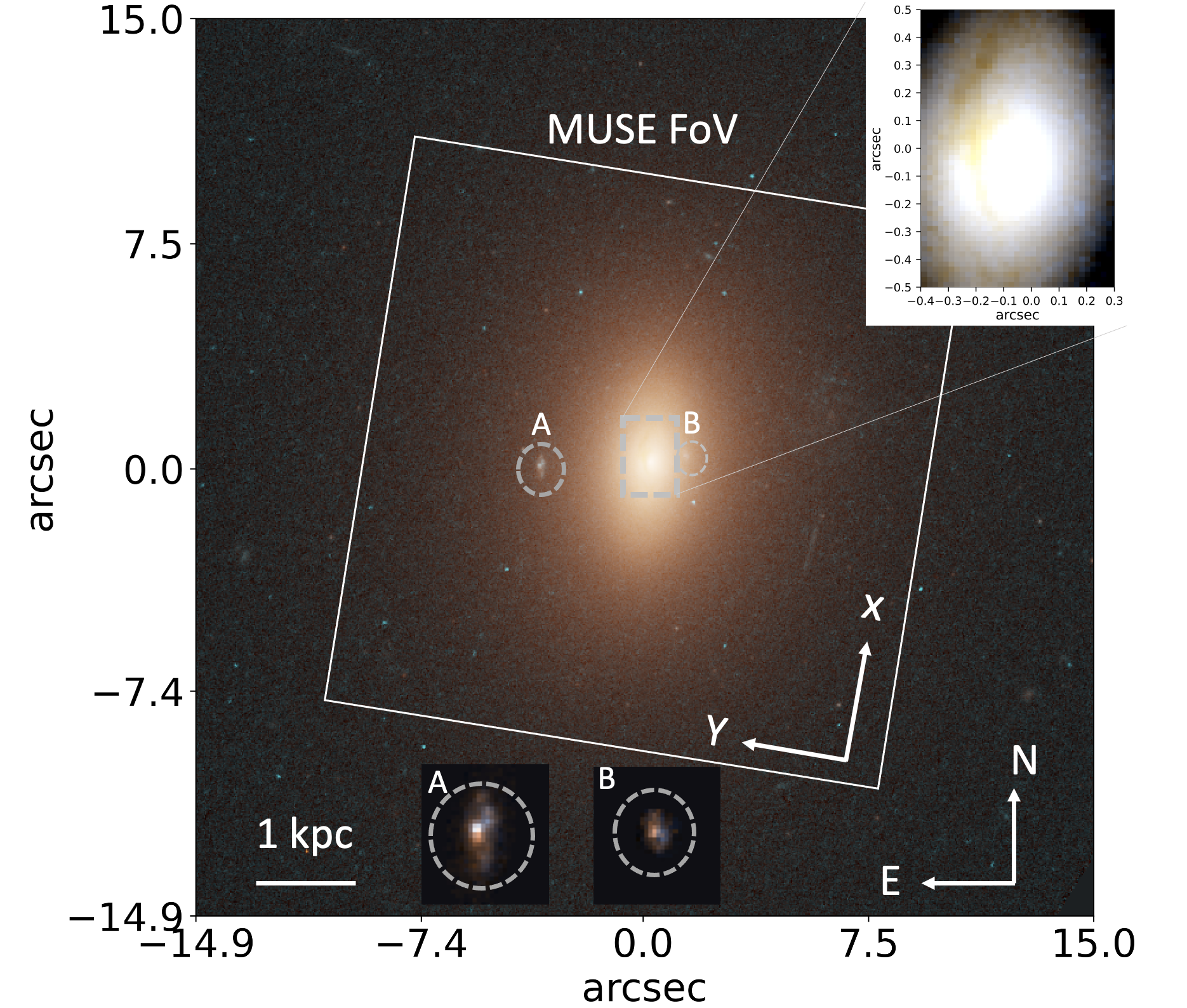}
    \includegraphics[width=0.46\linewidth]{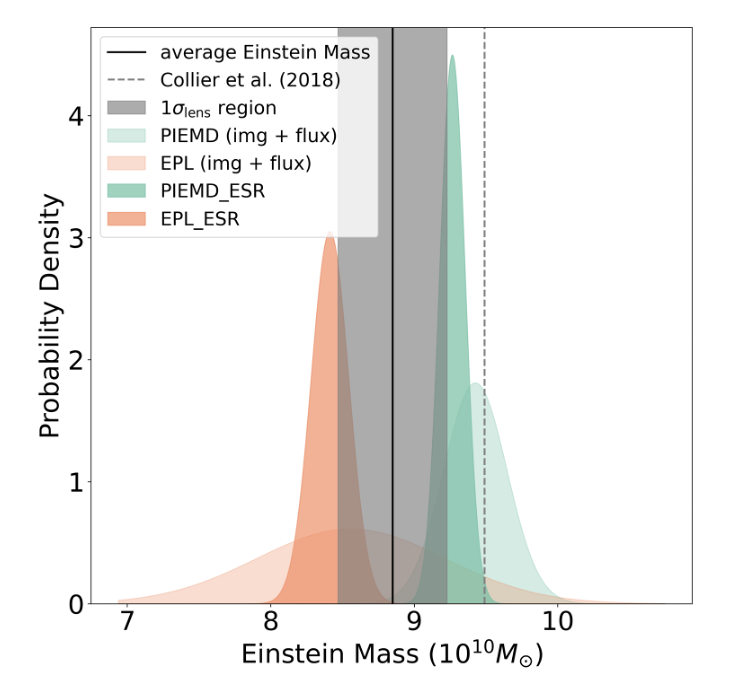}
    \caption{Photometric and kinematic constraints for the lens ESO0286. \textit{Left:} HST/F814W image of the lens overlaid with the MUSE WFM FoV (white rectangle). The coordinate system ($x, y$) aligns with the galaxy's photometric major and minor axes, respectively. The upper inset provides a zoom-in of the central region to highlight the dust structure, while dashed circles mark lensed images A and B (shown in the bottom insets). \textit{Right:} Predicted $M_{\rm Ein}$ in units of $10^{10}~\rm M_\odot$, derived from lensing and dynamical modeling. The more transparent green and orange regions show the probability density distribution of $M_{\rm Ein}^{\rm lens}$ constrained only by image positions and flux ratios, while the less transparent regions correspond to models constrained by the extended images in the F814W and F336W bands. The vertical black line and gray shaded region indicate the inferred value, $\overline{M_{\rm Ein}^{\rm lens}} = (8.85 \pm 0.38) \times 10^{10}~\rm M_\odot$, based on the best-fit EPL-ESR and PIEMD-ESR models. The gray dashed line marks the lens mass derived by \citetalias{2018Collier}.}
    \label{fig:ES0286_HST_imaging}
\end{figure*}

The imaging reveals a dust-obscured central region and clearly confirms a two-image lens configuration, with a half-image separation of $2.43 \pm 0.03\arcsec$ and a flux ratio of $A/B = 2.2 \pm 0.1$ (see \citetalias{2018Collier}). Image B lies close to the lens center, making it challenging to observe clearly in the HST images (see Fig.~\ref{fig:ES0286_HST_imaging}).

We adopt the F814W and F336W HST images for the lensing analysis. To prepare the images for modeling, residual background fluctuations caused by cosmic rays and scattered light are removed by subtracting a background level estimated via sigma-clipped statistics. Specifically, we use a $3\sigma$ clipping threshold to compute the background mean, median, and standard deviation, and then subtract the median value from the drizzled images. As the lens galaxy light drops to the noise level at $\sim 30\arcsec$, we adopt a $30\arcsec \times 30\arcsec$ field of view (FoV) for the strong lensing modeling in both bands. For the dynamical modeling, the three-dimensional (3D) light density profile is derived from the F814W image within the same FoV.



\subsection{Spectroscopy from MUSE}
\label{subsection:Spectroscopy from MUSE}

We obtained the spatially resolved kinematic information, that is required here as input for the dynamical models, from data taken with the Multi Unit Spectroscopic Explorer (MUSE) \citep{Bacon_2010}. This spectrograph, operated at the ESO VLT, has a spectral resolution R of 1750 at $4650\AA$ up to 3750 at $9300\AA$. For ESO0286, two MUSE data sets are currently available and readily reduced with the standard ESO pipelines: i) Our primary data set was obtained with MUSE operated in its wide-field mode (WFM) without adaptive optics (AO) under program ID 0100.B-0769 (PI: A.~Poci), and ii) the second MUSE data set was obtained in its narrow-field mode (NFM) under program ID 109.22X3.001 (PI: A.~Poci). The NFM observations are affected by issues with the background subtraction as suggested by \citet{2025Poci}, which significantly degrade their quality. Therefore, all subsequent analysis in this work is based exclusively on the WFM data set.

The WFM MUSE covers a $60\arcsec \times 60\arcsec$ field of view (FoV) with a pixel scale of $0.2\arcsec\,\mathrm{pix}^{-1}$, corresponding to $0.1~\kpc\,\mathrm{pix}^{-1}$. The WFM observations, taken without adaptive optics, are seeing-limited, and for ESO0286, the point spread function FWHM is estimated to be $1.0\arcsec$. 

To extract the stellar kinematics reliably, a higher signal-to-noise-ratio (S/N) spectrum is required than what is present in most of the individual MUSE pixels. Therefore, after masking pixels contaminated by point sources or dust, we binned the MUSE data cubes using the Voronoi tessellation method of \cite{Cappellari_2003}. We adopted a target S/N of 80 for the WFM, which is empirically sufficient (see \cite{2024Mehrgan}) to derive non-parametric line-of-sight velocity distributions (LOSVDs) using WINGFIT (J. Thomas, in prep.).

After this binning, our smallest resolution elements are $0.6\arcsec$ for the WFM. We further exclude Voronoi Bins (mostly in the outer regions of the FoV) that still had bad S/N or contaminated spectra from the final dynamical modeling.



We fit the binned spectra with the MILES stellar library \citep{Sanchez_2006,Falcon_Barroso_2011} using WINGFIT. This allows us to extract non-parametric LOSVDs with a degree of regularization optimized as described in \cite{2022Thomas}. To extract the LOSVDs for the dynamical models, we restricted the fit to the blue spectral range $\lambda \in [4850,5650]~\AA$, because this blue part of MUSE is less affected by noise and sky contamination. The final fitted spectrum contains the important $\mathrm{H}\beta$, $\mathrm{Mg}\ b$ triplet, and iron absorption lines. In App.~\ref{app:Kinematics fitting} we show an example of the WFM spectrum from one of the Voronoi Bins of ESO0286 and its corresponding WINGFIT model fit. The model spectrum is the result of a convolution of a superposition of MILES library templates and the non-parametric LOSVDs model that is later used as input for the dynamical models.


For illustrative purposes, we also fit the non-parametric LOSVDs using a Gauss-Hermite model \citep{van_der_Marel_1993} and show the resulting kinematic maps in App.~\ref{app:Best-fit kinematics}.

\section{Strong lensing modeling}
\label{Sect:Strong lensing modeling}
In this work, we adopt the GPU-accelerated Gravitational Lensing and Dynamics (\texttt{GLaD}) \citep{2025GLaD} for lens modeling. The lensing module of \texttt{GLaD} is built upon the \texttt{GLEE} software \citep{2010SuyuHalkola_Dpie2, Suyu2012}. The goal of this modeling is to leverage the information from the extended images in two HST bands to refine the lens model of \citetalias{2018Collier}. In Sect.~\ref{sect:parameterised mass profiles}, we present the adopted lens mass profiles and modeling procedure. In Sect.~\ref{sect:Measurements of the mass within Einstein radius}, we present the inferred Einstein mass (i.e., mass enclosed within the Einstein radius) from the two lens models and compare them with the previous measurements reported by \citetalias{2018Collier}.
\subsection{Lensing mass profiles and modeling procedure}
\label{sect:parameterised mass profiles}
We adopt two lens mass models. The first assumes an isothermal mass distribution, represented by a Pseudo-Isothermal Elliptical Mass Distribution (PIEMD) \citep{1993Kassiola}. In this model, the dimensionless surface mass density $\kappa$ is expressed in Cartesian coordinates $(x, y)$ for a source at redshift $z_{\rm source}$:

\begin{equation}
\kappa_{\rm PIEMD} = \frac{\frac{\theta_{\rm E,piemd}}{1+q_{\rm piemd}}}{\sqrt{\frac{4 w^2}{(1+q_{\rm piemd})^2} + x^2 + \frac{y^2}{q_{\rm piemd}^2}}} ,
\end{equation}
where \(w = 10^{-5}\)\arcsec is the core radius, \(q_{\rm piemd}\) the axis ratio, and $\theta_{\rm E,piemd}$ is the normalisation of the PIEMD profile.

The second model is an elliptical power-law (EPL) profile \citep{EPL}, defined as
\begin{equation}
\kappa_{\rm epl}  = \left(\frac{3-\gamma_{\rm epl}}{2}\right) \left(\frac{b}{\sqrt{x^2 + y^2/q_{\rm epl}^2 + w^2}} \right)^{\gamma_{\rm epl} - 1},
\end{equation}
with
\begin{equation}
b = \left( \frac{2}{1+q_{\rm epl}} \right)^{\frac{1}{\gamma_{\rm epl}-1}} \theta_{\rm E,epl},
\end{equation}
where $q_{\rm epl}$ is the axis ratio, $\theta_{\rm E,epl}$ is the normalisation of the EPl profile, and $\gamma_{\rm epl}$ is the slope of the profile; the isothermal case corresponds to $\gamma_{\rm epl} = 2$. In this case, we keep the core radius with \(w = 10^{-5}\)\arcsec. Both PIEMD and EPL can be rotated by a position angle $\varphi_{\rm PA, piemd}$ or $\varphi_{\rm PA, epl}$ to account for the mass distribution’s orientation. In both cases, $\Rein$ is defined as the radius at which the azimuthally averaged convergence $\bar{\kappa}$  over the 2D lens plane $\boldsymbol{\theta}$ equals unity:
\begin{equation}
\bar{\kappa}(<\Rein) \equiv 
\frac{1}{\pi \Rein^2} 
\int_{|\boldsymbol{\theta^{\prime}}| < \Rein} 
\kappa(\boldsymbol{\theta}^{\prime}) \, d^2\theta^{\prime} 
 = 1.
\label{eq:Einstein_radius}
\end{equation}

The lens mass distribution is modeled with both EPL and PIEMD profiles, hereafter referred to as the EPL-ESR and PIEMD-ESR models. Since image B lies near the lensing centroid, we first subtract the lens light from the HST images by fitting the isophotes in both bands with \texttt{Photutils} \citep{Photutils}, masking the central dust region and other noisy areas, and fitting up to sixth-order isophotes. 

We reconstruct the extended source (ESR) intensity distribution using the two lensed images in the F814W and F336W bands as constraints (see Fig.~\ref{fig:ES0286_HST_imaging}). This requires a model of the PSF in each band, which we build using the PSF reconstruction and deconvolution software package \textsc{starred}\footnote{\url{https://gitlab.com/cosmograil/starred}} \citep{2023JOSS....8.5340M,2024AJ....168...55M}. Our PSF model is composed of a Moffat profile with pixelated corrections regularized with wavelets, jointly constrained by $\sim10$ suitable stars (i.e., not saturated and reasonable signal-to-noise ratio) selected in the vicinity of ESO0286. Extended image modeling is then performed on the lens-subtracted images by optimizing the lensing image likelihood, which is the Bayesian evidence of the source intensity reconstruction (see Eq.~39 in \cite{2025GLaD}; \cite{2006esr}) using the Metropolis-Hastings MCMC method. For comparison with \citetalias{2018Collier}, we also build another model by minimizing the differences between the predicted and observed two-image positions and flux ratio, again using MCMC for sampling. In all lensing models, the centroid is fixed to the kinematic map centroid.

\subsection{Measurements of the mass within Einstein radius}
\label{sect:Measurements of the mass within Einstein radius}
We compute the projected two dimensional (2D) enclosed mass within the Einstein radius $\Rein$ by multiplying the critical surface mass density $\Sigma_{\rm crit}$, with the convergence $\kappa$:
\begin{equation}
\Sigma = \Sigma_{\rm crit} \, \kappa,
\end{equation}
where the critical surface mass density is defined as
\begin{equation}
\Sigma_{\rm crit} = \frac{c^2 D_{\rm s}}{4 \pi G D_{\rm d} D_{\rm ds}}.
\end{equation}
The enclosed 2D mass is then obtained by summing all $\Sigma$ values within the an aperture $\mathcal{A}_{\rm Ein}$ of radius $\Rein$:
\begin{equation}
M_{\rm Ein}^{\rm lens} \;=\; \int_{\mathcal{A}_{\rm Ein}} \Sigma(\boldsymbol{\theta}) \, d^{2}\theta ,
\label{eq:M_ein}
\end{equation}
We randomly draw 3000 samples from each MCMC chain and compute the Einstein mass, $M_{\rm Ein}^{\rm lens}$. The resulting probability distributions for all models are shown in Fig.~\ref{fig:ES0286_HST_imaging}.

The distributions of $M_{\rm Ein}^{\rm lens}$ obtained from ESR and image+flux modeling largely overlap, but the ESR distribution is significantly narrower, as it incorporates roughly 900 pixels across two HST-imaging bands, whereas the image+flux approach provides only five constraints. Consequently, the image+flux modeling exhibits stronger parameter degeneracies. The two ESR models, however, produce disjoint distributions. While they differ in their specific mass density parameterizations, both models reproduce the observed flux ratios within the 1-$\sigma$ uncertainties reported by \citetalias{2018Collier}.

{The best-fit EPL-ESR model yields a slope of $\gamma_{\rm epl} = 2.27$, which is steeper than the typical isothermal value reported in SLACS and ATLAS-3D \citep{2010Auger,2010Treu,2013Cap,2015Poscacki,2021Shajib,DinosII}. Because the system provides only two images, one of which is affected by lens-light subtraction uncertainties and possible HST PSF residuals, the exact mass density slope is highly sensitive to these systematics and to the chosen source grid resolution. Therefore, rather than strictly favoring one model over the other, we treat the total enclosed mass as a more robust, fundamental quantity. To account for the systematic uncertainties between the different parameterizations, we adopt the average Einstein mass from the EPL-ESR and PIEMD-ESR models, yielding $\overline{M_{\rm Ein}^{\rm lens}} = 8.85 \pm 0.38 \times 10^{10}~\rm M_\odot$. The 1-$\sigma$ uncertainty is computed as
\begin{equation}
\sigma_{\rm lens} = \sqrt{ \left( \frac{M_{\rm Ein,EPL}^{\rm lens} - M_{\rm Ein,PIEMD}^{\rm lens}}{2} \right)^{2} 
+ \left( \frac{\sigma_{\rm EPL} + \sigma_{\rm PIEMD}}{2} \right)^{2} },
\end{equation}
where $\sigma_{\rm EPL}$ and $\sigma_{\rm PIEMD}$ are derived from the 16th, 50th, and 84th percentiles of the respective ESR modeling distributions. This provides a conservative estimate of systematic uncertainty

In \citetalias{2018Collier}, the authors reported an Einstein mass of $9.49 \pm 0.15 \times 10^{10}~\rm M_\odot$ at an Einstein radius of $2.38\arcsec$, both of which are higher than our estimates. In our analysis, the Einstein radius, averaged over the two mass assumptions, is $\overline{\theta_{\rm Ein}} = 2.32\arcsec$, corresponding to a 6.7\% lower Einstein mass. Their analysis relied solely on image positions and flux ratios as constraints. This value is nevertheless consistent with our results when we restrict our modeling to the same constraints (see Fig.~\ref{fig:ES0286_HST_imaging}). However, incorporating the information from the arc provides additional constraining power, which reduces the allowed parameter space and leads to a lower preferred Einstein mass.
\begin{table*}
  \caption{Strong lensing model parameters with associated 1-$\sigma$ uncertainties.}
\centering
\begin{tabular}{lccccc}
 \toprule \\[-0.3em]
Model & $q$ & $\rm PA~[\circ]$ & $\varphi~[\arcsec]$ & $\gamma$ & Flux ratio\\ 
\hline \\[-0.3em]
PIEMD-ESR & $0.83_{-0.01}^{+0.01}$ & $106_{-1}^{+1}$ & $2.36_{-0.01}^{+0.01}$ &- & 2.1\\
\hline  \\[-0.3em]
EPL-ESR & $0.63_{-0.05}^{+0.04}$ & $99_{-1.0}^{+1.2}$ & $2.28_{-0.03}^{+0.04}$   & $-2.28_{-0.04}^{+0.04}$ &2.2 \\
\bottomrule
\end{tabular}
\tablefoot{We show the median value with the corresponding 1-$\sigma$ uncertainty derived from the models that use the extended images as constraints. The predicted flux ratios from the best-fit models are consistent with the measured value of $2.2 \pm 0.1$ reported in \citetalias{2018Collier}.}
\label{tab:model_parameters}
\end{table*}

We do not include external shear in our mass model. External shear represents the tidal stretching induced by nearby galaxies, but ESO0286 is most likely located in an isolated environment; in particular, we do not observe any clear overdense region in the FoV of the HST imaging. Moreover, external shear does not contribute to the surface mass density of the lens but instead primarily affects the shape of the observed images. This effect can be mimicked by the axis ratio of the lens galaxy, leading to a strong degeneracy. \citetalias{2018Collier} used flux ratios and image positions to help break this degeneracy. In contrast, our modeling of the extended images reproduces the observed flux ratios directly and matches them with high accuracy (see Tab.~\ref{tab:model_parameters}). Thus, the difference in the inferred $M_{\rm Ein}$ might arise from degeneracies in the mass parameter space under different lensing constraints.



\section{Dynamical modeling}
\label{sect:Dynamical modeling}
We investigate the internal structure of ESO0286 using orbit-based Schwarzschild modeling. Section~\ref{Sect:Axisymmetric Schwarschild Modeling} provides an overview of the method, including the specific axisymmetric and triaxial codes employed in this analysis. Section~\ref{Sect:Luminosity deprojection} describes the deprojection procedure used to derive the intrinsic 3D luminosity density required for the gravitational potential. Finally, Sect.~\ref{Sect:dynamical Mass density profile} presents the composite mass model we adopted in the modeling.

\subsection{The Schwarschild technique}
\label{Sect:Axisymmetric Schwarschild Modeling}



\begin{figure}
    \centering
    \includegraphics[width=\columnwidth]{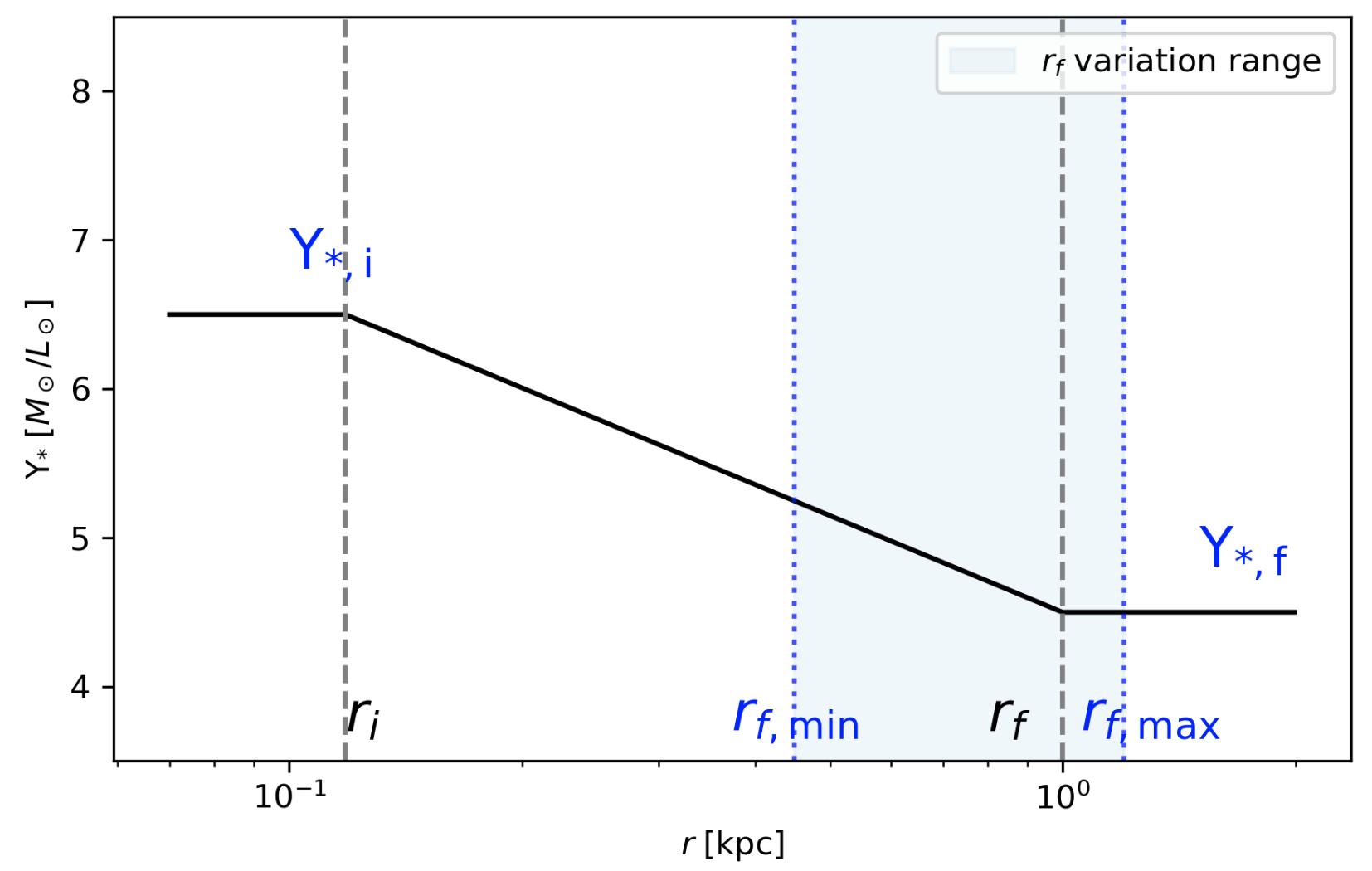}
    \caption{Setup for the mass distribution in the dynamical modeling, illustrating the implementation of the stellar mass-to-light ratio gradient. The inner radius $r_i$ is fixed at 0.1 kpc. The mass-to-light ratios $\Upsilon_{\ast,i}$ and $\Upsilon_{\ast,f}$, as well as the outer radius $r_f$, are allowed to vary during the dynamical modeling. As indicated by the shaded region and dotted lines, the variation range for $r_f$ is restricted to between $r_{f,\rm min} = 0.45$ kpc and $r_{f,\rm max} = 1.2$ kpc.}
    \label{fig:setup_dynmaics}
\end{figure}

We model ESO0286 with both axisymmetric and triaxial implementations of the Schwarzschild orbit-superposition method \citep{1979schwarzschild}. The axisymmetric version of the code that we use was first described in \citet{2004thomas}, and has been continuously improved since. For generalization to triaxiality, we employ the recently developed Schwarzschild code SMART, which was presented in \citet{Neureiter_2021} and validated in \citet{Neureiter_2023_B,deNicola_2022_B}. Both axisymmetric and triaxial implementations were further refined by \citet{2021Lipka,2022Thomas}, who introduced a data-driven scheme to optimize the level of regularization.  


Schwarzschild modeling offers a highly flexible and general framework for dynamical studies, underlying a minimal set of assumptions. These assumptions are well-justified for elliptical galaxies, which are typically gas-poor and dynamically relaxed over timescales much longer than the crossing time.

\begin{enumerate}
    \item The system is collisionless, satisfying the collisionless Boltzmann equation, i.e., the distribution function (DF) of stars in 6-dimensional phase space $(\mathbf{x}, \mathbf{v})$ obeys
    \[
    \frac{{\rm d} f(\mathbf{x}, \mathbf{v}, t)}{{\rm d}t} = 0,
    \]
    \item The system is in a steady state, such that
    \[
    \frac{\partial f}{\partial t} = 0,
    \]
    which implies that the DF depends only on the integrals of motion.
\end{enumerate}

Very briefly, the construction of a Schwarzschild model starts with establishing a candidate mass model of its 3D density distribution $\rho_{\rm tot}$ (Sect.~\ref{Sect:dynamical Mass density profile}), which embeds one's knowledge of the 3D light distribution that is inferred via deprojection (Sect.~\ref{Sect:Luminosity deprojection}). Given the candidate model density $\rho_{\rm tot}$, the corresponding gravitational potential $\Phi$ is derived and a representative set of orbits is integrated within $\Phi$. Each orbit in this set is assigned an adjustable weight $w_{i}$, and the weighted superposition of the orbits is fitted to the galaxy's observables. In our case, the observables are the 2D stellar kinematic information, such that the goodness-of-fit $\chi_{\rm kin}^2$ for a candidate model is evaluated by its deviation to the observed line-of-sight velocity distributions (LOSVDs). 
\begin{equation}
\chi_{\rm kin}^2 = \sum_{i=1}^{N_{\rm kin} \times N_{\rm vel}} 
\left( \frac{l_{{\rm obs,}\, i} - l_{{\rm mod,}\, i}}{\Delta l_{{\rm obs,}\, i}} \right)^2,
\label{Eq:chi2kin}
\end{equation}
where $N_{\rm kin}$ is the number of kinematic bins in the Voronoi bin map, and $N_{\rm vel}$ is the number of velocity bins per Voronoi bin. In this paper, we adopt $N_{\rm vel} = 33$ and $N_{\rm kin} = 62$ for the WFM kinematic data.

Since the number of orbits is usually larger than the number of constraints, the determination of the weights necessitates regularization, which is achieved by maximizing the entropy-like quantity:
\begin{equation}
    \hat{S} \equiv S - \alpha \chi_{\rm kin}^2,
    \label{Eq:entropy_kin}
\end{equation}
where $S$ is defined as:
\begin{equation}
  S =  - \sum_i w_i \ln \left( \frac{w_i}{\omega_i} \right).
  \label{eq:Boltzmann_aniso}
\end{equation}
Here $\alpha$ is the regularization parameter, $w_i$ is the weight of the $i$-th orbit, and $\omega_i$ are corresponding bias factors that can be used to bias the weights towards a specific solution. A well-motivated choice for the $\omega_i$ are the phase-space volumes $V_{i}$ of the orbit, such that $S$ becomes the Boltzmann entropy \citep{Thomas_2007_B}. This choice was made for the axisymmetric Schwarzschild implementation \citep{2004thomas}, whereas for the triaxial implementation, the $\omega_i$ are set to 1 \citep{Neureiter_2021} such that $S$ becomes the Shannon entropy. 


The number of free parameters in Schwarzschild modeling depends not only on the chosen mass parameters in $\rho_{\rm tot}$ but also on the weights assigned to the representative orbits. Since the number of orbits, and therefore the corresponding weights, typically exceeds the number of kinematic data points, the regularization parameter must be chosen carefully to avoid overfitting. To address this, \citet{2021Lipka} introduced an effective number of degrees of freedom, $m_{\rm eff}$, to penalize overly complex models. We therefore select the best-fit model by finding the minimal generalized Akaike Information Criterion ($\rm AIC$) across a list of $\alpha$ values \citep{2022Thomas}:

\begin{equation}
\mathrm{AIC}= \chi_{\rm kin}^2 + 2m_{\rm eff}.
\label{eq:AIC_kin}
\end{equation}

Schwarzschild modeling is computationally demanding; thus, MCMC sampling cannot be applied directly. Instead, to determine the best-fit $\rho_{\rm tot}$ in terms of the mass parameters, we construct a grid of candidate mass models spanning all probed parameters. Each mass parameter in $\rho_{\rm tot}$ is sampled with about 50–100 points. To search this grid efficiently, we employ the Nonlinear Optimization by Mesh Adaptive Direct Search (NOMAD) algorithm \citep[]{AuDe2006}. For a given set of mass parameters, the orbital weights are fitted to the observed LOSVD data. To determine the optimal regularization parameter $\alpha$, the Schwarzschild code evaluates a predefined list of $\alpha$ values and selects the model that minimizes the AIC. To ensure sufficient sampling of the mass parameters, we calculated at least 3000 models in the NOMAD grid for each model implementation listed in Tab.~\ref{tab:model_summary}.

Since the projected enclosed mass predicted at $\Rein$ should be consistent between strong lensing and dynamical modeling, we also explore the case where lensing constraints are explicitly included in Eq.~(\ref{eq:AIC_kin}). As a constraint, we adopt the predicted mass $\overline{M_{\rm Ein}^{\rm lens}} = 8.85 \times 10^{10}~\rm M_\odot$, averaged from the best-fit PIEMD-ESR and EPL-ESR mass models with uncertainty $\sigma_{\rm lens} = 3.8 \times 10^{9}~\rm M_\odot$. The modified $\rm AIC$ then becomes:

\begin{equation}
\mathrm{AIC}_{\rm tot}= \chi_{\rm kin}^2 + 2m_{\rm eff} + \frac{(M_{\rm Ein}^{\rm kin} - \overline{M_{\rm Ein}^{\rm lens}})^2}{\sigma_{\rm lens}^2},
\label{eq:AIC_tot}
\end{equation}
where $M_{\rm Ein}^{\rm kin}$ is the Einstein mass inferred from dynamical modeling, evaluated at the average Einstein radius $\overline{\theta_{\rm Ein}} = 2.32\arcsec$ of the best-fit EPL-ESR and PIEMD-ESR lens mass models. Following the standard 2D projected mass definition in strong lensing, we calculate the projected mass by integrating the 3D density $\rho_{\rm tot}$ along the line of sight $z$ using cylindrical coordinates:
\begin{equation}
    M_{\rm Ein}^{\rm kin} = 2\pi \int_{0}^{\overline{\theta_{\rm Ein}}} R \int_{-z_{\rm max}}^{z_{\rm max}} \rho_{\rm tot}(\sqrt{R^2 + z^2})~ dz\, dR,
    \label{eq:M_ein_kin}
\end{equation}
where $z_{\rm max} = \sqrt{r_{\rm max}^2 - R^2}$. Strictly speaking, $z_{\rm max}$ should extend to infinity; numerically, we adopt the 3D radius of  $r_{\rm max} = 300\arcsec$ in the galaxy frame, beyond which the contribution of $\rho_{\rm tot}$ is physically negligible.

\subsection{ Luminosity deprojection} 
\label{Sect:Luminosity deprojection}
In this section, we describe the derivation of the intrinsic 3D luminosity density from the observed surface brightness, performed first under the assumption of axisymmetry (see Sect.~\ref{Sect:Axisymmetric deprojection}) and subsequently within a triaxial framework (see Sect.~\ref{sect:Triaxial deprojection}).
\subsubsection{Axisymmetric deprojection}
\label{Sect:Axisymmetric deprojection}

We perform an axisymmetric deprojection of ESO0286 using the HST F814W-band surface brightness image. The 2D surface brightness distribution is modeled with isophotes, assuming a common centroid and position angle at all radii within the $30\arcsec \times 30\arcsec$ FoV, using \texttt{Photutils}. A total of 70 isophotes are fitted to the surface brightness over $\mathcal{O}(10^5)$ pixels. The region between $0.18\arcsec$ and $0.3\arcsec$ from the centroid, which is strongly affected by the dust lane, is excluded from direct fitting. Instead, we estimate the ellipticity and higher-order Fourier coefficients ($a_4$ and $a_6$) in this region by adopting a 1D linear interpolation.

Subsequently, we deproject the 2D surface brightness using the Metropolis algorithm of \citet{Magorrian1999}. The isophotes from \texttt{Photutils} are mapped onto an elliptical polar grid $(r, \theta)$ under the assumption of an oblate geometry. By penalizing deviations from smoothness across $r$ and $\theta$, as well as non-monotonic behavior toward the galaxy center, the algorithm identifies the 3D light density $\nu(r, \theta)$ that, when projected at a given inclination angle $i$, provides an acceptable fit to the observed isophotes.

We sample inclination angles $i$ such that the intrinsic flattening $q_{\rm intr}$ of ESO0286 is distributed linearly in stellar axis ratio. This strategy maximizes the efficiency of dynamical modeling by ensuring coverage of a representative range of intrinsic shapes of $\nu$. Specifically, we adopt six inclination angles: $66^\circ$, $70^\circ$, $73^\circ$, $76^\circ$, $80^\circ$, and $90^\circ$, corresponding to intrinsic flattenings $q_{\rm intr}$ increasing from 0.56 to 0.67 in steps of 0.02:
\begin{equation}
q_{\rm intr} = \sqrt{\frac{q'^{2} - \cos^{2}(i)}{\sin(i)}},
\end{equation}
where $q' = 0.67$ denotes the observed mean flattening of ESO0286 in the F814W band, obtained by averaging the fitted isophotal ellipticities at radii $>0.3\arcsec$, i.e., outside the region affected by the dust lane and the image PSF. At smaller radii, the surface brightness profile is strongly influenced by dust absorption and PSF effects, and is therefore excluded. We do not explore inclination angles below $66^\circ$, as this would require an intrinsic flattening of $q_{\rm intr} \lesssim 0.5$. This degree of flattening is physically unrealistic given the substantial total mass of $>10^{11}\ \mathrm{M}_\odot$ within the $\sim 5$~kpc FoV of ESO0286.
\begin{table*}
    \centering
    \caption{Summary of the dynamical mass models explored in this work.}
    \label{tab:model_summary}
    \renewcommand{\arraystretch}{1.3}
    \begin{tabular}{l l c c c l}
    \hline 
    \textbf{Model ID} & \textbf{Symmetry} & \textbf{$\Upsilon_{*,f}$} & \textbf{Constraints} & \textbf{DM Parameters} & \textbf{Motivation} \\
    \hline
    \textbf{I} & Axisymmetric & Free & K & Free $\gamma_{\mathrm{in}}$; $\gamma_{\mathrm{out}}=-3$ & Geometry test. \\
    \hline
    \textbf{\textcolor{model2}{II}} & Triaxial & Free & K & $\gamma_{\mathrm{in}} \in \{0, 0.5, 1, \mathrm{free}\}$; $\gamma_{\mathrm{out}}=-3$ & Intrinsic degeneracy, no lensing. \\
    \textbf{\textcolor{model3}{III}} & Triaxial & Free & K \& L & $\gamma_{\mathrm{in}} \in \{0, 0.5, 1, \mathrm{free}\}$; $\gamma_{\mathrm{out}}=-3$ & Lensing constraining power. \\
    \hline
    \textbf{\textcolor{model4}{IV}} & Triaxial & Fixed (4.5) & K & $\gamma_{\mathrm{in}} \in \{0, 1\}$; $\gamma_{\mathrm{out}}=-3$ & Test central SSP prediction \citep{2022Poci}. \\
    \textbf{\textcolor{model5}{V}} & Triaxial & Fixed (3.5) & K & $\gamma_{\mathrm{in}} \in \{0, 1\}$; $\gamma_{\mathrm{out}}=-3$ & Test outer SSP prediction \citep{2022Poci}. \\
    \hline
    \textbf{\textcolor{model6}{VI}} & Triaxial & Fixed (4.025) & K \& L & $\gamma_{\mathrm{in}} \in \{0, 1\}$; $\gamma_{\mathrm{out}}=-3$ & Standard halo with optimized $M_*/L$. \\
    \textbf{\textcolor{model7}{VII}} & Triaxial & Fixed (4.025) & K \& L & $\gamma_{\mathrm{in}}=0$; $\gamma_{\mathrm{out}}=-4$ & Stripped halo with optimized  $M_*/L$. \\
    \textbf{\textcolor[HTML]{5D1676}{VIII}} & Triaxial & Fixed (4.025) & K & $\gamma_{\mathrm{in}}=0$; $\gamma_{\mathrm{out}}=-4$ & Stripped halo with optimized $M_*/L$, no lensing. \\
    \hline
    \end{tabular}
    \tablefoot{The columns describe the assumed intrinsic symmetry, the treatment of the outer stellar mass to light ratio $\Upsilon_{*,f}$, the inclusion of strong lensing constraints $\overline{M_{\rm Ein}^{\rm lens}}$ (see Eq.~\ref{eq:AIC_tot}), the sampled DM halo parameters, and the primary motivation for each configuration. In the ``Constraints'' column, K denotes kinematic data from MUSE WFM, while K \& L indicates the inclusion of strong lensing constraints. The color of each Model ID corresponds to the individual AIC envelopes plotted in Figs.~\ref{fig:AIC_key_par} and \ref{fig:AIC_curves}.}
\end{table*}
\subsubsection{Triaxial deprojection}
\label{sect:Triaxial deprojection}

To obtain the triaxial form of the 3D luminosity distribution, which is required for the triaxial dynamical models, we again use the F814W-band surface brightness image, but now obtain the isophotes while allowing for a variable position angle. Strong isophote twists (a gradual radial change of the position angle) of $\gtrsim 20\degr$ are often a sign of triaxiality \citep{Franx_1991,Krajnovic_2011,Sanders_2015,Neureiter_2023}, since purely axisymmetric systems cannot produce such a twist. 

For the photometry of ESO0286, we did not find a strong sign of triaxiality in the photometry, as we only measure a very mild twist of $\sim4\degr$ between $3\arcsec$ and $9\arcsec$, which could also be produced by the projection of hidden secondary components such as disks or bars within an overall axisymmetric main body \citep{Scorza_1995,Jungwiert_1997,Barazza_2003}. 

To overcome the limitations of the axisymmetric approximation and capture the observed isophote twists, we performed a triaxial deprojection using the SHAPE3D code \citep{de_Nicola_2020}. Similar to the axisymmetric method, this algorithm allows us to probe different viewing angles ($\theta$, $\phi$, $\psi$) while accounting for the intrinsic deprojection degeneracy \citep{Rybicki_1987,Gerhard_1996,de_Nicola_2020} associated with a given orientation. Specifically, $\theta$ and $\phi$ determine the direction of the line of sight relative to the intrinsic principal axes of the galaxy. The third angle, $\psi$, defines the rotation of the galaxy around this line of sight, which dictates its projected position angle on the plane of the sky. In the constrained-shape implementation, the code assumes that the density is distributed on concentric ellipsoids:

\begin{equation}
m^{2-\xi(x)}=x^{2-\xi(x)}+\left(\frac{y}{p(x)}\right)^{2-\xi(x)}+\left(\frac{z}{q(x)}\right)^{2-\xi(x)}.
\label{eq:ellipsoid}
\end{equation}

These ellipsoids are fully defined by their density $\rho(x)$ along the major axis $x$, the intermediate and minor axis ratios $p(x)$ and $q(x)$, and the shape parameter $\xi(x)$ that quantifies diskiness or boxiness. Given a set of viewing angles and the surface brightness distribution, the code optimizes these four functions using a penalized non-parametric Metropolis approach.

Because the viewing angles recovered by \citetalias{2022Poci} ($\theta=88.05\degr$, $\phi=75.76\degr$, $\psi=75.76\degr$) imply ESO0286 is observed nearly along its intermediate axis, the standard triaxial deprojection has a specific intrinsic degeneracy \citep[e.g.,][]{de_Nicola_2020,de_Nicola_2022}. Specifically, variations in the line-of-sight flattening $p(x)$ leave the projected surface brightness nearly unchanged but can potentially have an impact on the inferred Einstein mass. Therefore, rather than blindly exploring various viewing angles, we restrict our test to the \citetalias{2022Poci} orientation. We instead marginalize over $p(x)$ in the range 0.8 to 0.95, with the goal of constraining it using our additional dynamical and lensing information.


\subsection{Mass density profile}
\label{Sect:dynamical Mass density profile}

We assume the total density of the lens galaxy $\rho_{\rm tot}(\boldsymbol{r})$,\footnote{For the axisymmetric modeling, $\boldsymbol{r}$ is expressed in terms of $r$ and $\theta$, while for the triaxial modeling, it is expressed in terms of $r$, $\theta$, and $\phi$.} which consists of the following components:
\begin{equation}
\rho_{\rm tot}(\boldsymbol{r}) = \Upsilon_\ast \, \nu(\boldsymbol{r}) + \rho_{\mathrm{DM}}(\boldsymbol{r}) + M_\bullet \, \delta(r)
\label{eq:rho_tot}
\end{equation}
where $\Upsilon_\ast$ is the stellar mass-to-light ratio, and $ \rho_{\mathrm{DM}}(\boldsymbol{r})$ presents the DM density, $M_\bullet$ is BH mass.  

In our modeling, we probe radially varying stellar mass-to-light ratios, and the constant stellar mass-to-light ratio is naturally a subset. Recently, the detailed study by \citet{2024Mehrgan} has shown that massive early-type galaxies (ETGs) display radial gradients in the stellar mass-to-light ratio on sub-kpc scales. Assuming a constant $\Upsilon_\ast$ can bias dynamical models, leading to stellar mass overestimates of up to $\sim$50\%. To address this, we allow for $\Upsilon_\ast$ gradients in our modeling. Specifically, we sample the inner and outer mass-to-light ratios, $\Upsilon_{\ast,i}$ and $\Upsilon_{\ast,f}$, independently at radii $r_i$ and $r_f$. Between these radii, $\Upsilon_\ast$ is interpolated log-linearly, while outside this range it is fixed to $\Upsilon_{\ast,i}$ and $\Upsilon_{\ast,f}$, respectively. We fix $r_i = 0.1~\rm kpc$ and allow $r_f$ to vary between $0.45~\rm kpc$ and $1.2~\rm kpc$, since \citet{2024Mehrgan} found that the gradients are confined to galaxy centers and occur on spatial scales of $r \approx 1~\rm kpc$ (see Fig.~\ref{fig:setup_dynmaics}).

We model the DM halo using a general parametrization based on the Zhao profile \citep[e.g., Eq. 1 in][]{1996Zhao} with the transition width parameter fixed at unity. The halo density profile at a given spatial position $\boldsymbol{r}$ is given by:
\begin{equation}
\rho_{\rm Zhao}(\boldsymbol{r}) = \frac{\rho_0}{\left( \frac{r}{\rs} \right)^{-\gamma_{\rm in}} \left( 1 + \frac{r}{\rs} \right)^{\gamma_{\rm in} - \gamma_{\rm out}} },
\label{Eq:zhao}
\end{equation}
where $\gamma_{\rm in}$ and $\gamma_{\rm out}$ denote the inner and outer logarithmic slopes respectively, $\rs$ is the scale radius marking the transition between these slopes, and $\rho_0$ is the density normalization parameter. The scalar variable $r$ represents the ellipsoidal radius. We consider the intrinsic shape of the halo in both axisymmetric and triaxial models. In the axisymmetric case, $q_{\rm DM}$ characterizes the intrinsic flattening of the halo. In the triaxial case, $q_{\rm DM}$ represents the flattening along the shortest axis while $p_{\rm DM}$ represents the flattening along the intermediate axis, such that the ellipsoidal radius is defined as $r^2 = x^2 + (y/p_{\rm DM})^2 + (z/q_{\rm DM})^2$.



\section{Model configuration and selection}
\label{sect:Model configuration and selection}
In this section, we describe the modeling strategy used to recover the intrinsic structure of ESO0286. In Sect.~\ref{sec:evidence_triaxiality}, we first present the kinematic evidence necessitating a triaxial modeling approach over standard axisymmetric assumptions. In Sect.~\ref{sec:mass_models}, we then describe the specific triaxial mass model configurations explored in this work, categorizing them by their treatment of the stellar and DM components. In Sect.~\ref{sec:bootstrap}, we outline the statistical framework, based on the AIC and bootstrap analysis, used to evaluate model performance and quantify uncertainties.
\subsection{Evidence for intrinsic triaxiality}
\label{sec:evidence_triaxiality}

\begin{figure}
  \includegraphics[width=1.\linewidth]{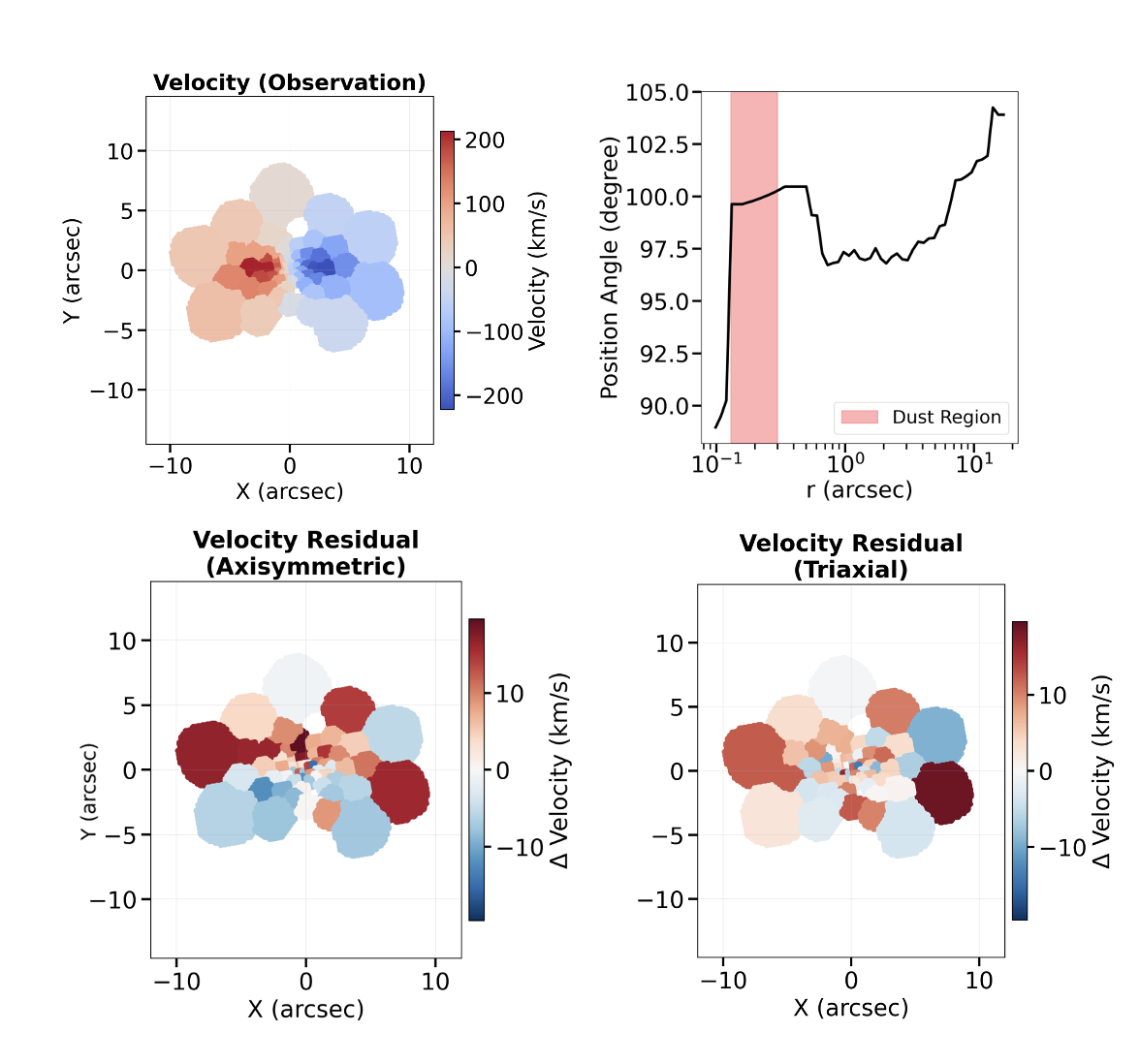}
  \caption{Kinematic data and modeling. The top row shows the observed MUSE rotation-velocity field (left) and the HST-derived isophotal position angle (right), with the red region corresponding to the dust-affected region. The bottom row displays velocity residuals for the best-fit axisymmetric model (left), revealing a systematic structure along the $y$-axis, and the triaxial model (right).}
  \label{fig:Vel_3D}
\end{figure}

To determine the intrinsic shape and mass distribution of ESO0286, we began by modeling the galaxy under the assumption of axisymmetry (Model I). We adopted this initial configuration because fast rotators are generally expected to possess an oblate intrinsic shape \citep{2014Weijmans, Cappellari2025}. For this setup, we employed an axisymmetric Schwarzschild code where we fixed $r_i = 0.1~\mathrm{kpc}$ and $\gamma_{\rm out} = -3$, while all other parameters were allowed to vary freely.

However, this preliminary analysis revealed significant systematic structures in the kinematic residuals. Specifically, we observed residual velocity features along the major axis that are indicative of rotation around the galaxy's long axis (see Fig.~\ref{fig:Vel_3D}). Such kinematic features are not possible in an axisymmetric geometry. Their presence in a fast rotator is rare and provides strong evidence that ESO0286 has a complex, intrinsically triaxial shape. This confirms the finding by \citetalias{2022Poci} that this galaxy is triaxial. Consequently, simple oblate models are insufficient, and a full triaxial modeling approach is required to accurately recover the mass distribution. Indeed, ESO0286 represents only the second such fast rotator reported in the literature to require this, following NGC 1453 \citep{2022Quenneville}.

\subsection{Overview of the triaxial mass models}
\label{sec:mass_models}

We extended our analysis to a comprehensive exploration of the parameter space using triaxial models. To systematically investigate the degeneracies between the stellar and DM halo mass distributions, we categorized our triaxial configurations (Models II--VII) into three main groups. The specific configurations are summarized in Tab~\ref{tab:model_summary} and defined as follows\footnote{Across all dynamical models, the parameter $\Upsilon_{*,i}$ is fixed to $0.1~\rm kpc$. Any subsequent parameters introduced without a specified fixed value are treated as free parameters during the modeling.}:

\begin{enumerate}
    \item \textbf{Exploratory triaxial models (Models II--III):} We explored the intrinsic parameter degeneracies within a fully triaxial framework. We treated $\Upsilon_{*,f}$ as a free parameter and sampled the DM halo inner slope from the set $\gamma_{\mathrm{in}} \in \{0, 0.5, 1.0, \mathrm{free}\}$. In all these configurations, the outer halo slope was fixed to the standard value of $\gamma_{\mathrm{out}} = 3$.  We ran these in two setups: without lensing constraints (\textbf{Model II}) and with lensing constraints (\textbf{Model III}), allowing us to measure how strongly the Einstein mass influences the dynamical solution.

  \item  \textbf{Exploratory SSP constraints (Models IV--V):} To explore the impact of SSP constraints on the dynamical mass breakdown, we ran triaxial models without lensing constraints using fixed $\Upsilon_{*, f}$ values directly taken from the spatially resolved SSP analysis of ESO0286 by \citetalias{2022Poci}. Their maps reveal a clear radial gradient in the stellar mass-to-light ratio, decreasing from $\approx 4.5$ in the central region down to $\approx 3.5$ in the outskirts. Fixing the outer stellar mass-to-light ratio, $\Upsilon_{*, f}$, allows us to break the inherent degeneracy between the stellar mass and the DM halo profile, particularly in the region where DM dominates. Without this constraint, different combinations of $\Upsilon_{*, f}$ and halo parameters can conspire to reproduce the same total mass distribution. For each fixed $\Upsilon_{*, f}$ assumption, we tested two representative DM halo inner slopes with $\gamma_{\mathrm{in}} \in \{0, 1\}$, keeping the outer slope fixed to $\gamma_{\mathrm{out}} = -3$:

\begin{itemize}
    \item \textbf{Model IV (Central SSP constraint):} $\Upsilon_{*, f}$ fixed to 4.5. This value is directly adopted from the higher stellar mass-to-light ratio measured in the central region of the galaxy by \citetalias{2022Poci}.
    \item \textbf{Model V (Outer SSP constraint):} $\Upsilon_{*, f}$ fixed to 3.5. This value is directly adopted from the lower mass-to-light ratio measured in the outskirts of the galaxy by \citetalias{2022Poci}.
\end{itemize}

 \item \textbf{Optimized halo geometry tests (Models VI--VIII):} Finally, to probe the detailed structure of the halo, we fixed the stellar component to an optimized value of $\Upsilon_{*,f} = 4.025$. This value corresponds to the average best-fit derived from the free stellar mass-to-light ratio configuration with lensing constraints (Model III). Within this optimized stellar mass setup, we explore specific variations in the halo geometry both with and without the inclusion of strong lensing data:
\begin{itemize}
    \item \textbf{Model VI (standard halo):} Includes lensing constraints and tests standard inner slopes ($\gamma_{\mathrm{in}} \in \{0, 1\}$) while fixing the outer slope to the canonical value ($\gamma_{\mathrm{out}} = -3$).
    \item \textbf{Model VII (steep outer slope):} Includes lensing constraints and tests a truncated halo scenario with a flat core ($\gamma_{\mathrm{in}} = 0$) and a steeper outer slope ($\gamma_{\mathrm{out}} = -4$), mimicking a more rapid density decline typical of stripped halos.
    \item \textbf{Model VIII (steep outer slope without lensing):} Identical to Model VII ($\gamma_{\mathrm{in}} = 0$; $\gamma_{\mathrm{out}} = -4$), but relies exclusively on kinematic constraints. Comparing this configuration to Model VII allows us to directly isolate the constraining power of the strong lensing data on the inferred halo truncation.
\end{itemize}
\end{enumerate}
After optimizing all different mass models, we found that the lowest AIC values cluster around 1000. When fitting both the kinematics and lensing data, Model VII emerges as the best model with $\rm AIC_{\rm tot} = 1002$. It yields a dynamical Einstein mass of $M_{\rm Ein}^{\rm kin} = 9.14 \times 10^{10}~\rm M_\odot$, which is consistent with the lensing prediction within $1\sigma$ uncertainty.
\begin{figure}
    \centering
    \includegraphics[width=\columnwidth]{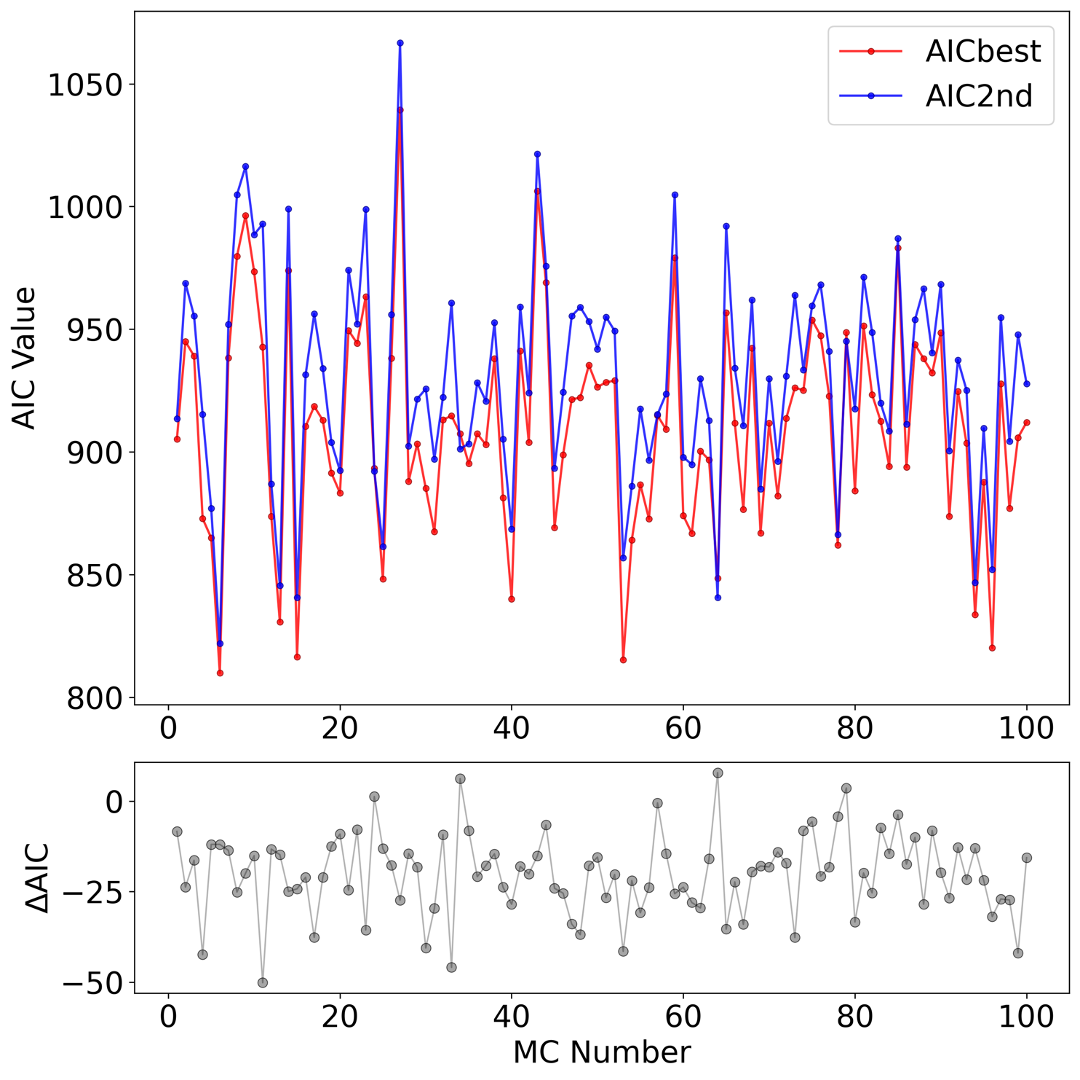}
    \caption{ Statistical robustness check using bootstrapping. \textit{Top panel:} The distribution of absolute AIC values for the global best-fit model (red line) and the second-best model (blue line) across 100 Monte Carlo noise realizations. The x-axis represents the individual Monte Carlo runs, and the y-axis shows the resulting AIC value for each fit. \textit{Bottom panel:} The difference in AIC values ($\Delta \mathrm{AIC} = \mathrm{AIC}_{\mathrm{best}} - \mathrm{AIC}_{\mathrm{2nd}}$) calculated for each corresponding realization in the top panel. The grey points indicate the specific $\Delta \mathrm{AIC}$ value obtained for each noise iteration.}
    \label{fig:mc_noise}
\end{figure}


\begin{figure*}
    \centering
    \includegraphics[width=1.7\columnwidth]{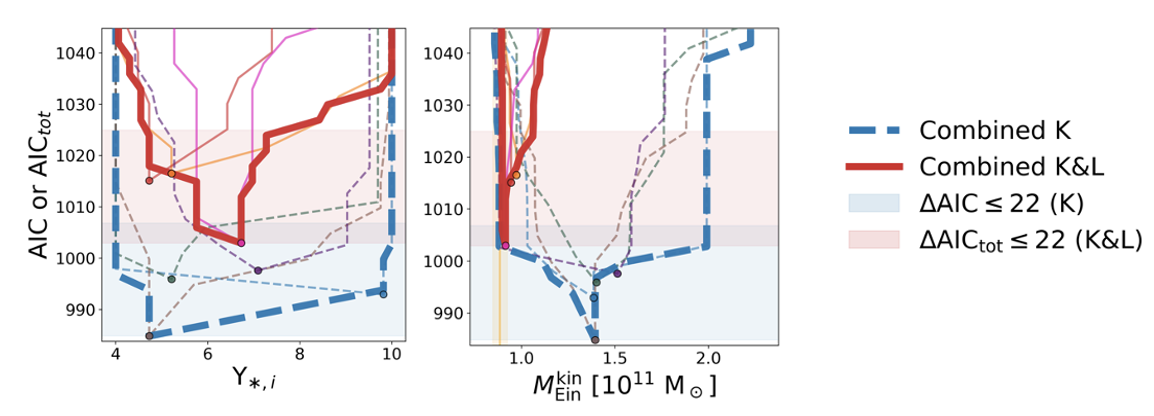}
    \caption{AIC envelopes for $\Upsilon_{*,i}$ and $M_{\rm Ein}^{\rm kin}$ based on all triaxial mass models. The thin colored lines show the AIC envelopes for individual model classes, with solid points marking the global best-fit value for each specific model family (in the corresponding color of Model ID in Tab.~\ref{tab:model_summary}). The thick dashed blue and thick solid red lines represent the overall combined envelopes for the kinematics-only (K) and kinematics+lensing (K$\&$L) categories, respectively. Shaded horizontal regions denote the $2\sigma$ statistical uncertainty threshold derived from bootstrapping for the K (light blue) and K$\&$L (light red) models. The vertical orange line and its corresponding shaded region in the $M_{\rm Ein}^{\rm kin}$ panel represent the independent mass measurement and uncertainty from the lensing-only model.}
    \label{fig:AIC_key_par}
\end{figure*}

\subsection{Statistical criterion and model selection}
\label{sec:bootstrap}

Since our grid-based Schwarzschild modeling does not involve direct sampling of the posterior (e.g., via MCMC), we employ a parametric bootstrap approach to assess the statistical robustness of our model selection and estimate uncertainties.

To quantify the expected scatter in the goodness-of-fit statistics, we generated mock kinematic datasets based on our two best-performing mass configurations: the global best-fit model for the class Model II and the second-best model in the same class\footnote{We evaluated these statistics using the dynamics-only models. This noise floor is directly applicable to the models with lensing constraints, as the additional data simply contributes a constant offset to both $\mathrm{AIC}_{\mathrm{best}}$ and $\mathrm{AIC}_{\mathrm{2nd}}$, preserving the relative scatter.}. For each configuration, we created 200 Monte Carlo (MC) realizations by perturbing the best-fit kinematic maps with random Gaussian noise consistent with the observational errors. We then re-fitted these 200 mock datasets using the same mass parameters of the original fit, but optimise the model in terms of the orbits weight and the regularization parameter. We then analyzed the distribution of the resulting AIC values as shown in Fig.~\ref{fig:mc_noise}.

We observe that the scatter of the absolute AIC values for both datasets is approximately $\sigma_{\mathrm{AIC}} \approx 42$. This empirical scatter is consistent with theoretical expectations, where the scatter is given by $\sigma = \sqrt{2 N_{\mathrm{dof}}}$. Here, $N_{\mathrm{dof}}$ represents the effective degrees of freedom, which approximates the mean $\chi^2$ of the fit. In our case, the average $\chi^2$ is $\approx 670$, yielding a theoretical expectation of $\sigma_{\mathrm{AIC}} \approx \sqrt{2 \times 670} \approx 37$. The close agreement between the observed and expected scatter confirms that our error budget is realistic. To establish a robust threshold for model selection, we compute the dispersion of the AIC differences between the best and second-best configurations across our MC realizations. We find a significantly reduced scatter of $\sigma_{\Delta \mathrm{AIC}} \approx 11$. This marked reduction from the absolute scatter is expected: models evaluated against the same noise realization are highly correlated. While a specific noise pattern shifts the absolute goodness-of-fit systematically, the relative differences between models remain highly stable \citep{2022Thomas}. Consequently, we interpret our empirically derived $\Delta \mathrm{AIC} \approx 11$ as an approximate $1\sigma$-equivalent scale that characterizes the dominant statistical scatter in our model selection. Accounting for the discrete sampling of the parameter grid and for changes in orbit construction induced by this discreteness, we adopt a more conservative $2\sigma$-equivalent threshold ($\approx 95\%$) to define our uncertainty range. Specifically, we select models satisfying $\Delta \mathrm{AIC} < 22$ for models with kinematics-only constraints or $\Delta \mathrm{AIC}_{\mathrm{tot}} < 22$ if lensing constraints are included. We emphasize that this $2\sigma$ value does not represent a posterior uncertainty derived from an MCMC chain; rather, it corresponds to the empirical dispersion of AIC differences measured from MC realizations of perturbed kinematic data and serves as an effective $\sigma$-equivalent criterion for model selection.

To apply this threshold, we separate the triaxial mass models into two categories: those constrained by kinematics alone (K) and those that incorporate both kinematic and lensing constraints (K$\&$L). For each category, we identify the absolute minimum AIC value independently and consider all models falling within $\Delta \mathrm{AIC} \le 22$ or $\Delta \mathrm{AIC}_{\mathrm{tot}} \le 22$ of their respective minima to be statistically equivalent (see Fig.~\ref{fig:AIC_key_par}, and detailed in App.~\ref{app:The AIC envelope for all parameters across the models}). As expected, the minimum AIC value obtained from the kinematic-only fits is lower than $\mathrm{AIC}_{\mathrm{tot}}$, since the latter includes the additional contribution from the lensing data and their associated measurement uncertainties. Applying these independent thresholds yields a final robust sample of 266 K models and 31 K$\&$L models, which we use to derive the resulting mass distributions.

Fig.~\ref{fig:AIC_key_par} illustrates the statistical power gained by combining dynamical and lensing constraints through the resulting AIC envelopes of the mass parameters. When using kinematic data alone, the AIC envelopes for key parameters such as the inner stellar mass-to-light ratio $\Upsilon_{\ast,i}$ remain relatively flat over a wide parameter range, indicating substantial degeneracies and weak constraints. The inclusion of lensing information significantly reduces these degeneracies and leads to much tighter AIC envelopes. For example, Models II and III differ only in the applied constraints, yet the model that includes lensing yields a markedly narrower allowed range for $\Upsilon_{\ast,i}$. The envelopes of $\Upsilon_{\ast,i}$ further indicate that fixing the outer stellar mass-to-light ratio $\Upsilon_{\ast,f}$ is well motivated (e.g., Models IV to VII). In the outer regions, where the mass distribution is dominated by DM, $\Upsilon_{\ast,f}$ becomes highly degenerate with the DM halo parameters and is therefore expected to correlate with $\Upsilon_{\ast,i}$. Consistent with this expectation, models with fixed $\Upsilon_{\ast,f}$ yield significantly tighter AIC envelopes than those where this parameter is left free. As shown in Fig.~\ref{fig:AIC_key_par}, the Einstein mass exhibits significant scatter in models without lensing constraints, indicating that it is not-well constrained. A detailed discussion of the Einstein mass constraints is provided in Sect.~\ref{subsect:total_mass}.

\section{Results and discussion}
\label{sect:Results}

In this section, we discuss the results from all models displayed in Tab.~\ref{tab:model_summary}. We treat them essentially as two categories: K and K$\&$L. We begin by analyzing the total density distribution and Einstein mass estimates in Sect.~\ref{subsect:total_mass}, followed by a discussion of the stellar mass-to-light ratio and IMF constraints in Sect.~\ref{subsect:imf} Finally, we discuss the properties of the DM halo in Sect.~\ref{subsect:dm}.

\subsection{Total density distribution and Einstein mass}
\label{subsect:total_mass}

ESO0286 has a measured Einstein radius of $R_{\rm Ein} = 1.44$ kpc (see Sect.~\ref{sect:Measurements of the mass within Einstein radius}) and an effective radius of $R_{\rm eff} = 2.15$ kpc (see \citetalias{2022Poci}). The WFM IFU data extend to approximately $2.5\,R_{\rm eff}$.

\begin{figure}[t!]
    \centering
\includegraphics[width=0.8\columnwidth]{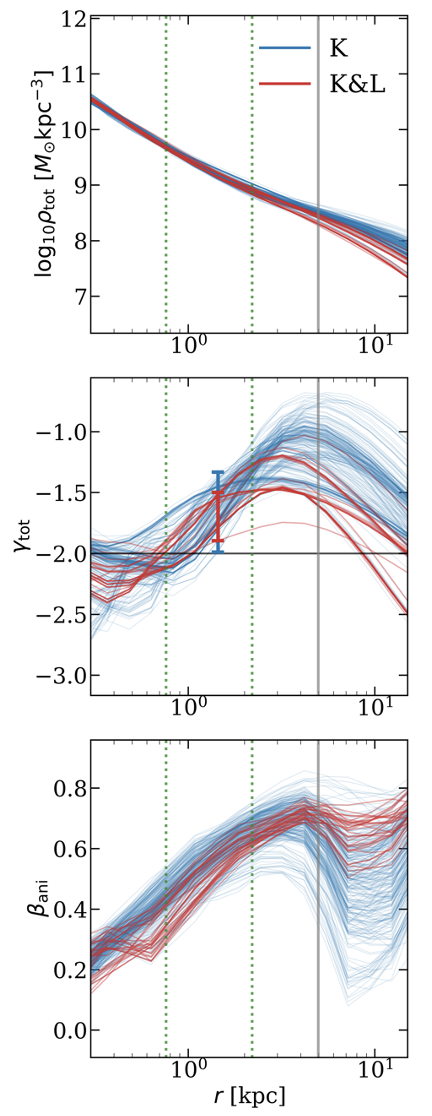}
    \caption{Radial profiles of the total density (top), total density slope (middle), and anisotropy (bottom) for ESO0286. The vertical error bars in the middle panel indicate the range of slope values at $R_{\rm Ein}$. In all three panels, models selected using $\Delta \mathrm{AIC} < 22$ from the combined model categories based on kinematic constraints only are shown by the blue curves. Models satisfying $\Delta \mathrm{AIC}_{\rm tot} < 22$, including both kinematic and lensing constraints, are shown by the red curves. The green dotted vertical lines indicate the lensed image positions, while the gray vertical line marks the spatial extent $\sim 5$~kpc of the IFU FoV.
  }
    \label{fig:total_mass}
\end{figure}
As shown in Fig.~\ref{fig:total_mass}, the total density profiles, $\rho_{\rm tot}$, of ESO0286 predicted by the K and K\&L models are in excellent agreement out to the radial extent of the IFU data ($\sim 5$ kpc). Within this region, the K\&L models form a clear subset of the broader K model family, indicating that the inferred mass distribution is largely insensitive to the inclusion of lensing constraints. The total density slope, $\gamma_{\rm tot}$, exhibits a pronounced radial variation, transitioning from a steep inner value of $\sim -2.2$ to a shallower value of $\sim -1.5$ at larger radii. For both model families, the predicted slopes are on average slightly shallower than the isothermal case, with $\langle \gamma_{\rm tot} \rangle \sim -1.7$ at $R_{\rm Ein}$. This behaviour suggests that ESO0286 departs from a simple isothermal or single power-law profile and instead exhibits a radially varying mass distribution that is robust to the choice of dynamical model.

We note that the value of $\gamma_{\rm tot}$ at $R_{\rm Ein}$ inferred from the dynamical modelling differs from the SL-only prediction, $\gamma_{\rm EPL} = -2.28^{+0.04}_{-0.04}$. However, ESO0286 produces only two lensed images, and in such systems, the SL constraints primarily fix the enclosed mass within the Einstein radius rather than the radial density slope. The apparently tight uncertainty on $\gamma_{\rm EPL}$ therefore reflects the restrictive nature of the assumed parametric mass model, rather than a robust measurement of the true total density slope. This is further supported by the fact that, in our analysis, the SL information is used only through the Einstein mass as a complementary constraint to the kinematic data when probing the inner mass structure of ESO0286.

While the mass distribution and orbital structure are well constrained within the region covered by the IFU data, the two model classes diverge at larger radii. The K models 
exhibit a large scatter in their predicted Einstein masses, demonstrating that the kinematic FoV is not sufficient to tightly reproduce the independent SL constraint in this galaxy (Fig.~\ref{fig:AIC_key_par}, right panel). More specifically, without the SL constraint, the majority of the K models tend to place additional mass in the halo. This preference for large outer masses could reflect slight inaccuracies in the kinematic data, the sparse sampling of the posterior distribution  of the model parameters or a bias related to the maximum-entropy regularisation. Excess mass at large radii drives the outer orbital structure toward isotropy ($\beta_{\rm ani}\approx 0$).

The K\&L models, by contrast, favor radially dominated orbits in the outskirts and successfully recover the lensing-derived Einstein mass. These results underscore the value of joint constraints: strong lensing provides a crucial complementary anchor for the total mass at large radii, which in turn narrows the range of allowed orbital anisotropies. The poor constraint on the Einstein mass in the K-only models also suggests that IFU data extending to $\sim 2.5~R_{\rm eff}$ may simply not reach far enough to robustly pin down the outer mass profile. One additional source of uncertainty in our modeling is how kinematic errors are treated. By assuming that each spatial and velocity bin is independent, we neglect the covariance matrix and any spatial correlations, which leads to an underestimate of $\chi^2_{\rm kin}$ and can consequently bias the AIC-based model selection.

This galaxy was also studied by \citetalias{2022Poci}, who successfully recovered the Einstein mass to within $5\%$ using kienmatic data only which is a reassuring point of consistency with our results, given that our reference Einstein mass is $6.7\%$ lower than the value they adopted. The two frameworks are broadly complementary rather than directly comparable: whereas 
\citetalias{2022Poci} constrained their models using the kinematic moments ($V_{\rm rot}$, 
$\sigma_{\rm vel}$, $h_3$, $h_4$), our approach fits the full LOSVDs and explores a 
considerably broader dynamical parameter space, spanning variations in the stellar mass-to-light gradient, the inner and outer DM halo slopes, and the inclusion of lensing constraints 
(see Table~\ref{tab:model_summary}).

\subsection{Stellar IMF and its implication}
\label{subsect:imf}
\begin{figure*}[t!]
    \centering
\includegraphics[width=2.0\columnwidth]{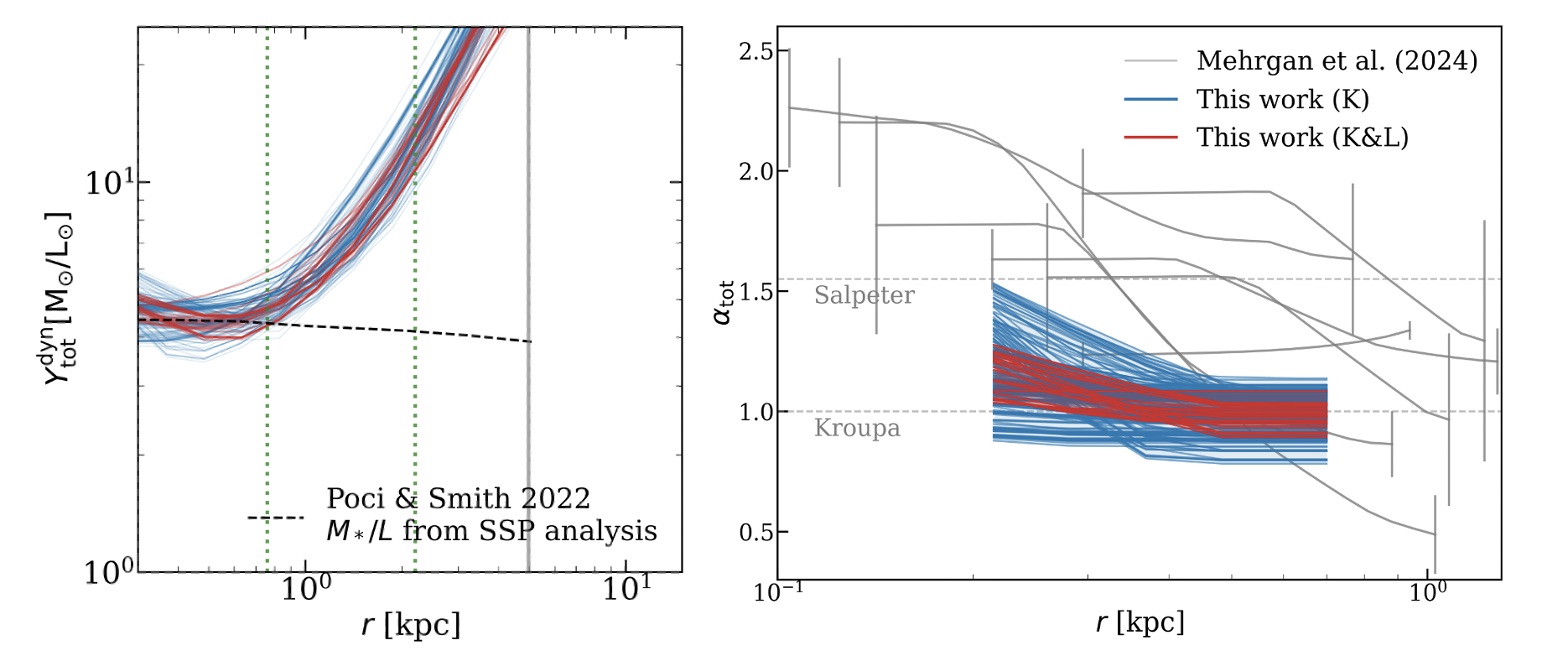}
    \caption{Constraints on the stellar IMF. \textit{Left:} Total mass-to-light ratios derived from the K (blue) and K\&L (red) models. The black dashed line indicates the stellar mass-to-light ratio $\Upsilon_{\ast}^{\rm SSP}$ in the F814W band, as measured by \citetalias{2022Poci} via SSP modeling assuming a Kroupa IMF. \textit{Right:} Comparison of the mismatch parameter $\alpha_{\rm tot}$ for ESO0286 against a sample of nearby ETGs from \citet{2024Mehrgan}. For ESO0286, the central velocity dispersion is $\sigma_{\rm vel} = 310$ km/s. At $r = 0.7$ kpc, our K models yield a mean $\alpha_{\rm tot} = 1.0 \pm 0.08$, while the  K\&L models constrain this to $\alpha_{\rm tot} = 1.0 \pm 0.05$.}
    \label{fig:IMF}
\end{figure*}

In this section, we present constraints on the stellar IMF and compare our results with those of other ETGs. The most stringent constraint on the IMF is provided by the minimum value of the spherically averaged total dynamical mass-to-light ratio, $\Upsilon_{\rm tot}^{\rm dyn}(r)$ \citep{2024Mehrgan}. This minimum establishes a strict upper boundary for the stellar mass-to-light ratio, $\Upsilon_*$. To derive $\Upsilon_{\rm tot}^{\rm dyn}(r)$ strictly as a function of the intrinsic radius $r$, we compute the spherical average of the 3D mass and luminosity distributions. By rearranging Eq.~(\ref{eq:rho_tot}) and dividing by the spherically averaged luminosity density $\nu$ we obtain\footnote{We note that our models include a BH in addition to the stars and DM halo. Given the relatively poor central resolution of the kinematic data, uncertainties in the BH mass can interfere with the detailed inner profiles of the other two components. However, the sphere of influence of a $5\times10^{9} M_{\odot}$ BH is $< 0.2$~kpc and, thus, small compared to the radii of interest for the IMF measurement.}:

\begin{equation}
    \frac{\rho_{\rm tot}}{\nu} = \Upsilon_\ast + \frac{\rho_{\rm DM}}{\nu}.
    \label{eq:total_mass_to_light}
\end{equation}
Since the DM density must be physically non-negative ($\rho_{\rm DM} \ge 0$), the term $\rho_{\rm DM}/\nu$ is always non-negative. This implies that the spherically averaged total dynamical mass-to-light ratio must strictly exceed the stellar component, yielding the inequality:
\begin{equation}
    \Upsilon_{\rm tot}^{\rm dyn}(r) \ge \Upsilon_*.
\end{equation}
As shown in the left panel of Fig.~\ref{fig:IMF}, the modeled $\Upsilon_{\rm tot}^{\rm dyn}$ from both K and K\&L model families exhibit a `V'-shaped trend. The distributions reach a minimum at $r_{\rm min} \approx 0.5-0.7$~kpc. This radius marks the point where the total dynamical mass most closely traces the luminous matter, providing the tightest constraint on the stellar mass-to-light ratio in the extreme limiting case. The rise in $\Upsilon_{\rm tot}^{\rm dyn}$ at smaller radii may be driven by a central DM cusp or a bottom-heavy IMF, while the increase in the outer regions reflects the expected transition to a DM-dominated halo. Furthermore, the minimum of our modeled $\Upsilon_{\rm tot}^{\rm dyn}$, which sets the upper limit for $\Upsilon_*$, is consistent with the value inferred from SSP modeling by \citetalias{2022Poci} where a Kroupa IMF was assumed. This agreement suggests that the stellar population in the main body of ESO0286 (at $r > r_{\rm min}$) is consistent with a Kroupa-like IMF \citep{2024Mehrgan}.

We quantify the radial variation of the stellar IMF by calculating the total mismatch parameter, $\alpha_{\rm tot} = \Upsilon_{\rm tot}^{\rm dyn}/\Upsilon_{\ast}^{\rm SSP}$. We trace this profile out to the radius of minimum mass-to-light ratio, $r_{\min}$, beyond which we assume the parameter remains constant. We adopt $\Upsilon_{\rm tot}^{\rm dyn}$ rather than $\Upsilon_{\ast}$ for this calculation because it provides the most robust diagnostic of the IMF upper boundary. Because the total dynamical mass must be greater than or equal to the stellar mass (i.e., the DM fraction cannot be physically negative), $\alpha_{\rm tot}$ represents a strict upper limit on the true stellar IMF normalization. As illustrated in the right panel of Fig.~\ref{fig:IMF}, evaluating this strict upper boundary at $0.7$~kpc strongly favors a Kroupa IMF. The slightly higher upper limit in the central regions accommodates a slightly more bottom-heavy IMF, aligning with the Salpeter-like IMF favored by the stellar population analysis of \citet{2025Poci} in the innermost region based on NFM data.

To place ESO0286 in context, we compare our results with the sample of non-lensing elliptical galaxies analyzed by \citet{2024Mehrgan}. In that study, seven systems at $z < 0.02$ with central velocity dispersions in the range $\sigma_{\rm vel} = 230$--$320~\mathrm{km\,s^{-1}}$ were modeled using axisymmetric Schwarzschild methods applied to both SINFONI adaptive-optics and wide-field MUSE data. The central $\sigma_{\rm vel}$ of ESO0286 falls within this same range. As shown in the right panel of Fig.~\ref{fig:IMF}, the radial trend of the IMF in ESO0286 is generally consistent with the ETGs analyzed by \citet{2024Mehrgan}: the galactic center exhibits a bottom-heavy IMF which transitions to a lighter IMF in the main body. However, the radial gradient in ESO0286 is notably shallower, and the transition radius $r_{\rm min}$ is smaller than in the comparison galaxies. We note that the kinematic data of the sample of \citet{2024Mehrgan} has about ten times higher spatial resolution near the center. The weaker evidence for a central increase in the mass-to-light ratio of ESO0286 could also reflect the relatively poor central resolution of the kinematic data.

However, for lensing galaxies at intermediate redshifts from the Sloan Lens ACS Survey \citep[SLACS;][]{2006Bolton} and the SL Legacy Survey \citep[SL2S;][]{2012More, 2012Gavazzi}, where the lens galaxies are located in 
the range of $0.1<z<1$, analyses generally favor a bottom-heavy IMF, consistent with a Salpeter form, based on combined lensing and dynamical modeling \citep[e.g.,][]{2010Auger, 2010Treu, 2021Shajib, DinosII}. This raises the question of whether the apparent tension in the favored IMF for elliptical galaxies is driven by redshift evolution or by differences in velocity dispersion, since the majority of SLACS and SL2S lens galaxies have $\sigma_{\rm vel} \sim 250~\rm km\,s^{-1}$, whereas ESO0286 lies at the high end of this population with $\sigma_{\rm vel} = 310~\rm km\,s^{-1}$.  

From the observational side, \citet{DinosII} studied 56 lens galaxies at $0.09 < z < 0.844$ from SLACS and SL2S and found a negative evolution trend in the stellar-to-total mass ratio with redshift. To infer the IMF, they compared stellar masses derived from lensing and dynamical modeling with those estimated from multi-band photometry under different IMF assumptions \citep{2009Auger}. Due to the redshift dependence of the stellar mass-to-light ratio, the preferred IMF in their sample also evolves with redshift, becoming increasingly heavier at lower redshift. This trend stands in tension with our findings for ESO0286 at $z_{\rm lens} = 0.0312$, as well as for another nearby lens ES0325-G004 at $z = 0.035$ \citep{2018Collett} and the seven ETGs studied by \citet{2024Mehrgan}, all of which favor a Kroupa IMF in the region $r \gtrsim 1~\rm kpc$.

In contrast, \citet{2017Sonnenfeld} used a simple dry merger evolution model based on cosmological N-body simulations, applied to mock galaxy samples, and found that the IMF normalization of individual galaxies becomes lighter with time. This matches the intuitive expectation that the outer stellar envelopes of ellipticals are built up via dry mergers with surrounding satellites, which themselves typically have a Kroupa IMF.

Regarding the dependence of the IMF on $\sigma_{\rm vel}$, our measurement of ESO0286 further confirms the tension noted by \citet{2024Mehrgan} and observed across all nearby lenses from the SNELLS survey: the commonly reported trend that higher $\sigma_{\rm vel}$ correlates with a more bottom-heavy IMF in elliptical galaxies \citep[e.g.,][]{2010Auger, 2010Treu, 2011Thomas, 2013Cap, 2015Poscacki} may not hold universally.

Interestingly, the distinctive `V'-shaped radial behavior of the total mass-to-light ratio is not exclusive to massive ETGs. Recent orbit-superposition modeling of dwarf ellipticals (dEs) by \citet{2024Lipka} and \citet{2024Lipka2} revealed remarkably similar radially resolved gradients at an entirely different mass scale, exhibiting the exact same pattern: a central enhancement, a minimum at intermediate radii, and a subsequent rise in the outskirts. While the matching outer rise in both giant ETGs and dEs is fully expected, as it is universally driven by the increasing dominance of the DM halo, it is the striking similarity of the central increase that is particularly surprising. Although a bottom-heavy IMF is a plausible explanation for the central peak in massive ETGs like ESO0286, universally applying this interpretation to dwarf galaxies is less physically motivated given their distinct formation histories. Instead, the recovery of a Kroupa-like IMF from our combined K\&L models strengthens the case that the stellar initial mass function resembles that of the Milky Way throughout the main body of even massive early-type galaxies. If the IMF does indeed vary, our results suggest these variations are strictly confined to the innermost regions. However, whether a true physical variation in the IMF exists at the very center ultimately remains an open question.

In summary, we find that a subset of nearby elliptical galaxies does not follow the previously reported correlation between the IMF, redshift, and $\sigma_{\rm vel}$ inferred from lensing and dynamical studies. One possible explanation is that differences in mass-model assumptions, particularly whether a radially varying $\Upsilon_{\ast}(r)$ is adopted, can strongly affect the inferred trend \citep{2024Mehrgan}. In addition, our constraints on $\Upsilon_{\ast}(r)$ are driven by the innermost regions ($< 1~\rm kpc$), a scale below the effective resolution of intermediate-redshift lensing galaxies, where both lensing sensitivity and kinematic constraints are weaker. Consistently, \citet{DinosII}, who introduced a stellar mass-to-light gradient parameter in a hierarchical analysis of the SLACS and SL2S samples, reported negligible evidence for such a gradient at the population level. Interestingly, however, one of their strongest outliers, SDSSJ2302$-$0840, favors a Chabrier or even lighter IMF, and this system lies at the lowest redshift at $z = 0.09$ in the sample.

\subsection{Dark matter halo properties} 
\label{subsect:dm}

\begin{figure}[t!]
    \centering
\includegraphics[width=0.8\columnwidth]{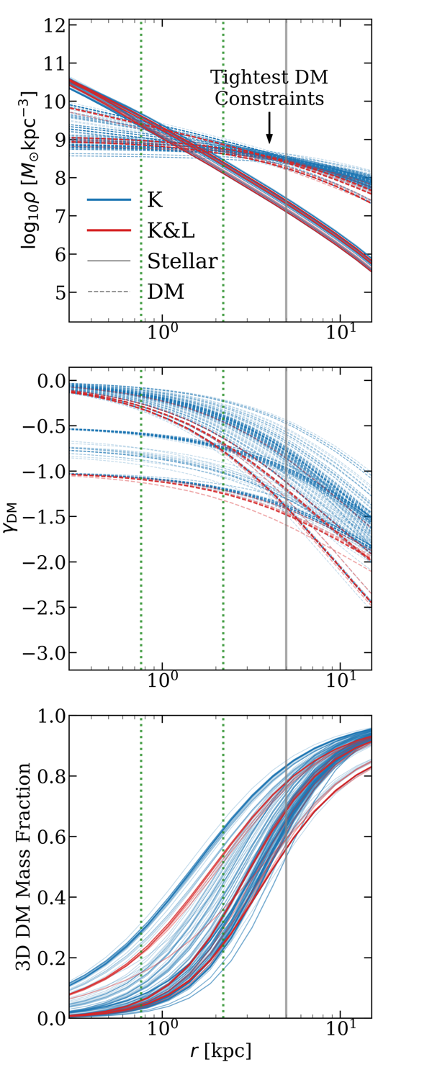}
    \caption{Radial profiles of the stellar and DM density (top), DM density slope (middle), and 3D DM mass fraction (bottom) for ESO0286. In the top panel, solid curves represent the stellar density profiles, while dashed curves represent the DM density profiles. In all three panels, models based on kinematic constraints only (K) are shown by the blue curves, while models satisfying both kinematic and lensing constraints (K\&L) are shown by the red curves. 
The green dotted vertical lines indicate the lensed image positions, while the gray vertical line marks the spatial extent $\sim 5$~kpc of the IFU FoV.}
\label{fig:dm_profiles}
    \label{fig:DM}
\end{figure}

In Fig.~\ref{fig:DM}, we present the decomposition of the stellar and DM distributions for ESO0286. For the stellar mass component, the two classes of models (K and K\&L) are consistent with each other. This agreement indicates that the stellar mass distribution can be robustly constrained by dynamics-only modeling, independent of lensing data, which further supports the robustness of our IMF conclusion.

Regarding the DM halo, our models suggest a relatively compact structure. 
The K\&L models constrain the scale radius, $r_{\rm s}$, to a range of $9$--$22$\,kpc. 
While this value is $2$--$5$ smaller than predictions from simulations \citep{2007Macci, 2014Dutton}, it is consistent with the findings of \citetalias{2022Poci}. 
The inner density slope, $\gamma_{\rm in}$, remains poorly constrained, exhibiting significant scatter between core-like and cusp-like profiles even when lensing constraints are included for the inner region ($r < 0.8$\,kpc). 
This limitation is significant, as the inner slope is critical for distinguishing between fundamental DM scenarios.

Beyond the IFU coverage ($r > 5$\,kpc), the scatter in the DM density distribution naturally increases; however, the inclusion of lensing constraints (K\&L) helps tighten the profile significantly in these outer regions. We find that the DM distribution is most tightly constrained at $r \sim 4$\,kpc, where the cumulative 3D DM fraction exceeds 40\% across all models. This indicates that halo properties are best recovered in the regime where DM begins to dominate the mass budget. In this transition regime, lensing provides essential additional leverage to reduce the degeneracies between the dark and stellar mass. Driven by this joint constraint, the K\&L models strongly favor a steeper DM slope at $r \sim 4$\,kpc compared to the K-only models. However, because ESO0286 is associated with only two lensed images, the lensing constraints remain limited. Future studies targeting systems with extended arcs or quadruply imaged configurations, combined with dynamical modeling, should provide significantly stronger constraints on the inner DM distribution.

For the shape of the DM halo in ESO0286, the K\&L models favor a specific subset of the $q_{\rm dm}$ and $p_{\rm dm}$ parameter space compared to the K models. This preference indicates that $p_{\rm dm}$ is slightly smaller than $q_{\rm dm}$, which would physically imply that the DM halo is misaligned with the baryonic component, with the intermediate and minor axes swapped. However, because our current dynamical models assume a single, fixed viewing angle, this apparent misalignment may largely be a consequence of the inherent degeneracy between the intrinsic 3D shape and the projection angle. Therefore, while this structural twist is intriguing, we cannot definitively conclude that ESO0286 is an outlier in the typical formation and evolution history of fast-rotating ETGs without first exploring a broader range of viewing angles.

\section{Summary and outlook}
\label{sect:Summary and outlook}
In this paper, we study the inner structure of the nearby lensing galaxy ESO0286 at $z_{\rm lens} = 0.0312$ using HST imaging and MUSE WFM kinematic data. On the lensing modeling side, we improve upon previous measurements, which were based on image positions and fluxes, by utilizing the extended arcs in two imaging bands. Subsequently, we perform state-of-the-art Schwarzschild modeling to investigate the IMF and DM distribution of ESO0286. Our findings are as follows:
\begin{itemize}
\item We obtain $\overline{M_{\rm Ein}^{\rm lens}} = (8.85 \pm 0.38) \times 10^{10}~\rm M_\odot$ at $\overline{\theta_{\rm Ein}} = 2.32\arcsec$ for ESO0286, based on extended image modeling with HST F814W and F336W imaging. This value is 6.7\% lower than the estimate reported in \citetalias{2018Collier}.

\item ESO0286 is a confirmed fast-rotating ETG but exhibits clear signatures of triaxiality in its rotation map, specifically rotation along both the minor and major axes (see Fig.~\ref{fig:ES0286_HST_imaging}). 

\item The inclusion of lensing data particularly helps to reduce the uncertainty in the outer mass profile and implicitly allows to extrapolate the orbital anisotropy beyond the IFU FoV: only models with large outer radial anisotropy are consistent with the lensing constraints (see Fig.~\ref{fig:total_mass}).

\item Evaluating the total mass density as an upper limit for the stellar mass, we find that the IMF of ESO0286 is Kroupa (or lighter) at $r \sim 0.7$~kpc, a result consistent with findings for other nearby ETGs of similar velocity dispersion \citep[see][]{2024Mehrgan} (see Fig.~\ref{fig:IMF}). It is plausible to assume that the galaxy retains a Kroupa IMF at larger radii as well. In the very center, the total mass allows for a very mildly bottom-heavy IMF; this radially varying profile is consistent with \citet{2025Poci}, who favored a Salpeter-like IMF in the innermost region based on their stellar population analysis of NFM data. 

\item Based on our fiducial mass decomposition, the inner DM density slope cannot be well constrained and shows significant scatter, independent of the inclusion of lensing data. Reliable constraints on the halo profile are only obtained in the outer regions where the cumulative DM mass fraction exceeds $40\%$ (see Fig.~\ref{fig:DM}).

\end{itemize}
This work demonstrates the power of combining SL and stellar kinematics to probe the inner structure of a nearby elliptical galaxy. While the stellar mass and IMF are robustly constrained, the inner DM halo slope remains weakly constrained for an individual system. Extending this joint analysis to larger samples will be essential for achieving tighter constraints on halo properties. Future wide-field surveys such as \textit{Euclid} \citep{2022Euclid, 2025sl} and Rubin Legacy Survey of Space and Time \citep{2019Ivezi, 2025Shajib} will enable this population-level approach and
overcome the limitations of modeling individual systems.

\begin{acknowledgements}
HW and SHS thank the Max Planck Society for support through the Max Planck Fellowship for SHS. 
This work is supported in part by the Deutsche Forschungsgemeinschaft (DFG, German Research Foundation) under Germany's Excellence Strategy -- EXC-2094 -- 390783311. Computations were performed on the HPC systems Raven and Viper at the Max Planck Computing and Data Facility.
\end{acknowledgements}
%




\bibliography{reference}

@ARTICLE{2023JOSS....8.5340M,
       author = {{Michalewicz}, Kevin and {Millon}, Martin and {Dux}, Fr{\'e}d{\'e}ric and {Courbin}, Fr{\'e}d{\'e}ric},
        title = "{STARRED: a two-channel deconvolution method with Starlet regularization}",
      journal = {The Journal of Open Source Software},
         year = 2023,
        month = may,
       volume = {8},
       number = {85},
          eid = {5340},
        pages = {5340},
          doi = {10.21105/joss.05340},
archivePrefix = {arXiv},
       eprint = {2305.18526},
 primaryClass = {astro-ph.IM},
       adsurl = {https://ui.adsabs.harvard.edu/abs/2023JOSS....8.5340M}
}

@ARTICLE{2024AJ....168...55M,
       author = {{Millon}, Martin and {Michalewicz}, Kevin and {Dux}, Fr{\'e}d{\'e}ric and {Courbin}, Fr{\'e}d{\'e}ric and {Marshall}, Philip J.},
        title = "{Image Deconvolution and Point-spread Function Reconstruction with STARRED: A Wavelet-based Two-channel Method Optimized for Light-curve Extraction}",
      journal = {\aj},
         year = 2024,
        month = aug,
       volume = {168},
       number = {2},
          eid = {55},
        pages = {55},
          doi = {10.3847/1538-3881/ad4da7},
archivePrefix = {arXiv},
       eprint = {2402.08725},
 primaryClass = {astro-ph.IM},
       adsurl = {https://ui.adsabs.harvard.edu/abs/2024AJ....168...55M}
}

@ARTICLE{DinosII,
       author = {{Sheu}, William and {Shajib}, Anowar J. and {Treu}, Tommaso and {Sonnenfeld}, Alessandro and {Birrer}, Simon and {Cappellari}, Michele and {Oldham}, Lindsay J. and {Tan}, Chin Yi},
        title = "{Project Dinos II: redshift evolution of dark and luminous matter density profiles in strong-lensing elliptical galaxies across 0.1 < z < 0.9}",
      journal = {\mnras},
         year = 2025,
        month = jul,
       volume = {541},
       number = {1},
        pages = {1-27},
          doi = {10.1093/mnras/staf976},
archivePrefix = {arXiv},
       eprint = {2408.10316},
 primaryClass = {astro-ph.GA},
       adsurl = {https://ui.adsabs.harvard.edu/abs/2025MNRAS.541....1S}
}

@ARTICLE{TDCOSMO13,
       author = {{Y{\i}ld{\i}r{\i}m}, A. and {Suyu}, S.~H. and {Chen}, G.~C. -F. and {Komatsu}, E.},
        title = "{TDCOSMO. XIII. Cosmological distance measurements in light of the mass-sheet degeneracy: Forecasts from strong lensing and integral field unit stellar kinematics}",
      journal = {\aap},
         year = 2023,
        month = jul,
       volume = {675},
          eid = {A21},
        pages = {A21},
          doi = {10.1051/0004-6361/202142318},
archivePrefix = {arXiv},
       eprint = {2109.14615},
 primaryClass = {astro-ph.CO},
       adsurl = {https://ui.adsabs.harvard.edu/abs/2023A&A...675A..21Y}
}

@ARTICLE{ShajibRXJ1131,
       author = {{Shajib}, Anowar J. and {Mozumdar}, Pritom and {Chen}, Geoff C. -F. and {Treu}, Tommaso and {Cappellari}, Michele and {Knabel}, Shawn and {Suyu}, Sherry H. and {Bennert}, Vardha N. and {Frieman}, Joshua A. and {Sluse}, Dominique and {Birrer}, Simon and {Courbin}, Frederic and {Fassnacht}, Christopher D. and {Villafa{\~n}a}, Lizvette and {Williams}, Peter R.},
        title = "{TDCOSMO. XII. Improved Hubble constant measurement from lensing time delays using spatially resolved stellar kinematics of the lens galaxy}",
      journal = {\aap},
         year = 2023,
        month = may,
       volume = {673},
          eid = {A9},
        pages = {A9},
          doi = {10.1051/0004-6361/202345878},
archivePrefix = {arXiv},
       eprint = {2301.02656},
 primaryClass = {astro-ph.CO},
       adsurl = {https://ui.adsabs.harvard.edu/abs/2023A&A...673A...9S}
}

@ARTICLE{TDCOSMO4,
       author = {{Birrer}, S. and {Shajib}, A.~J. and {Galan}, A. and {Millon}, M. and {Treu}, T. and {Agnello}, A. and {Auger}, M. and {Chen}, G.~C. -F. and {Christensen}, L. and {Collett}, T. and {Courbin}, F. and {Fassnacht}, C.~D. and {Koopmans}, L.~V.~E. and {Marshall}, P.~J. and {Park}, J. -W. and {Rusu}, C.~E. and {Sluse}, D. and {Spiniello}, C. and {Suyu}, S.~H. and {Wagner-Carena}, S. and {Wong}, K.~C. and {Barnab{\`e}}, M. and {Bolton}, A.~S. and {Czoske}, O. and {Ding}, X. and {Frieman}, J.~A. and {Van de Vyvere}, L.},
        title = "{TDCOSMO. IV. Hierarchical time-delay cosmography - joint inference of the Hubble constant and galaxy density profiles}",
      journal = {\aap},
         year = 2020,
        month = nov,
       volume = {643},
          eid = {A165},
        pages = {A165},
          doi = {10.1051/0004-6361/202038861},
archivePrefix = {arXiv},
       eprint = {2007.02941},
 primaryClass = {astro-ph.CO},
       adsurl = {https://ui.adsabs.harvard.edu/abs/2020A&A...643A.165B}
}

@ARTICLE{Holicow_wong,
       author = {{Wong}, Kenneth C. and {Suyu}, Sherry H. and {Chen}, Geoff C. -F. and {Rusu}, Cristian E. and {Millon}, Martin and {Sluse}, Dominique and {Bonvin}, Vivien and {Fassnacht}, Christopher D. and {Taubenberger}, Stefan and {Auger}, Matthew W. and {Birrer}, Simon and {Chan}, James H.~H. and {Courbin}, Frederic and {Hilbert}, Stefan and {Tihhonova}, Olga and {Treu}, Tommaso and {Agnello}, Adriano and {Ding}, Xuheng and {Jee}, Inh and {Komatsu}, Eiichiro and {Shajib}, Anowar J. and {Sonnenfeld}, Alessandro and {Blandford}, Roger D. and {Koopmans}, L{\'e}on V.~E. and {Marshall}, Philip J. and {Meylan}, Georges},
        title = "{H0LiCOW - XIII. A 2.4 per cent measurement of H0 from lensed quasars: 5.3{\ensuremath{\sigma}} tension between early- and late-Universe probes}",
      journal = {\mnras},
         year = 2020,
        month = oct,
       volume = {498},
       number = {1},
        pages = {1420-1439},
          doi = {10.1093/mnras/stz3094},
archivePrefix = {arXiv},
       eprint = {1907.04869},
 primaryClass = {astro-ph.CO},
       adsurl = {https://ui.adsabs.harvard.edu/abs/2020MNRAS.498.1420W}
}

@ARTICLE{2020Akin,
       author = {{Y{\i}ld{\i}r{\i}m}, Ak{\i}n and {Suyu}, Sherry H. and {Halkola}, Aleksi},
        title = "{Time-delay cosmographic forecasts with strong lensing and JWST stellar kinematics}",
      journal = {\mnras},
         year = 2020,
        month = apr,
       volume = {493},
       number = {4},
        pages = {4783-4807},
          doi = {10.1093/mnras/staa498},
archivePrefix = {arXiv},
       eprint = {1904.07237},
 primaryClass = {astro-ph.CO},
       adsurl = {https://ui.adsabs.harvard.edu/abs/2020MNRAS.493.4783Y}
}

@ARTICLE{2010SuyuHalkola_Dpie2,
       author = {{Suyu}, S.~H. and {Halkola}, A.},
        title = "{The halos of satellite galaxies: the companion of the massive elliptical lens SL2S J08544-0121}",
      journal = {\aap},
         year = 2010,
        month = dec,
       volume = {524},
          eid = {A94},
        pages = {A94},
          doi = {10.1051/0004-6361/201015481},
archivePrefix = {arXiv},
       eprint = {1007.4815},
 primaryClass = {astro-ph.CO},
       adsurl = {https://ui.adsabs.harvard.edu/abs/2010A&A...524A..94S}
}

@ARTICLE{2006esr,
       author = {{Suyu}, S.~H. and {Marshall}, P.~J. and {Hobson}, M.~P. and {Blandford}, R.~D.},
        title = "{A Bayesian analysis of regularized source inversions in gravitational lensing}",
      journal = {\mnras},
         year = 2006,
        month = sep,
       volume = {371},
       number = {2},
        pages = {983-998},
          doi = {10.1111/j.1365-2966.2006.10733.x},
archivePrefix = {arXiv},
       eprint = {astro-ph/0601493},
 primaryClass = {astro-ph},
       adsurl = {https://ui.adsabs.harvard.edu/abs/2006MNRAS.371..983S}
}

@ARTICLE{TDCOSMO1,
       author = {{Millon}, M. and {Galan}, A. and {Courbin}, F. and {Treu}, T. and {Suyu}, S.~H. and {Ding}, X. and {Birrer}, S. and {Chen}, G.~C. -F. and {Shajib}, A.~J. and {Sluse}, D. and {Wong}, K.~C. and {Agnello}, A. and {Auger}, M.~W. and {Buckley-Geer}, E.~J. and {Chan}, J.~H.~H. and {Collett}, T. and {Fassnacht}, C.~D. and {Hilbert}, S. and {Koopmans}, L.~V.~E. and {Motta}, V. and {Mukherjee}, S. and {Rusu}, C.~E. and {Sonnenfeld}, A. and {Spiniello}, C. and {Van de Vyvere}, L.},
        title = "{TDCOSMO. I. An exploration of systematic uncertainties in the inference of H$_{0}$ from time-delay cosmography}",
      journal = {\aap},
         year = 2020,
        month = jul,
       volume = {639},
          eid = {A101},
        pages = {A101},
          doi = {10.1051/0004-6361/201937351},
archivePrefix = {arXiv},
       eprint = {1912.08027},
 primaryClass = {astro-ph.CO},
       adsurl = {https://ui.adsabs.harvard.edu/abs/2020A&A...639A.101M}
}

@ARTICLE{2021Shajib,
       author = {{Shajib}, Anowar J. and {Treu}, Tommaso and {Birrer}, Simon and {Sonnenfeld}, Alessandro},
        title = "{Dark matter haloes of massive elliptical galaxies at z {\ensuremath{\sim}} 0.2 are well described by the Navarro-Frenk-White profile}",
      journal = {\mnras},
         year = 2021,
        month = may,
       volume = {503},
       number = {2},
        pages = {2380-2405},
          doi = {10.1093/mnras/stab536},
archivePrefix = {arXiv},
       eprint = {2008.11724},
 primaryClass = {astro-ph.GA},
       adsurl = {https://ui.adsabs.harvard.edu/abs/2021MNRAS.503.2380S}
}

@ARTICLE{EPL,
       author = {{Tessore}, Nicolas and {Metcalf}, R. Benton},
        title = "{The elliptical power law profile lens}",
      journal = {\aap},
         year = 2015,
        month = aug,
       volume = {580},
          eid = {A79},
        pages = {A79},
          doi = {10.1051/0004-6361/201526773},
archivePrefix = {arXiv},
       eprint = {1507.01819},
 primaryClass = {astro-ph.CO},
       adsurl = {https://ui.adsabs.harvard.edu/abs/2015A&A...580A..79T}
}

@ARTICLE{Suyu2012,
       author = {{Suyu}, S.~H. and {Hensel}, S.~W. and {McKean}, J.~P. and {Fassnacht}, C.~D. and {Treu}, T. and {Halkola}, A. and {Norbury}, M. and {Jackson}, N. and {Schneider}, P. and {Thompson}, D. and {Auger}, M.~W. and {Koopmans}, L.~V.~E. and {Matthews}, K.},
        title = "{Disentangling Baryons and Dark Matter in the Spiral Gravitational Lens B1933+503}",
      journal = {\apj},
         year = 2012,
        month = may,
       volume = {750},
       number = {1},
          eid = {10},
        pages = {10},
          doi = {10.1088/0004-637X/750/1/10},
archivePrefix = {arXiv},
       eprint = {1110.2536},
 primaryClass = {astro-ph.CO},
       adsurl = {https://ui.adsabs.harvard.edu/abs/2012ApJ...750...10S}
}

@Article{Cappellari2025,
  author        = {Cappellari, Michele},
  title         = {Early-Type Galaxies: Elliptical and S0 Galaxies, or Fast and Slow Rotators},
  note         = {Elsevier, in press},
  year          = {2025},
  month         = mar,
  pages         = {arXiv:2503.02746},
  adsurl        = {https://ui.adsabs.harvard.edu/abs/2025arXiv250302746C},
  archiveprefix = {arXiv},
  doi           = {10.48550/arXiv.2503.02746},
  eid           = {arXiv:2503.02746},
  eprint        = {2503.02746},
  primaryclass  = {astro-ph.GA},
}

@ARTICLE{2016Cappellari,
       author = {{Cappellari}, Michele},
        title = "{Structure and Kinematics of Early-Type Galaxies from Integral Field Spectroscopy}",
      journal = {\araa},
         year = 2016,
        month = sep,
       volume = {54},
        pages = {597-665},
          doi = {10.1146/annurev-astro-082214-122432},
archivePrefix = {arXiv},
       eprint = {1602.04267},
 primaryClass = {astro-ph.GA},
       adsurl = {https://ui.adsabs.harvard.edu/abs/2016ARA&A..54..597C}
}

@ARTICLE{2022Poci,
       author = {{Poci}, Adriano and {Smith}, Russell J.},
        title = "{Comparing lensing and stellar orbital models of a nearby massive strong-lens galaxy}",
      journal = {\mnras},
         year = 2022,
        month = jun,
       volume = {512},
       number = {4},
        pages = {5298-5310},
          doi = {10.1093/mnras/stac776},
archivePrefix = {arXiv},
       eprint = {2203.09309},
 primaryClass = {astro-ph.GA},
       adsurl = {https://ui.adsabs.harvard.edu/abs/2022MNRAS.512.5298P}
}

@ARTICLE{SINFONI2015,
       author = {{Smith}, Russell J. and {Lucey}, John R. and {Conroy}, Charlie},
        title = "{The SINFONI Nearby Elliptical Lens Locator Survey: discovery of two new low-redshift strong lenses and implications for the initial mass function in giant early-type galaxies}",
      journal = {\mnras},
         year = 2015,
        month = jun,
       volume = {449},
       number = {4},
        pages = {3441-3457},
          doi = {10.1093/mnras/stv518},
archivePrefix = {arXiv},
       eprint = {1503.02661},
 primaryClass = {astro-ph.GA},
       adsurl = {https://ui.adsabs.harvard.edu/abs/2015MNRAS.449.3441S}
}

@ARTICLE{2018Collier,
       author = {{Collier}, William P. and {Smith}, Russell J. and {Lucey}, John R.},
        title = "{Improved mass constraints for two nearby strong-lensing elliptical galaxies from Hubble Space Telescope imaging}",
      journal = {\mnras},
         year = 2018,
        month = jan,
       volume = {473},
       number = {1},
        pages = {1103-1107},
          doi = {10.1093/mnras/stx2297},
archivePrefix = {arXiv},
       eprint = {1709.01931},
 primaryClass = {astro-ph.GA},
       adsurl = {https://ui.adsabs.harvard.edu/abs/2018MNRAS.473.1103C}
}

@ARTICLE{2025GLaD,
       author = {{Wang}, Han and {Suyu}, Sherry H. and {Galan}, Aymeric and {Halkola}, Aleksi and {Cappellari}, Michele and {Shajib}, Anowar J. and {Cernetic}, Miha},
        title = "{GPU-Accelerated Gravitational Lensing \& Dynamical (GLaD) Modeling for Cosmology and Galaxies}",
      journal = {arXiv e-prints},
         year = 2025,
        month = apr,
          eid = {arXiv:2504.01302},
        pages = {arXiv:2504.01302},
          doi = {10.48550/arXiv.2504.01302},
archivePrefix = {arXiv},
       eprint = {2504.01302},
 primaryClass = {astro-ph.CO},
       adsurl = {https://ui.adsabs.harvard.edu/abs/2025arXiv250401302W}
}

@software{Photutils,
  author       = {Larry Bradley and
                  Brigitta Sip{\H o}cz and
                  Thomas Robitaille and
                  Erik Tollerud and
                  Z\`e Vin{\'{\i}}cius and
                  Christoph Deil and
                  Kyle Barbary and
                  Tom J Wilson and
                  Ivo Busko and
                  Axel Donath and
                  Hans Moritz G{\"u}nther and
                  Mihai Cara and
                  P. L. Lim and
                  Sebastian Me{\ss}linger and
                  Simon Conseil and
                  Zach Burnett and
                  Azalee Bostroem and
                  Michael Droettboom and
                  E. M. Bray and
                  Lars Andersen Bratholm and
                  Adam Ginsburg and
                  William Jamieson and
                  Geert Barentsen and
                  Matt Craig and
                  Brett M. Morris and
                  Marshall Perrin and
                  Shivangee Rathi and
                  Sergio Pascual and
                  Iskren Y. Georgiev},
  title        = {astropy/photutils: 2.0.2},
  month        = oct,
  year         = 2024,
  publisher    = {Zenodo},
  version      = {2.0.2},
  doi          = {10.5281/zenodo.13989456},
  url          = {https://doi.org/10.5281/zenodo.13989456},
}

@ARTICLE{2004thomas,
       author = {{Thomas}, J. and {Saglia}, R.~P. and {Bender}, R. and {Thomas}, D. and {Gebhardt}, K. and {Magorrian}, J. and {Richstone}, D.},
        title = "{Mapping stationary axisymmetric phase-space distribution functions by orbit libraries}",
      journal = {\mnras},
         year = 2004,
        month = sep,
       volume = {353},
       number = {2},
        pages = {391-404},
          doi = {10.1111/j.1365-2966.2004.08072.x},
archivePrefix = {arXiv},
       eprint = {astro-ph/0406014},
 primaryClass = {astro-ph},
       adsurl = {https://ui.adsabs.harvard.edu/abs/2004MNRAS.353..391T}
}

@ARTICLE{1979schwarzschild,
       author = {{Schwarzschild}, M.},
        title = "{A numerical model for a triaxial stellar system in dynamical equilibrium.}",
      journal = {\apj},
         year = 1979,
        month = aug,
       volume = {232},
        pages = {236-247},
          doi = {10.1086/157282},
       adsurl = {https://ui.adsabs.harvard.edu/abs/1979ApJ...232..236S}
}

@ARTICLE{2021Lipka,
       author = {{Lipka}, Mathias and {Thomas}, Jens},
        title = "{A novel approach to optimize the regularization and evaluation of dynamical models using a model selection framework}",
      journal = {\mnras},
         year = 2021,
        month = jul,
       volume = {504},
       number = {3},
        pages = {4599-4625},
          doi = {10.1093/mnras/stab1092},
archivePrefix = {arXiv},
       eprint = {2104.10168},
 primaryClass = {astro-ph.GA},
       adsurl = {https://ui.adsabs.harvard.edu/abs/2021MNRAS.504.4599L}
}

@ARTICLE{2022Thomas,
       author = {{Thomas}, Jens and {Lipka}, Mathias},
        title = "{A simple data-driven method to optimize the penalty strengths of penalized models and its application to non-parametric smoothing}",
      journal = {\mnras},
         year = 2022,
        month = aug,
       volume = {514},
       number = {4},
        pages = {6203-6214},
          doi = {10.1093/mnras/stac1581},
archivePrefix = {arXiv},
       eprint = {2206.04067},
 primaryClass = {stat.ME},
       adsurl = {https://ui.adsabs.harvard.edu/abs/2022MNRAS.514.6203T}
}

@article{AuDe2006,

  Author   = {C. Audet and J.E. {Dennis, Jr.}},

  Title    = {Mesh Adaptive Direct Search Algorithms for Constrained

              Optimization},

  Journal  = {SIAM Journal on Optimization},

  Volume   = {17},

  Number   = {1},

  Pages    = {188--217},

  Doi      = {doi:10.1137/040603371},

  Url      = {http://dx.doi.org/doi:10.1137/040603371},

  Year     = {2006}

}

@ARTICLE{Magorrian1999,
       author = {{Magorrian}, John},
        title = "{Kinematical signatures of hidden stellar discs}",
      journal = {\mnras},
         year = 1999,
        month = jan,
       volume = {302},
       number = {3},
        pages = {530-536},
          doi = {10.1046/j.1365-8711.1999.02135.x},
archivePrefix = {arXiv},
       eprint = {astro-ph/9902033},
 primaryClass = {astro-ph},
       adsurl = {https://ui.adsabs.harvard.edu/abs/1999MNRAS.302..530M},
}

@ARTICLE{2010Auger,
       author = {{Auger}, M.~W. and {Treu}, T. and {Bolton}, A.~S. and {Gavazzi}, R. and {Koopmans}, L.~V.~E. and {Marshall}, P.~J. and {Moustakas}, L.~A. and {Burles}, S.},
        title = "{The Sloan Lens ACS Survey. X. Stellar, Dynamical, and Total Mass Correlations of Massive Early-type Galaxies}",
      journal = {\apj},
         year = 2010,
        month = nov,
       volume = {724},
       number = {1},
        pages = {511-525},
          doi = {10.1088/0004-637X/724/1/511},
archivePrefix = {arXiv},
       eprint = {1007.2880},
 primaryClass = {astro-ph.CO},
       adsurl = {https://ui.adsabs.harvard.edu/abs/2010ApJ...724..511A}
}

@ARTICLE{2010Treu,
       author = {{Treu}, Tommaso and {Auger}, Matthew W. and {Koopmans}, L{\'e}on V.~E. and {Gavazzi}, Rapha{\"e}l and {Marshall}, Philip J. and {Bolton}, Adam S.},
        title = "{The Initial Mass Function of Early-Type Galaxies}",
      journal = {\apj},
         year = 2010,
        month = feb,
       volume = {709},
       number = {2},
        pages = {1195-1202},
          doi = {10.1088/0004-637X/709/2/1195},
archivePrefix = {arXiv},
       eprint = {0911.3392},
 primaryClass = {astro-ph.CO},
       adsurl = {https://ui.adsabs.harvard.edu/abs/2010ApJ...709.1195T}
}

@ARTICLE{2015Poscacki,
       author = {{Posacki}, Silvia and {Cappellari}, Michele and {Treu}, Tommaso and {Pellegrini}, Silvia and {Ciotti}, Luca},
        title = "{The stellar initial mass function of early-type galaxies from low to high stellar velocity dispersion: homogeneous analysis of ATLAS$^{3D}$ and Sloan Lens ACS galaxies}",
      journal = {\mnras},
         year = 2015,
        month = jan,
       volume = {446},
       number = {1},
        pages = {493-509},
          doi = {10.1093/mnras/stu2098},
archivePrefix = {arXiv},
       eprint = {1407.5633},
 primaryClass = {astro-ph.GA},
       adsurl = {https://ui.adsabs.harvard.edu/abs/2015MNRAS.446..493P}
}

@ARTICLE{2013Cap,
       author = {{Cappellari}, Michele and {McDermid}, Richard M. and {Alatalo}, Katherine and {Blitz}, Leo and {Bois}, Maxime and {Bournaud}, Fr{\'e}d{\'e}ric and {Bureau}, M. and {Crocker}, Alison F. and {Davies}, Roger L. and {Davis}, Timothy A. and {de Zeeuw}, P.~T. and {Duc}, Pierre-Alain and {Emsellem}, Eric and {Khochfar}, Sadegh and {Krajnovi{\'c}}, Davor and {Kuntschner}, Harald and {Morganti}, Raffaella and {Naab}, Thorsten and {Oosterloo}, Tom and {Sarzi}, Marc and {Scott}, Nicholas and {Serra}, Paolo and {Weijmans}, Anne-Marie and {Young}, Lisa M.},
        title = "{The ATLAS$^{3D}$ project - XX. Mass-size and mass-{\ensuremath{\sigma}} distributions of early-type galaxies: bulge fraction drives kinematics, mass-to-light ratio, molecular gas fraction and stellar initial mass function}",
      journal = {\mnras},
         year = 2013,
        month = jul,
       volume = {432},
       number = {3},
        pages = {1862-1893},
          doi = {10.1093/mnras/stt644},
archivePrefix = {arXiv},
       eprint = {1208.3523},
 primaryClass = {astro-ph.CO},
       adsurl = {https://ui.adsabs.harvard.edu/abs/2013MNRAS.432.1862C}
}

@ARTICLE{2024Mehrgan,
       author = {{Mehrgan}, Kianusch and {Thomas}, Jens and {Saglia}, Roberto and {Parikh}, Taniya and {Neureiter}, Bianca and {Erwin}, Peter and {Bender}, Ralf},
        title = "{Dynamical Stellar Mass-to-light Ratio Gradients: Evidence for Very Centrally Concentrated IMF Variations in ETGs?}",
      journal = {\apj},
         year = 2024,
        month = jan,
       volume = {961},
       number = {1},
          eid = {127},
        pages = {127},
          doi = {10.3847/1538-4357/acfe09},
archivePrefix = {arXiv},
       eprint = {2309.15911},
 primaryClass = {astro-ph.GA},
       adsurl = {https://ui.adsabs.harvard.edu/abs/2024ApJ...961..127M}
}

@ARTICLE{2025Poci,
       author = {{Poci}, Adriano and {Smith}, Russell J.},
        title = "{SNELLS-HD I: a first look at the stellar properties of the massive strong-lens galaxy SNL-1 with 50 pc resolution}",
      journal = {\mnras},
         year = 2025,
        month = sep,
          doi = {10.1093/mnras/staf1491},
archivePrefix = {arXiv},
       eprint = {2509.01732},
 primaryClass = {astro-ph.GA},
       adsurl = {https://ui.adsabs.harvard.edu/abs/2025MNRAS.tmp.1438P}
}

@ARTICLE{2006Bolton,
       author = {{Bolton}, Adam S. and {Burles}, Scott and {Koopmans}, L{\'e}on V.~E. and {Treu}, Tommaso and {Moustakas}, Leonidas A.},
        title = "{The Sloan Lens ACS Survey. I. A Large Spectroscopically Selected Sample of Massive Early-Type Lens Galaxies}",
      journal = {\apj},
         year = 2006,
        month = feb,
       volume = {638},
       number = {2},
        pages = {703-724},
          doi = {10.1086/498884},
archivePrefix = {arXiv},
       eprint = {astro-ph/0511453},
 primaryClass = {astro-ph},
       adsurl = {https://ui.adsabs.harvard.edu/abs/2006ApJ...638..703B}
}

@ARTICLE{2012Gavazzi,
       author = {{Gavazzi}, Rapha{\"e}l and {Treu}, Tommaso and {Marshall}, Philip J. and {Brault}, Florence and {Ruff}, Andrea},
        title = "{The SL2S Galaxy-scale Gravitational Lens Sample. I. The Alignment of Mass and Light in Massive Early-type Galaxies at z = 0.2-0.9}",
      journal = {\apj},
         year = 2012,
        month = dec,
       volume = {761},
       number = {2},
          eid = {170},
        pages = {170},
          doi = {10.1088/0004-637X/761/2/170},
archivePrefix = {arXiv},
       eprint = {1202.3852},
 primaryClass = {astro-ph.CO},
       adsurl = {https://ui.adsabs.harvard.edu/abs/2012ApJ...761..170G}
}

@ARTICLE{2012More,
       author = {{More}, A. and {Cabanac}, R. and {More}, S. and {Alard}, C. and {Limousin}, M. and {Kneib}, J. -P. and {Gavazzi}, R. and {Motta}, V.},
        title = "{The CFHTLS-Strong Lensing Legacy Survey (SL2S): Investigating the Group-scale Lenses with the SARCS Sample}",
      journal = {\apj},
         year = 2012,
        month = apr,
       volume = {749},
       number = {1},
          eid = {38},
        pages = {38},
          doi = {10.1088/0004-637X/749/1/38},
archivePrefix = {arXiv},
       eprint = {1109.1821},
 primaryClass = {astro-ph.CO},
       adsurl = {https://ui.adsabs.harvard.edu/abs/2012ApJ...749...38M}
}

@ARTICLE{2017Sonnenfeld,
       author = {{Sonnenfeld}, Alessandro and {Nipoti}, Carlo and {Treu}, Tommaso},
        title = "{Merger-driven evolution of the effective stellar initial mass function of massive early-type galaxies}",
      journal = {\mnras},
         year = 2017,
        month = feb,
       volume = {465},
       number = {2},
        pages = {2397-2410},
          doi = {10.1093/mnras/stw2919},
archivePrefix = {arXiv},
       eprint = {1607.01394},
 primaryClass = {astro-ph.GA},
       adsurl = {https://ui.adsabs.harvard.edu/abs/2017MNRAS.465.2397S}
}

@ARTICLE{2009Auger,
       author = {{Auger}, M.~W. and {Treu}, T. and {Bolton}, A.~S. and {Gavazzi}, R. and {Koopmans}, L.~V.~E. and {Marshall}, P.~J. and {Bundy}, K. and {Moustakas}, L.~A.},
        title = "{The Sloan Lens ACS Survey. IX. Colors, Lensing, and Stellar Masses of Early-Type Galaxies}",
      journal = {\apj},
         year = 2009,
        month = nov,
       volume = {705},
       number = {2},
        pages = {1099-1115},
          doi = {10.1088/0004-637X/705/2/1099},
archivePrefix = {arXiv},
       eprint = {0911.2471},
 primaryClass = {astro-ph.CO},
       adsurl = {https://ui.adsabs.harvard.edu/abs/2009ApJ...705.1099A}
}

@ARTICLE{2025Sonnenfeld,
       author = {{Sonnenfeld}, Alessandro},
        title = "{The SLACS strong lens sample, debiased: II. Lensing-only constraints on the stellar initial mass function and dark matter contraction in early-type galaxies}",
      journal = {\aap},
         year = 2025,
        month = may,
       volume = {697},
          eid = {A95},
        pages = {A95},
          doi = {10.1051/0004-6361/202452846},
archivePrefix = {arXiv},
       eprint = {2501.02054},
 primaryClass = {astro-ph.GA},
       adsurl = {https://ui.adsabs.harvard.edu/abs/2025A&A...697A..95S}
}

@ARTICLE{2022Quenneville,
       author = {{Quenneville}, Matthew E. and {Liepold}, Emily R. and {Ma}, Chung-Pei},
        title = "{Triaxial Orbit-based Dynamical Modeling of Galaxies with Supermassive Black Holes and an Application to Massive Elliptical Galaxy NGC 1453}",
      journal = {\apj},
         year = 2022,
        month = feb,
       volume = {926},
       number = {1},
          eid = {30},
        pages = {30},
          doi = {10.3847/1538-4357/ac3e68},
archivePrefix = {arXiv},
       eprint = {2111.06904},
 primaryClass = {astro-ph.GA},
       adsurl = {https://ui.adsabs.harvard.edu/abs/2022ApJ...926...30Q}
}

@ARTICLE{2015Newman,
       author = {{Newman}, Andrew B. and {Ellis}, Richard S. and {Treu}, Tommaso},
        title = "{Luminous and Dark Matter Profiles from Galaxies to Clusters: Bridging the Gap with Group-scale Lenses}",
      journal = {\apj},
         year = 2015,
        month = nov,
       volume = {814},
       number = {1},
          eid = {26},
        pages = {26},
          doi = {10.1088/0004-637X/814/1/26},
archivePrefix = {arXiv},
       eprint = {1503.05282},
 primaryClass = {astro-ph.GA},
       adsurl = {https://ui.adsabs.harvard.edu/abs/2015ApJ...814...26N}
}

@ARTICLE{2014Dutton,
       author = {{Dutton}, Aaron A. and {Macci{\`o}}, Andrea V.},
        title = "{Cold dark matter haloes in the Planck era: evolution of structural parameters for Einasto and NFW profiles}",
      journal = {\mnras},
         year = 2014,
        month = jul,
       volume = {441},
       number = {4},
        pages = {3359-3374},
          doi = {10.1093/mnras/stu742},
archivePrefix = {arXiv},
       eprint = {1402.7073},
 primaryClass = {astro-ph.CO},
       adsurl = {https://ui.adsabs.harvard.edu/abs/2014MNRAS.441.3359D}
}

@ARTICLE{2007Macci,
       author = {{Macci{\`o}}, Andrea V. and {Dutton}, Aaron A. and {van den Bosch}, Frank C. and {Moore}, Ben and {Potter}, Doug and {Stadel}, Joachim},
        title = "{Concentration, spin and shape of dark matter haloes: scatter and the dependence on mass and environment}",
      journal = {\mnras},
         year = 2007,
        month = jun,
       volume = {378},
       number = {1},
        pages = {55-71},
          doi = {10.1111/j.1365-2966.2007.11720.x},
archivePrefix = {arXiv},
       eprint = {astro-ph/0608157},
 primaryClass = {astro-ph},
       adsurl = {https://ui.adsabs.harvard.edu/abs/2007MNRAS.378...55M}
}

@ARTICLE{2010deBlok,
       author = {{de Blok}, W.~J.~G.},
        title = "{The Core-Cusp Problem}",
      journal = {Advances in Astronomy},
         year = 2010,
        month = jan,
       volume = {2010},
          eid = {789293},
        pages = {789293},
          doi = {10.1155/2010/789293},
archivePrefix = {arXiv},
       eprint = {0910.3538},
 primaryClass = {astro-ph.CO},
       adsurl = {https://ui.adsabs.harvard.edu/abs/2010AdAst2010E...5D}
}

@ARTICLE{2018Genina,
       author = {{Genina}, Anna and {Ben{\'\i}tez-Llambay}, Alejandro and {Frenk}, Carlos S. and {Cole}, Shaun and {Fattahi}, Azadeh and {Navarro}, Julio F. and {Oman}, Kyle A. and {Sawala}, Till and {Theuns}, Tom},
        title = "{The core-cusp problem: a matter of perspective}",
      journal = {\mnras},
         year = 2018,
        month = feb,
       volume = {474},
       number = {1},
        pages = {1398-1411},
          doi = {10.1093/mnras/stx2855},
archivePrefix = {arXiv},
       eprint = {1707.06303},
 primaryClass = {astro-ph.GA},
       adsurl = {https://ui.adsabs.harvard.edu/abs/2018MNRAS.474.1398G}
}

@ARTICLE{1994Flores,
       author = {{Flores}, Ricardo A. and {Primack}, Joel R.},
        title = "{Observational and Theoretical Constraints on Singular Dark Matter Halos}",
      journal = {\apjl},
         year = 1994,
        month = may,
       volume = {427},
        pages = {L1},
          doi = {10.1086/187350},
archivePrefix = {arXiv},
       eprint = {astro-ph/9402004},
 primaryClass = {astro-ph},
       adsurl = {https://ui.adsabs.harvard.edu/abs/1994ApJ...427L...1F}
}

@ARTICLE{2007Thomas,
       author = {{Thomas}, J. and {Saglia}, R.~P. and {Bender}, R. and {Thomas}, D. and {Gebhardt}, K. and {Magorrian}, J. and {Corsini}, E.~M. and {Wegner}, G.},
        title = "{Dynamical modelling of luminous and dark matter in 17 Coma early-type galaxies}",
      journal = {\mnras},
         year = 2007,
        month = dec,
       volume = {382},
       number = {2},
        pages = {657-684},
          doi = {10.1111/j.1365-2966.2007.12434.x},
archivePrefix = {arXiv},
       eprint = {0709.0691},
 primaryClass = {astro-ph},
       adsurl = {https://ui.adsabs.harvard.edu/abs/2007MNRAS.382..657T}
}

@ARTICLE{2011Thomas,
       author = {{Thomas}, J. and {Saglia}, R.~P. and {Bender}, R. and {Thomas}, D. and {Gebhardt}, K. and {Magorrian}, J. and {Corsini}, E.~M. and {Wegner}, G. and {Seitz}, S.},
        title = "{Dynamical masses of early-type galaxies: a comparison to lensing results and implications for the stellar initial mass function and the distribution of dark matter}",
      journal = {\mnras},
         year = 2011,
        month = jul,
       volume = {415},
       number = {1},
        pages = {545-562},
          doi = {10.1111/j.1365-2966.2011.18725.x},
archivePrefix = {arXiv},
       eprint = {1103.3414},
 primaryClass = {astro-ph.CO},
       adsurl = {https://ui.adsabs.harvard.edu/abs/2011MNRAS.415..545T}
}

@ARTICLE{2022Euclid,
       author = {{Euclid Collaboration} and {Scaramella}, R. and {Amiaux}, J. and {Mellier}, Y. and {Burigana}, C. and {Carvalho}, C.~S. and {Cuillandre}, J. -C. and {Da Silva}, A. and {Derosa}, A. and {Dinis}, J. and {Maiorano}, E. and {Maris}, M. and {Tereno}, I. and {Laureijs}, R. and {Boenke}, T. and {Buenadicha}, G. and {Dupac}, X. and {Gaspar Venancio}, L.~M. and {G{\'o}mez-{\'A}lvarez}, P. and {Hoar}, J. and {Lorenzo Alvarez}, J. and {Racca}, G.~D. and {Saavedra-Criado}, G. and {Schwartz}, J. and {Vavrek}, R. and {Schirmer}, M. and {Aussel}, H. and {Azzollini}, R. and {Cardone}, V.~F. and {Cropper}, M. and {Ealet}, A. and {Garilli}, B. and {Gillard}, W. and {Granett}, B.~R. and {Guzzo}, L. and {Hoekstra}, H. and {Jahnke}, K. and {Kitching}, T. and {Maciaszek}, T. and {Meneghetti}, M. and {Miller}, L. and {Nakajima}, R. and {Niemi}, S.~M. and {Pasian}, F. and {Percival}, W.~J. and {Pottinger}, S. and {Sauvage}, M. and {Scodeggio}, M. and {Wachter}, S. and {Zacchei}, A. and {Aghanim}, N. and {Amara}, A. and {Auphan}, T. and {Auricchio}, N. and {Awan}, S. and {Balestra}, A. and {Bender}, R. and {Bodendorf}, C. and {Bonino}, D. and {Branchini}, E. and {Brau-Nogue}, S. and {Brescia}, M. and {Candini}, G.~P. and {Capobianco}, V. and {Carbone}, C. and {Carlberg}, R.~G. and {Carretero}, J. and {Casas}, R. and {Castander}, F.~J. and {Castellano}, M. and {Cavuoti}, S. and {Cimatti}, A. and {Cledassou}, R. and {Congedo}, G. and {Conselice}, C.~J. and {Conversi}, L. and {Copin}, Y. and {Corcione}, L. and {Costille}, A. and {Courbin}, F. and {Degaudenzi}, H. and {Douspis}, M. and {Dubath}, F. and {Duncan}, C.~A.~J. and {Dusini}, S. and {Farrens}, S. and {Ferriol}, S. and {Fosalba}, P. and {Fourmanoit}, N. and {Frailis}, M. and {Franceschi}, E. and {Franzetti}, P. and {Fumana}, M. and {Gillis}, B. and {Giocoli}, C. and {Grazian}, A. and {Grupp}, F. and {Haugan}, S.~V.~H. and {Holmes}, W. and {Hormuth}, F. and {Hudelot}, P. and {Kermiche}, S. and {Kiessling}, A. and {Kilbinger}, M. and {Kohley}, R. and {Kubik}, B. and {K{\"u}mmel}, M. and {Kunz}, M. and {Kurki-Suonio}, H. and {Lahav}, O. and {Ligori}, S. and {Lilje}, P.~B. and {Lloro}, I. and {Mansutti}, O. and {Marggraf}, O. and {Markovic}, K. and {Marulli}, F. and {Massey}, R. and {Maurogordato}, S. and {Melchior}, M. and {Merlin}, E. and {Meylan}, G. and {Mohr}, J.~J. and {Moresco}, M. and {Morin}, B. and {Moscardini}, L. and {Munari}, E. and {Nichol}, R.~C. and {Padilla}, C. and {Paltani}, S. and {Peacock}, J. and {Pedersen}, K. and {Pettorino}, V. and {Pires}, S. and {Poncet}, M. and {Popa}, L. and {Pozzetti}, L. and {Raison}, F. and {Rebolo}, R. and {Rhodes}, J. and {Rix}, H. -W. and {Roncarelli}, M. and {Rossetti}, E. and {Saglia}, R. and {Schneider}, P. and {Schrabback}, T. and {Secroun}, A. and {Seidel}, G. and {Serrano}, S. and {Sirignano}, C. and {Sirri}, G. and {Skottfelt}, J. and {Stanco}, L. and {Starck}, J.~L. and {Tallada-Cresp{\'\i}}, P. and {Tavagnacco}, D. and {Taylor}, A.~N. and {Teplitz}, H.~I. and {Toledo-Moreo}, R. and {Torradeflot}, F. and {Trifoglio}, M. and {Valentijn}, E.~A. and {Valenziano}, L. and {Verdoes Kleijn}, G.~A. and {Wang}, Y. and {Welikala}, N. and {Weller}, J. and {Wetzstein}, M. and {Zamorani}, G. and {Zoubian}, J. and {Andreon}, S. and {Baldi}, M. and {Bardelli}, S. and {Boucaud}, A. and {Camera}, S. and {Di Ferdinando}, D. and {Fabbian}, G. and {Farinelli}, R. and {Galeotta}, S. and {Graci{\'a}-Carpio}, J. and {Maino}, D. and {Medinaceli}, E. and {Mei}, S. and {Neissner}, C. and {Polenta}, G. and {Renzi}, A. and {Romelli}, E. and {Rosset}, C. and {Sureau}, F. and {Tenti}, M. and {Vassallo}, T. and {Zucca}, E. and {Baccigalupi}, C. and {Balaguera-Antol{\'\i}nez}, A. and {Battaglia}, P. and {Biviano}, A. and {Borgani}, S. and {Bozzo}, E. and {Cabanac}, R. and {Cappi}, A.},
        title = "{Euclid preparation. I. The Euclid Wide Survey}",
      journal = {\aap},
         year = 2022,
        month = jun,
       volume = {662},
          eid = {A112},
        pages = {A112},
          doi = {10.1051/0004-6361/202141938},
archivePrefix = {arXiv},
       eprint = {2108.01201},
 primaryClass = {astro-ph.CO},
       adsurl = {https://ui.adsabs.harvard.edu/abs/2022A&A...662A.112E}
}

@ARTICLE{2025sl,
       author = {{Euclid Collaboration} and {Walmsley}, M. and {Holloway}, P. and {Lines}, N.~E.~P. and {Rojas}, K. and {Collett}, T.~E. and {Verma}, A. and {Li}, T. and {Nightingale}, J.~W. and {Despali}, G. and {Schuldt}, S. and {Gavazzi}, R. and {Melo}, A. and {Metcalf}, R.~B. and {Andika}, I.~T. and {Leuzzi}, L. and {Manj{\'o}n-Garc{\'\i}a}, A. and {Pearce-Casey}, R. and {Vincken}, S.~H. and {Wilde}, J. and {Busillo}, V. and {Tortora}, C. and {Acevedo Barroso}, J.~A. and {Dole}, H. and {Ecker}, L.~R. and {Pearson}, J. and {Marshall}, P.~J. and {More}, A. and {Saifollahi}, T. and {Gracia-Carpio}, J. and {Baeten}, E. and {Cornen}, C. and {Johnson}, L.~C. and {Macmillan}, C. and {Kruk}, S. and {Remmelgas}, K.~A. and {Cl{\'e}ment}, B. and {Degaudenzi}, H. and {Courbin}, F. and {Bovy}, J. and {Casas}, S. and {Dannerbauer}, H. and {Diego}, J.~M. and {Finner}, K. and {Galan}, A. and {Giocoli}, C. and {Hogg}, N.~B. and {Jahnke}, K. and {Katona}, J. and {Kov{\'a}cs}, A. and {De Leo}, C. and {Mahler}, G. and {Millon}, M. and {Nagam}, B.~C. and {Nugent}, P. and {Sainz de Murieta}, A. and {O'Riordan}, C.~M. and {Sluse}, D. and {Sonnenfeld}, A. and {Spiniello}, C. and {Serjeant}, S. and {Thai}, T.~T. and {Ulivi}, L. and {Walth}, G.~L. and {Weisenbach}, L. and {Zumalacarregui}, M. and {Aghanim}, N. and {Altieri}, B. and {Amara}, A. and {Andreon}, S. and {Auricchio}, N. and {Aussel}, H. and {Baccigalupi}, C. and {Baldi}, M. and {Balestra}, A. and {Bardelli}, S. and {Battaglia}, P. and {Bernardeau}, F. and {Biviano}, A. and {Bonchi}, A. and {Bonino}, D. and {Branchini}, E. and {Brescia}, M. and {Brinchmann}, J. and {Camera}, S. and {Ca{\~n}as-Herrera}, G. and {Capobianco}, V. and {Carbone}, C. and {Cardone}, V.~F. and {Carretero}, J. and {Castander}, F.~J. and {Castellano}, M. and {Castignani}, G. and {Cavuoti}, S. and {Chambers}, K.~C. and {Cimatti}, A. and {Colodro-Conde}, C. and {Congedo}, G. and {Conselice}, C.~J. and {Conversi}, L. and {Copin}, Y. and {Corcione}, L. and {Courtois}, H.~M. and {Cropper}, M. and {Da Silva}, A. and {De Lucia}, G. and {Di Giorgio}, A.~M. and {Dolding}, C. and {Dubath}, F. and {Duncan}, C.~A.~J. and {Dupac}, X. and {Ealet}, A. and {Escoffier}, S. and {Fabricius}, M. and {Farina}, M. and {Farinelli}, R. and {Faustini}, F. and {Finelli}, F. and {Fotopoulou}, S. and {Frailis}, M. and {Franceschi}, E. and {Fumana}, M. and {Galeotta}, S. and {George}, K. and {Gillard}, W. and {Gillis}, B. and {G{\'o}mez-Alvarez}, P. and {Granett}, B.~R. and {Grazian}, A. and {Grupp}, F. and {Guzzo}, L. and {Gwyn}, S. and {Haugan}, S.~V.~H. and {Hoekstra}, H. and {Holmes}, W. and {Hook}, I.~M. and {Hormuth}, F. and {Hornstrup}, A. and {Hudelot}, P. and {Jhabvala}, M. and {Joachimi}, B. and {Keih{\"a}nen}, E. and {Kermiche}, S. and {Kiessling}, A. and {Kubik}, B. and {K{\"u}mmel}, M. and {Kunz}, M. and {Kurki-Suonio}, H. and {Lahav}, O. and {Le Boulc'h}, Q. and {Le Brun}, A.~M.~C. and {Le Mignant}, D. and {Ligori}, S. and {Lilje}, P.~B. and {Lindholm}, V. and {Lloro}, I. and {Mainetti}, G. and {Maino}, D. and {Maiorano}, E. and {Mansutti}, O. and {Marcin}, S. and {Marggraf}, O. and {Martinelli}, M. and {Martinet}, N. and {Marulli}, F. and {Massey}, R. and {Maurogordato}, S. and {McCracken}, H.~J. and {Medinaceli}, E. and {Mei}, S. and {Mellier}, Y. and {Meneghetti}, M. and {Merlin}, E. and {Meylan}, G. and {Mora}, A. and {Moresco}, M. and {Moscardini}, L. and {Nakajima}, R. and {Neissner}, C. and {Nichol}, R.~C. and {Niemi}, S. -M. and {Padilla}, C. and {Paltani}, S. and {Pasian}, F. and {Pedersen}, K. and {Percival}, W.~J. and {Pettorino}, V. and {Pires}, S. and {Polenta}, G. and {Poncet}, M. and {Popa}, L.~A. and {Pozzetti}, L. and {Raison}, F. and {Rebolo}, R. and {Renzi}, A. and {Rhodes}, J. and {Riccio}, G. and {Romelli}, E. and {Roncarelli}, M. and {Saglia}, R.},
        title = "{Euclid Quick Data Release (Q1): The Strong Lensing Discovery Engine A -- System overview and lens catalogue}",
      journal = {arXiv e-prints},
         year = 2025,
        month = mar,
          eid = {arXiv:2503.15324},
        pages = {arXiv:2503.15324},
          doi = {10.48550/arXiv.2503.15324},
archivePrefix = {arXiv},
       eprint = {2503.15324},
 primaryClass = {astro-ph.GA},
       adsurl = {https://ui.adsabs.harvard.edu/abs/2025arXiv250315324E}
}

@ARTICLE{2025Shajib,
       author = {{Shajib}, Anowar J. and {Smith}, Graham P. and {Birrer}, Simon and {Verma}, Aprajita and {Arendse}, Nikki and {Collett}, Thomas and {Daylan}, Tansu and {Serjeant}, Stephen and {LSST Strong Lensing Science Collaboration}},
        title = "{Strong gravitational lenses from the Vera C. Rubin Observatory}",
      journal = {Philosophical Transactions of the Royal Society of London Series A},
         year = 2025,
        month = may,
       volume = {383},
       number = {2295},
          eid = {20240117},
        pages = {20240117},
          doi = {10.1098/rsta.2024.0117},
archivePrefix = {arXiv},
       eprint = {2406.08919},
 primaryClass = {astro-ph.CO},
       adsurl = {https://ui.adsabs.harvard.edu/abs/2025RSPTA.38340117S}
}

@ARTICLE{2016Leier,
       author = {{Leier}, Dominik and {Ferreras}, Ignacio and {Saha}, Prasenjit and {Charlot}, St{\'e}phane and {Bruzual}, Gustavo and {La Barbera}, Francesco},
        title = "{Strong gravitational lensing and the stellar IMF of early-type galaxies}",
      journal = {\mnras},
         year = 2016,
        month = jul,
       volume = {459},
       number = {4},
        pages = {3677-3692},
          doi = {10.1093/mnras/stw885},
archivePrefix = {arXiv},
       eprint = {1512.00462},
 primaryClass = {astro-ph.GA},
       adsurl = {https://ui.adsabs.harvard.edu/abs/2016MNRAS.459.3677L}
}

@ARTICLE{2025TDCOSMO,
       author = {{TDCOSMO Collaboration} and {Birrer}, Simon and {Buckley-Geer}, Elizabeth J. and {Cappellari}, Michele and {Courbin}, Fr{\'e}d{\'e}ric and {Dux}, Fr{\'e}d{\'e}ric and {Fassnacht}, Christopher D. and {Frieman}, Joshua A. and {Galan}, Aymeric and {Gilman}, Daniel and {Huang}, Xiang-Yu and {Knabel}, Shawn and {Langeroodi}, Danial and {Lin}, Huan and {Millon}, Martin and {Morishita}, Takahiro and {Motta}, Veronica and {Mozumdar}, Pritom and {Paic}, Eric and {Shajib}, Anowar J. and {Sheu}, William and {Sluse}, Dominique and {Sonnenfeld}, Alessandro and {Spiniello}, Chiara and {Stiavelli}, Massimo and {Suyu}, Sherry H. and {Tan}, Chin Yi and {Treu}, Tommaso and {Van de Vyvere}, Lyne and {Wang}, Han and {Wells}, Patrick and {Williams}, Devon M. and {Wong}, Kenneth C.},
        title = "{TDCOSMO 2025: Cosmological constraints from strong lensing time delays}",
      journal = {arXiv e-prints},
         year = 2025,
        month = jun,
          eid = {arXiv:2506.03023},
        pages = {arXiv:2506.03023},
          doi = {10.48550/arXiv.2506.03023},
archivePrefix = {arXiv},
       eprint = {2506.03023},
 primaryClass = {astro-ph.CO},
       adsurl = {https://ui.adsabs.harvard.edu/abs/2025arXiv250603023T}
}

@ARTICLE{2018Collett,
       author = {{Collett}, Thomas E. and {Oldham}, Lindsay J. and {Smith}, Russell J. and {Auger}, Matthew W. and {Westfall}, Kyle B. and {Bacon}, David and {Nichol}, Robert C. and {Masters}, Karen L. and {Koyama}, Kazuya and {van den Bosch}, Remco},
        title = "{A precise extragalactic test of General Relativity}",
      journal = {Science},
         year = 2018,
        month = jun,
       volume = {360},
       number = {6395},
        pages = {1342-1346},
          doi = {10.1126/science.aao2469},
archivePrefix = {arXiv},
       eprint = {1806.08300},
 primaryClass = {astro-ph.CO},
       adsurl = {https://ui.adsabs.harvard.edu/abs/2018Sci...360.1342C}
}

@ARTICLE{2001Gerhard,
       author = {{Gerhard}, Ortwin and {Kronawitter}, Andi and {Saglia}, R.~P. and {Bender}, Ralf},
        title = "{Dynamical Family Properties and Dark Halo Scaling Relations of Giant Elliptical Galaxies}",
      journal = {\aj},
         year = 2001,
        month = apr,
       volume = {121},
       number = {4},
        pages = {1936-1951},
          doi = {10.1086/319940},
archivePrefix = {arXiv},
       eprint = {astro-ph/0012381},
 primaryClass = {astro-ph},
       adsurl = {https://ui.adsabs.harvard.edu/abs/2001AJ....121.1936G}
}

@ARTICLE{2003Koopmans,
       author = {{Koopmans}, L{\'e}on V.~E. and {Treu}, Tommaso},
        title = "{The Structure and Dynamics of Luminous and Dark Matter in the Early-Type Lens Galaxy of 0047-281 at z = 0.485}",
      journal = {\apj},
         year = 2003,
        month = feb,
       volume = {583},
       number = {2},
        pages = {606-615},
          doi = {10.1086/345423},
archivePrefix = {arXiv},
       eprint = {astro-ph/0205281},
 primaryClass = {astro-ph},
       adsurl = {https://ui.adsabs.harvard.edu/abs/2003ApJ...583..606K}
}

@ARTICLE{2004Treu,
       author = {{Treu}, Tommaso and {Koopmans}, L{\'e}on V.~E.},
        title = "{Massive Dark Matter Halos and Evolution of Early-Type Galaxies to z \raisebox{-0.5ex}\textasciitilde 1}",
      journal = {\apj},
         year = 2004,
        month = aug,
       volume = {611},
       number = {2},
        pages = {739-760},
          doi = {10.1086/422245},
archivePrefix = {arXiv},
       eprint = {astro-ph/0401373},
 primaryClass = {astro-ph},
       adsurl = {https://ui.adsabs.harvard.edu/abs/2004ApJ...611..739T}
}

@ARTICLE{2006Humphrey,
       author = {{Humphrey}, Philip J. and {Buote}, David A. and {Gastaldello}, Fabio and {Zappacosta}, Luca and {Bullock}, James S. and {Brighenti}, Fabrizio and {Mathews}, William G.},
        title = "{A Chandra View of Dark Matter in Early-Type Galaxies}",
      journal = {\apj},
         year = 2006,
        month = aug,
       volume = {646},
       number = {2},
        pages = {899-918},
          doi = {10.1086/505019},
archivePrefix = {arXiv},
       eprint = {astro-ph/0601301},
 primaryClass = {astro-ph},
       adsurl = {https://ui.adsabs.harvard.edu/abs/2006ApJ...646..899H}
}

@ARTICLE{2013Grillo,
       author = {{Grillo}, C. and {Christensen}, L. and {Gallazzi}, A. and {Rasmussen}, J.},
        title = "{Measuring the total and baryonic mass profiles of the very massive CASSOWARY 31 strong lens. A fossil system at z ≃ 0.7?}",
      journal = {\mnras},
         year = 2013,
        month = aug,
       volume = {433},
       number = {3},
        pages = {2604-2612},
          doi = {10.1093/mnras/stt930},
archivePrefix = {arXiv},
       eprint = {1305.5844},
 primaryClass = {astro-ph.CO},
       adsurl = {https://ui.adsabs.harvard.edu/abs/2013MNRAS.433.2604G}
}

@ARTICLE{2005Mamon,
       author = {{Mamon}, Gary A. and {{\L}okas}, Ewa L.},
        title = "{Dark matter in elliptical galaxies - II. Estimating the mass within the virial radius}",
      journal = {\mnras},
         year = 2005,
        month = nov,
       volume = {363},
       number = {3},
        pages = {705-722},
          doi = {10.1111/j.1365-2966.2005.09400.x},
archivePrefix = {arXiv},
       eprint = {astro-ph/0405491},
 primaryClass = {astro-ph},
       adsurl = {https://ui.adsabs.harvard.edu/abs/2005MNRAS.363..705M}
}

@ARTICLE{2007Diehl,
       author = {{Diehl}, Steven and {Statler}, Thomas S.},
        title = "{The Hot Interstellar Medium of Normal Elliptical Galaxies. I. A Chandra Gas Gallery and Comparison of X-Ray and Optical Morphology}",
      journal = {\apj},
         year = 2007,
        month = oct,
       volume = {668},
       number = {1},
        pages = {150-167},
          doi = {10.1086/521009},
archivePrefix = {arXiv},
       eprint = {astro-ph/0606215},
 primaryClass = {astro-ph},
       adsurl = {https://ui.adsabs.harvard.edu/abs/2007ApJ...668..150D}
}

@ARTICLE{2018Dries,
       author = {{Dries}, M. and {Trager}, S.~C. and {Koopmans}, L.~V.~E. and {Popping}, G. and {Somerville}, R.~S.},
        title = "{Hierarchical Bayesian inference of the initial mass function in composite stellar populations}",
      journal = {\mnras},
         year = 2018,
        month = mar,
       volume = {474},
       number = {3},
        pages = {3500-3515},
          doi = {10.1093/mnras/stx2979},
archivePrefix = {arXiv},
       eprint = {1711.07192},
 primaryClass = {astro-ph.GA},
       adsurl = {https://ui.adsabs.harvard.edu/abs/2018MNRAS.474.3500D}
}

@ARTICLE{2024Navarro,
       author = {{Mart{\'\i}n-Navarro}, Ignacio and {de Lorenzo-C{\'a}ceres}, Adriana and {Gadotti}, Dimitri A. and {M{\'e}ndez-Abreu}, Jairo and {Falc{\'o}n-Barroso}, Jes{\'u}s and {S{\'a}nchez-Bl{\'a}zquez}, Patricia and {Coelho}, Paula and {Neumann}, Justus and {van de Ven}, Glenn and {P{\'e}rez}, Isabel},
        title = "{The universal variability of the stellar initial mass function probed by the TIMER survey}",
      journal = {\aap},
         year = 2024,
        month = apr,
       volume = {684},
          eid = {A110},
        pages = {A110},
          doi = {10.1051/0004-6361/202348060},
archivePrefix = {arXiv},
       eprint = {2312.13355},
 primaryClass = {astro-ph.GA},
       adsurl = {https://ui.adsabs.harvard.edu/abs/2024A&A...684A.110M}
}

@ARTICLE{2009Conroy,
       author = {{Conroy}, Charlie and {Gunn}, James E. and {White}, Martin},
        title = "{The Propagation of Uncertainties in Stellar Population Synthesis Modeling. I. The Relevance of Uncertain Aspects of Stellar Evolution and the Initial Mass Function to the Derived Physical Properties of Galaxies}",
      journal = {\apj},
         year = 2009,
        month = jul,
       volume = {699},
       number = {1},
        pages = {486-506},
          doi = {10.1088/0004-637X/699/1/486},
archivePrefix = {arXiv},
       eprint = {0809.4261},
 primaryClass = {astro-ph},
       adsurl = {https://ui.adsabs.harvard.edu/abs/2009ApJ...699..486C}
}

@ARTICLE{2005Smith,
       author = {{Smith}, Russell J. and {Blakeslee}, John P. and {Lucey}, John R. and {Tonry}, John},
        title = "{Discovery of Strong Lensing by an Elliptical Galaxy at z = 0.0345}",
      journal = {\apjl},
         year = 2005,
        month = jun,
       volume = {625},
       number = {2},
        pages = {L103-L106},
          doi = {10.1086/431240},
archivePrefix = {arXiv},
       eprint = {astro-ph/0504453},
 primaryClass = {astro-ph},
       adsurl = {https://ui.adsabs.harvard.edu/abs/2005ApJ...625L.103S}
}

@ARTICLE{2025Riordan,
       author = {{O'Riordan}, C.~M. and {Oldham}, L.~J. and {Nersesian}, A. and {Li}, T. and {Collett}, T.~E. and {Sluse}, D. and {Altieri}, B. and {Cl{\'e}ment}, B. and {Vasan}, K.~G.~C. and {Rhoades}, S. and {Chen}, Y. and {Jones}, T. and {Adami}, C. and {Gavazzi}, R. and {Vegetti}, S. and {Powell}, D.~M. and {Acevedo Barroso}, J.~A. and {Andika}, I.~T. and {Bhatawdekar}, R. and {Cooray}, A.~R. and {Despali}, G. and {Diego}, J.~M. and {Ecker}, L.~R. and {Galan}, A. and {G{\'o}mez-Alvarez}, P. and {Leuzzi}, L. and {Meneghetti}, M. and {Metcalf}, R.~B. and {Schirmer}, M. and {Serjeant}, S. and {Tortora}, C. and {Vaccari}, M. and {Vernardos}, G. and {Walmsley}, M. and {Amara}, A. and {Andreon}, S. and {Auricchio}, N. and {Aussel}, H. and {Baccigalupi}, C. and {Baldi}, M. and {Balestra}, A. and {Bardelli}, S. and {Basset}, A. and {Battaglia}, P. and {Bender}, R. and {Bonino}, D. and {Branchini}, E. and {Brescia}, M. and {Brinchmann}, J. and {Caillat}, A. and {Camera}, S. and {Capobianco}, V. and {Carbone}, C. and {Carretero}, J. and {Casas}, S. and {Castander}, F.~J. and {Castellano}, M. and {Castignani}, G. and {Cavuoti}, S. and {Cimatti}, A. and {Colodro-Conde}, C. and {Congedo}, G. and {Conselice}, C.~J. and {Conversi}, L. and {Copin}, Y. and {Corcione}, L. and {Courbin}, F. and {Courtois}, H.~M. and {Cropper}, M. and {Da Silva}, A. and {Degaudenzi}, H. and {De Lucia}, G. and {Di Giorgio}, A.~M. and {Dinis}, J. and {Dubath}, F. and {Duncan}, C.~A.~J. and {Dupac}, X. and {Dusini}, S. and {Farina}, M. and {Farrens}, S. and {Faustini}, F. and {Ferriol}, S. and {Fourmanoit}, N. and {Frailis}, M. and {Franceschi}, E. and {Fumana}, M. and {Galeotta}, S. and {Gillard}, W. and {Gillis}, B. and {Giocoli}, C. and {Granett}, B.~R. and {Grazian}, A. and {Grupp}, F. and {Guzzo}, L. and {Haugan}, S.~V.~H. and {Hoar}, J. and {Hoekstra}, H. and {Holmes}, W. and {Hook}, I. and {Hormuth}, F. and {Hornstrup}, A. and {Hudelot}, P. and {Jahnke}, K. and {Jhabvala}, M. and {Joachimi}, B. and {Keih{\"a}nen}, E. and {Kermiche}, S. and {Kiessling}, A. and {Kilbinger}, M. and {Kohley}, R. and {Kubik}, B. and {K{\"u}mmel}, M. and {Kunz}, M. and {Kurki-Suonio}, H. and {Lahav}, O. and {Laureijs}, R. and {Le Mignant}, D. and {Ligori}, S. and {Lilje}, P.~B. and {Lindholm}, V. and {Lloro}, I. and {Mainetti}, G. and {Maiorano}, E. and {Mansutti}, O. and {Marggraf}, O. and {Markovic}, K. and {Martinelli}, M. and {Martinet}, N. and {Marulli}, F. and {Massey}, R. and {Medinaceli}, E. and {Mei}, S. and {Melchior}, M. and {Mellier}, Y. and {Merlin}, E. and {Meylan}, G. and {Moresco}, M. and {Moscardini}, L. and {Nakajima}, R. and {Nichol}, R.~C. and {Niemi}, S. -M. and {Nightingale}, J.~W. and {Padilla}, C. and {Paltani}, S. and {Pasian}, F. and {Pedersen}, K. and {Percival}, W.~J. and {Pettorino}, V. and {Pires}, S. and {Polenta}, G. and {Poncet}, M. and {Popa}, L.~A. and {Pozzetti}, L. and {Raison}, F. and {Rebolo}, R. and {Renzi}, A. and {Rhodes}, J. and {Riccio}, G. and {Rix}, H. -W. and {Romelli}, E. and {Roncarelli}, M. and {Rossetti}, E. and {Rusholme}, B. and {Saglia}, R. and {Sakr}, Z. and {S{\'a}nchez}, A.~G. and {Sapone}, D. and {Sartoris}, B. and {Schneider}, P. and {Schrabback}, T. and {Secroun}, A. and {Seidel}, G. and {Serrano}, S. and {Sirignano}, C. and {Sirri}, G. and {Stanco}, L. and {Steinwagner}, J. and {Tallada-Cresp{\'\i}}, P. and {Tereno}, I. and {Toledo-Moreo}, R. and {Torradeflot}, F. and {Tutusaus}, I. and {Valenziano}, L. and {Vassallo}, T. and {Verdoes Kleijn}, G. and {Veropalumbo}, A. and {Wang}, Y. and {Weller}, J. and {Zacchei}, A. and {Zamorani}, G. and {Zucca}, E. and {Burigana}, C. and {Casenove}, P. and {Mora}, A. and {Scottez}, V. and {Viel}, M. and {Jauzac}, M. and {Dannerbauer}, H.},
        title = "{Euclid: A complete Einstein ring in NGC 6505}",
      journal = {\aap},
         year = 2025,
        month = feb,
       volume = {694},
          eid = {A145},
        pages = {A145},
          doi = {10.1051/0004-6361/202453014},
archivePrefix = {arXiv},
       eprint = {2502.06505},
 primaryClass = {astro-ph.GA},
       adsurl = {https://ui.adsabs.harvard.edu/abs/2025A&A...694A.145O}
}

@INPROCEEDINGS{Bacon_2010,
       author = {{Bacon}, R. and {Accardo}, M. and {Adjali}, L. and {Anwand}, H. and {Bauer}, S. and {Biswas}, I. and {Blaizot}, J. and {Boudon}, D. and {Brau-Nogue}, S. and {Brinchmann}, J. and {Caillier}, P. and {Capoani}, L. and {Carollo}, C.~M. and {Contini}, T. and {Couderc}, P. and {Daguis{\'e}}, E. and {Deiries}, S. and {Delabre}, B. and {Dreizler}, S. and {Dubois}, J. and {Dupieux}, M. and {Dupuy}, C. and {Emsellem}, E. and {Fechner}, T. and {Fleischmann}, A. and {Fran{\c{c}}ois}, M. and {Gallou}, G. and {Gharsa}, T. and {Glindemann}, A. and {Gojak}, D. and {Guiderdoni}, B. and {Hansali}, G. and {Hahn}, T. and {Jarno}, A. and {Kelz}, A. and {Koehler}, C. and {Kosmalski}, J. and {Laurent}, F. and {Le Floch}, M. and {Lilly}, S.~J. and {Lizon}, J. -L. and {Loupias}, M. and {Manescau}, A. and {Monstein}, C. and {Nicklas}, H. and {Olaya}, J. -C. and {Pares}, L. and {Pasquini}, L. and {P{\'e}contal-Rousset}, A. and {Pell{\'o}}, R. and {Petit}, C. and {Popow}, E. and {Reiss}, R. and {Remillieux}, A. and {Renault}, E. and {Roth}, M. and {Rupprecht}, G. and {Serre}, D. and {Schaye}, J. and {Soucail}, G. and {Steinmetz}, M. and {Streicher}, O. and {Stuik}, R. and {Valentin}, H. and {Vernet}, J. and {Weilbacher}, P. and {Wisotzki}, L. and {Yerle}, N.},
        title = "{The MUSE second-generation VLT instrument}",
    booktitle = {Ground-based and Airborne Instrumentation for Astronomy III},
         year = 2010,
       editor = {{McLean}, Ian S. and {Ramsay}, Suzanne K. and {Takami}, Hideki},
       series = {Society of Photo-Optical Instrumentation Engineers (SPIE) Conference Series},
       volume = {7735},
        month = jul,
          eid = {773508},
        pages = {773508},
          doi = {10.1117/12.856027},
archivePrefix = {arXiv},
       eprint = {2211.16795},
 primaryClass = {astro-ph.IM},
       adsurl = {https://ui.adsabs.harvard.edu/abs/2010SPIE.7735E..08B}
}

@ARTICLE{Cappellari_2003,
       author = {{Cappellari}, Michele and {Copin}, Yannick},
        title = "{Adaptive spatial binning of integral-field spectroscopic data using Voronoi tessellations}",
      journal = {\mnras},
         year = 2003,
        month = jun,
       volume = {342},
       number = {2},
        pages = {345-354},
          doi = {10.1046/j.1365-8711.2003.06541.x},
archivePrefix = {arXiv},
       eprint = {astro-ph/0302262},
 primaryClass = {astro-ph},
       adsurl = {https://ui.adsabs.harvard.edu/abs/2003MNRAS.342..345C}
}

@ARTICLE{Falcon_Barroso_2011,
       author = {{Falc{\'o}n-Barroso}, J. and {S{\'a}nchez-Bl{\'a}zquez}, P. and {Vazdekis}, A. and {Ricciardelli}, E. and {Cardiel}, N. and {Cenarro}, A.~J. and {Gorgas}, J. and {Peletier}, R.~F.},
        title = "{An updated MILES stellar library and stellar population models}",
      journal = {\aap},
         year = 2011,
        month = aug,
       volume = {532},
          eid = {A95},
        pages = {A95},
          doi = {10.1051/0004-6361/201116842},
archivePrefix = {arXiv},
       eprint = {1107.2303},
 primaryClass = {astro-ph.CO},
       adsurl = {https://ui.adsabs.harvard.edu/abs/2011A&A...532A..95F}
}

@ARTICLE{Sanchez_2006,
       author = {{S{\'a}nchez-Bl{\'a}zquez}, P. and {Peletier}, R.~F. and {Jim{\'e}nez-Vicente}, J. and {Cardiel}, N. and {Cenarro}, A.~J. and {Falc{\'o}n-Barroso}, J. and {Gorgas}, J. and {Selam}, S. and {Vazdekis}, A.},
        title = "{Medium-resolution Isaac Newton Telescope library of empirical spectra}",
      journal = {\mnras},
         year = 2006,
        month = sep,
       volume = {371},
       number = {2},
        pages = {703-718},
          doi = {10.1111/j.1365-2966.2006.10699.x},
archivePrefix = {arXiv},
       eprint = {astro-ph/0607009},
 primaryClass = {astro-ph},
       adsurl = {https://ui.adsabs.harvard.edu/abs/2006MNRAS.371..703S}
}

@ARTICLE{van_der_Marel_1993,
       author = {{van der Marel}, Roeland P. and {Franx}, Marijn},
        title = "{A New Method for the Identification of Non-Gaussian Line Profiles in Elliptical Galaxies}",
      journal = {\apj},
         year = 1993,
        month = apr,
       volume = {407},
        pages = {525},
          doi = {10.1086/172534},
       adsurl = {https://ui.adsabs.harvard.edu/abs/1993ApJ...407..525V}
}

@ARTICLE{Sanders_2015,
       author = {{Sanders}, Jason L. and {Evans}, N. Wyn},
        title = "{Self-consistent triaxial models}",
      journal = {\mnras},
         year = 2015,
        month = nov,
       volume = {454},
       number = {1},
        pages = {299-314},
          doi = {10.1093/mnras/stv1898},
archivePrefix = {arXiv},
       eprint = {1507.04129},
 primaryClass = {astro-ph.GA},
       adsurl = {https://ui.adsabs.harvard.edu/abs/2015MNRAS.454..299S}
}

@ARTICLE{Krajnovic_2011,
       author = {{Krajnovi{\'c}}, Davor and {Emsellem}, Eric and {Cappellari}, Michele and {Alatalo}, Katherine and {Blitz}, Leo and {Bois}, Maxime and {Bournaud}, Fr{\'e}d{\'e}ric and {Bureau}, Martin and {Davies}, Roger L. and {Davis}, Timothy A. and {de Zeeuw}, P.~T. and {Khochfar}, Sadegh and {Kuntschner}, Harald and {Lablanche}, Pierre-Yves and {McDermid}, Richard M. and {Morganti}, Raffaella and {Naab}, Thorsten and {Oosterloo}, Tom and {Sarzi}, Marc and {Scott}, Nicholas and {Serra}, Paolo and {Weijmans}, Anne-Marie and {Young}, Lisa M.},
        title = "{The ATLAS$^{3D}$ project - II. Morphologies, kinemetric features and alignment between photometric and kinematic axes of early-type galaxies}",
      journal = {\mnras},
         year = 2011,
        month = jul,
       volume = {414},
       number = {4},
        pages = {2923-2949},
          doi = {10.1111/j.1365-2966.2011.18560.x},
archivePrefix = {arXiv},
       eprint = {1102.3801},
 primaryClass = {astro-ph.CO},
       adsurl = {https://ui.adsabs.harvard.edu/abs/2011MNRAS.414.2923K}
}

@ARTICLE{Scorza_1995,
       author = {{Scorza}, C. and {Bender}, R.},
        title = "{The internal structure of disky elliptical galaxies.}",
      journal = {\aap},
         year = 1995,
        month = jan,
       volume = {293},
        pages = {20-43},
       adsurl = {https://ui.adsabs.harvard.edu/abs/1995A&A...293...20S}
}

@ARTICLE{Jungwiert_1997,
       author = {{Jungwiert}, B. and {Combes}, F. and {Axon}, D.~J.},
        title = "{Near-IR photometry of disk galaxies: Search for nuclear isophotal twist and double bars}",
      journal = {\aaps},
         year = 1997,
        month = nov,
       volume = {125},
        pages = {479-496},
          doi = {10.1051/aas:1997236},
archivePrefix = {arXiv},
       eprint = {astro-ph/9705175},
 primaryClass = {astro-ph},
       adsurl = {https://ui.adsabs.harvard.edu/abs/1997A&AS..125..479J}
}

@ARTICLE{Franx_1991,
       author = {{Franx}, Marijn and {Illingworth}, Garth and {de Zeeuw}, Tim},
        title = "{The Ordered Nature of Elliptical Galaxies: Implications for Their Intrinsic Angular Momenta and Shapes}",
      journal = {\apj},
         year = 1991,
        month = dec,
       volume = {383},
        pages = {112},
          doi = {10.1086/170769},
       adsurl = {https://ui.adsabs.harvard.edu/abs/1991ApJ...383..112F}
}

@ARTICLE{de_Nicola_2020,
       author = {{de Nicola}, Stefano and {Saglia}, Roberto P. and {Thomas}, Jens and {Dehnen}, Walter and {Bender}, Ralf},
        title = "{Non-parametric triaxial deprojection of elliptical galaxies}",
      journal = {\mnras},
         year = 2020,
        month = aug,
       volume = {496},
       number = {3},
        pages = {3076-3100},
          doi = {10.1093/mnras/staa1703},
archivePrefix = {arXiv},
       eprint = {2006.05971},
 primaryClass = {astro-ph.GA},
       adsurl = {https://ui.adsabs.harvard.edu/abs/2020MNRAS.496.3076D}
}

@INPROCEEDINGS{Rybicki_1987,
       author = {{Rybicki}, G.~B.},
        title = "{Deprojection of Galaxies - how much can BE Learned}",
    booktitle = {Structure and Dynamics of Elliptical Galaxies},
         year = 1987,
       editor = {{de Zeeuw}, Pieter Timotheus},
       series = {IAU Symposium},
       volume = {127},
        month = jan,
        pages = {397},
          doi = {10.1007/978-94-009-3971-4_41},
       adsurl = {https://ui.adsabs.harvard.edu/abs/1987IAUS..127..397R}
}

@ARTICLE{Gerhard_1996,
       author = {{Gerhard}, Ortwin E. and {Binney}, James J.},
        title = "{On the deprojection of axisymmetric bodies}",
      journal = {MNRAS},
         year = "1996",
        month = "Apr",
       volume = {279},
        pages = {993},
          doi = {10.1093/mnras/279.3.993},
archivePrefix = {arXiv},
       eprint = {astro-ph/9508116},
 primaryClass = {astro-ph},
       adsurl = {https://ui.adsabs.harvard.edu/abs/1996MNRAS.279..993G}
}

@ARTICLE{de_Nicola_2022,
       author = {{de Nicola}, Stefano and {Saglia}, Roberto P. and {Thomas}, Jens and {Pulsoni}, Claudia and {Kluge}, Matthias and {Bender}, Ralf and {Valenzuela}, Lucas M. and {Remus}, Rhea-Silvia},
        title = "{Intrinsic Shapes of Brightest Cluster Galaxies}",
      journal = {\apj},
         year = 2022,
        month = jul,
       volume = {933},
       number = {2},
          eid = {215},
        pages = {215},
          doi = {10.3847/1538-4357/ac7463},
archivePrefix = {arXiv},
       eprint = {2205.15672},
 primaryClass = {astro-ph.GA},
       adsurl = {https://ui.adsabs.harvard.edu/abs/2022ApJ...933..215D}
}

@ARTICLE{Neureiter_2023,
       author = {{Neureiter}, Bianca and {Thomas}, Jens and {Rantala}, Antti and {Naab}, Thorsten and {Mehrgan}, Kianusch and {Saglia}, Roberto and {de Nicola}, Stefano and {Bender}, Ralf},
        title = "{The Isotropic Center of NGC 5419-A Core in Formation?}",
      journal = {\apj},
         year = 2023,
        month = jun,
       volume = {950},
       number = {1},
          eid = {15},
        pages = {15},
          doi = {10.3847/1538-4357/accffa},
archivePrefix = {arXiv},
       eprint = {2305.03078},
 primaryClass = {astro-ph.GA},
       adsurl = {https://ui.adsabs.harvard.edu/abs/2023ApJ...950...15N}
}

@ARTICLE{Barazza_2003,
       author = {{Barazza}, F.~D. and {Binggeli}, B. and {Jerjen}, H.},
        title = "{VLT surface photometry and isophotal analysis of early-type dwarf galaxies in the Virgo cluster}",
      journal = {\aap},
         year = 2003,
        month = aug,
       volume = {407},
        pages = {121-135},
          doi = {10.1051/0004-6361:20030872},
archivePrefix = {arXiv},
       eprint = {astro-ph/0306339},
 primaryClass = {astro-ph},
       adsurl = {https://ui.adsabs.harvard.edu/abs/2003A&A...407..121B}
}

@ARTICLE{Neureiter_2021,
       author = {{Neureiter}, B. and {Thomas}, J. and {Saglia}, R. and {Bender}, R. and {Finozzi}, F. and {Krukau}, A. and {Naab}, T. and {Rantala}, A. and {Frigo}, M.},
        title = "{SMART: a new implementation of Schwarzschild's Orbit Superposition technique for triaxial galaxies and its application to an N-body merger simulation}",
      journal = {\mnras},
         year = 2021,
        month = jan,
       volume = {500},
       number = {1},
        pages = {1437-1465},
          doi = {10.1093/mnras/staa3014},
archivePrefix = {arXiv},
       eprint = {2009.08979},
 primaryClass = {astro-ph.GA},
       adsurl = {https://ui.adsabs.harvard.edu/abs/2021MNRAS.500.1437N}
}

@ARTICLE{Neureiter_2023_B,
       author = {{Neureiter}, B. and {de Nicola}, S. and {Thomas}, J. and {Saglia}, R. and {Bender}, R. and {Rantala}, A.},
        title = "{Accuracy and precision of triaxial orbit models I: SMBH mass, stellar mass, and dark-matter halo}",
      journal = {\mnras},
         year = 2023,
        month = feb,
       volume = {519},
       number = {2},
        pages = {2004-2016},
          doi = {10.1093/mnras/stac3652},
archivePrefix = {arXiv},
       eprint = {2212.06173},
 primaryClass = {astro-ph.GA},
       adsurl = {https://ui.adsabs.harvard.edu/abs/2023MNRAS.519.2004N}
}

@ARTICLE{deNicola_2022_B,
       author = {{de Nicola}, Stefano and {Neureiter}, Bianca and {Thomas}, Jens and {Saglia}, Roberto P. and {Bender}, Ralf},
        title = "{Accuracy and precision of triaxial orbit models - II. Viewing angles, shape, and orbital structure}",
      journal = {\mnras},
         year = 2022,
        month = dec,
       volume = {517},
       number = {3},
        pages = {3445-3458},
          doi = {10.1093/mnras/stac2852},
archivePrefix = {arXiv},
       eprint = {2210.01893},
 primaryClass = {astro-ph.GA},
       adsurl = {https://ui.adsabs.harvard.edu/abs/2022MNRAS.517.3445D}
}

@ARTICLE{Thomas_2007_B,
       author = {{Thomas}, J. and {Jesseit}, R. and {Naab}, T. and {Saglia}, R.~P. and {Burkert}, A. and {Bender}, R.},
        title = "{Axisymmetric orbit models of N-body merger remnants: a dependency of reconstructed mass on viewing angle}",
      journal = {\mnras},
         year = 2007,
        month = nov,
       volume = {381},
       number = {4},
        pages = {1672-1696},
          doi = {10.1111/j.1365-2966.2007.12343.x},
archivePrefix = {arXiv},
       eprint = {0708.2205},
 primaryClass = {astro-ph},
       adsurl = {https://ui.adsabs.harvard.edu/abs/2007MNRAS.381.1672T}
}

@ARTICLE{1996Zhao,
       author = {{Zhao}, Hongsheng},
        title = "{Analytical models for galactic nuclei}",
      journal = {\mnras},
         year = 1996,
        month = jan,
       volume = {278},
       number = {2},
        pages = {488-496},
          doi = {10.1093/mnras/278.2.488},
archivePrefix = {arXiv},
       eprint = {astro-ph/9509122},
 primaryClass = {astro-ph},
       adsurl = {https://ui.adsabs.harvard.edu/abs/1996MNRAS.278..488Z}
}

@ARTICLE{2014Weijmans,
       author = {{Weijmans}, Anne-Marie and {de Zeeuw}, P.~T. and {Emsellem}, Eric and {Krajnovi{\'c}}, Davor and {Lablanche}, Pierre-Yves and {Alatalo}, Katherine and {Blitz}, Leo and {Bois}, Maxime and {Bournaud}, Fr{\'e}d{\'e}ric and {Bureau}, Martin and {Cappellari}, Michele and {Crocker}, Alison F. and {Davies}, Roger L. and {Davis}, Timothy A. and {Duc}, Pierre-Alain and {Khochfar}, Sadegh and {Kuntschner}, Harald and {McDermid}, Richard M. and {Morganti}, Raffaella and {Naab}, Thorsten and {Oosterloo}, Tom and {Sarzi}, Marc and {Scott}, Nicholas and {Serra}, Paolo and {Verdoes Kleijn}, Gijs and {Young}, Lisa M.},
        title = "{The ATLAS $^{3D}$ project - XXIV. The intrinsic shape distribution of early-type galaxies}",
      journal = {\mnras},
         year = 2014,
        month = nov,
       volume = {444},
       number = {4},
        pages = {3340-3356},
          doi = {10.1093/mnras/stu1603},
archivePrefix = {arXiv},
       eprint = {1408.1099},
 primaryClass = {astro-ph.GA},
       adsurl = {https://ui.adsabs.harvard.edu/abs/2014MNRAS.444.3340W}
}

@ARTICLE{2015Spiniello,
       author = {{Spiniello}, C. and {Barnab{\`e}}, M. and {Koopmans}, L.~V.~E. and {Trager}, S.~C.},
        title = "{Are the total mass density and the low-mass end slope of the IMF anticorrelated?}",
      journal = {\mnras},
         year = 2015,
        month = sep,
       volume = {452},
       number = {1},
        pages = {L21-L25},
          doi = {10.1093/mnrasl/slv079},
archivePrefix = {arXiv},
       eprint = {1505.07450},
 primaryClass = {astro-ph.GA},
       adsurl = {https://ui.adsabs.harvard.edu/abs/2015MNRAS.452L..21S}
}

@ARTICLE{Navarro1996,
       author = {{Navarro}, Julio F. and {Frenk}, Carlos S. and {White}, Simon D.~M.},
        title = "{The Structure of Cold Dark Matter Halos}",
      journal = {\apj},
         year = 1996,
        month = may,
       volume = {462},
        pages = {563},
          doi = {10.1086/177173},
archivePrefix = {arXiv},
       eprint = {astro-ph/9508025},
 primaryClass = {astro-ph},
       adsurl = {https://ui.adsabs.harvard.edu/abs/1996ApJ...462..563N}
}

@ARTICLE{Navarro1997,
       author = {{Navarro}, Julio F. and {Frenk}, Carlos S. and {White}, Simon D.~M.},
        title = "{A Universal Density Profile from Hierarchical Clustering}",
      journal = {\apj},
         year = 1997,
        month = dec,
       volume = {490},
       number = {2},
        pages = {493-508},
          doi = {10.1086/304888},
archivePrefix = {arXiv},
       eprint = {astro-ph/9611107},
 primaryClass = {astro-ph},
       adsurl = {https://ui.adsabs.harvard.edu/abs/1997ApJ...490..493N}
}

@ARTICLE{2013Barnabe,
       author = {{Barnab{\`e}}, Matteo and {Spiniello}, Chiara and {Koopmans}, L{\'e}on V.~E. and {Trager}, Scott C. and {Czoske}, Oliver and {Treu}, Tommaso},
        title = "{A low-mass cut-off near the hydrogen burning limit for Salpeter-like initial mass functions in early-type galaxies}",
      journal = {\mnras},
         year = 2013,
        month = nov,
       volume = {436},
       number = {1},
        pages = {253-258},
          doi = {10.1093/mnras/stt1727},
archivePrefix = {arXiv},
       eprint = {1306.2635},
 primaryClass = {astro-ph.CO},
       adsurl = {https://ui.adsabs.harvard.edu/abs/2013MNRAS.436..253B}
}

@ARTICLE{2024Lipka,
       author = {{Lipka}, Mathias and {Thomas}, Jens and {Saglia}, Roberto and {Bender}, Ralf and {Fabricius}, Maximilian and {Hill}, Gary J. and {Kluge}, Matthias and {Landriau}, Martin and {Mazzalay}, Ximena and {Noyola}, Eva and {Parikh}, Taniya and {Snigula}, Jan},
        title = "{The VIRUS-dE Survey. I. Stars in Dwarf Elliptical Galaxies{\textemdash}3D Dynamics and Radially Resolved Stellar Initial Mass Functions}",
      journal = {\apj},
         year = 2024,
        month = nov,
       volume = {976},
       number = {1},
          eid = {16},
        pages = {16},
          doi = {10.3847/1538-4357/ad7bac},
archivePrefix = {arXiv},
       eprint = {2409.10518},
 primaryClass = {astro-ph.GA},
       adsurl = {https://ui.adsabs.harvard.edu/abs/2024ApJ...976...16L}
}

@ARTICLE{2024Lipka2,
       author = {{Lipka}, Mathias and {Thomas}, Jens and {Saglia}, Roberto and {Bender}, Ralf and {Fabricius}, Maximilian and {Partmann}, Christian},
        title = "{The VIRUS-dE Survey. II. Cuspy and Round Halos in Dwarf Ellipticals{\textemdash}A Result of Early Assembly?}",
      journal = {\apj},
         year = 2024,
        month = nov,
       volume = {976},
       number = {1},
          eid = {17},
        pages = {17},
          doi = {10.3847/1538-4357/ad7baa},
archivePrefix = {arXiv},
       eprint = {2409.11458},
 primaryClass = {astro-ph.GA},
       adsurl = {https://ui.adsabs.harvard.edu/abs/2024ApJ...976...17L}
}

@ARTICLE{1993Kassiola,
       author = {{Kassiola}, Aggeliki and {Kovner}, Israel},
        title = "{Elliptic Mass Distributions versus Elliptic Potentials in Gravitational Lenses}",
      journal = {\apj},
         year = 1993,
        month = nov,
       volume = {417},
        pages = {450},
          doi = {10.1086/173325},
       adsurl = {https://ui.adsabs.harvard.edu/abs/1993ApJ...417..450K}
}

@ARTICLE{2019Ivezi,
       author = {{Ivezi{\'c}}, {\v{Z}}eljko and {Kahn}, Steven M. and {Tyson}, J. Anthony and {Abel}, Bob and {Acosta}, Emily and {Allsman}, Robyn and {Alonso}, David and {AlSayyad}, Yusra and {Anderson}, Scott F. and {Andrew}, John and {Angel}, James Roger P. and {Angeli}, George Z. and {Ansari}, Reza and {Antilogus}, Pierre and {Araujo}, Constanza and {Armstrong}, Robert and {Arndt}, Kirk T. and {Astier}, Pierre and {Aubourg}, {\'E}ric and {Auza}, Nicole and {Axelrod}, Tim S. and {Bard}, Deborah J. and {Barr}, Jeff D. and {Barrau}, Aurelian and {Bartlett}, James G. and {Bauer}, Amanda E. and {Bauman}, Brian J. and {Baumont}, Sylvain and {Bechtol}, Ellen and {Bechtol}, Keith and {Becker}, Andrew C. and {Becla}, Jacek and {Beldica}, Cristina and {Bellavia}, Steve and {Bianco}, Federica B. and {Biswas}, Rahul and {Blanc}, Guillaume and {Blazek}, Jonathan and {Blandford}, Roger D. and {Bloom}, Josh S. and {Bogart}, Joanne and {Bond}, Tim W. and {Booth}, Michael T. and {Borgland}, Anders W. and {Borne}, Kirk and {Bosch}, James F. and {Boutigny}, Dominique and {Brackett}, Craig A. and {Bradshaw}, Andrew and {Brandt}, William Nielsen and {Brown}, Michael E. and {Bullock}, James S. and {Burchat}, Patricia and {Burke}, David L. and {Cagnoli}, Gianpietro and {Calabrese}, Daniel and {Callahan}, Shawn and {Callen}, Alice L. and {Carlin}, Jeffrey L. and {Carlson}, Erin L. and {Chandrasekharan}, Srinivasan and {Charles-Emerson}, Glenaver and {Chesley}, Steve and {Cheu}, Elliott C. and {Chiang}, Hsin-Fang and {Chiang}, James and {Chirino}, Carol and {Chow}, Derek and {Ciardi}, David R. and {Claver}, Charles F. and {Cohen-Tanugi}, Johann and {Cockrum}, Joseph J. and {Coles}, Rebecca and {Connolly}, Andrew J. and {Cook}, Kem H. and {Cooray}, Asantha and {Covey}, Kevin R. and {Cribbs}, Chris and {Cui}, Wei and {Cutri}, Roc and {Daly}, Philip N. and {Daniel}, Scott F. and {Daruich}, Felipe and {Daubard}, Guillaume and {Daues}, Greg and {Dawson}, William and {Delgado}, Francisco and {Dellapenna}, Alfred and {de Peyster}, Robert and {de Val-Borro}, Miguel and {Digel}, Seth W. and {Doherty}, Peter and {Dubois}, Richard and {Dubois-Felsmann}, Gregory P. and {Durech}, Josef and {Economou}, Frossie and {Eifler}, Tim and {Eracleous}, Michael and {Emmons}, Benjamin L. and {Fausti Neto}, Angelo and {Ferguson}, Henry and {Figueroa}, Enrique and {Fisher-Levine}, Merlin and {Focke}, Warren and {Foss}, Michael D. and {Frank}, James and {Freemon}, Michael D. and {Gangler}, Emmanuel and {Gawiser}, Eric and {Geary}, John C. and {Gee}, Perry and {Geha}, Marla and {Gessner}, Charles J.~B. and {Gibson}, Robert R. and {Gilmore}, D. Kirk and {Glanzman}, Thomas and {Glick}, William and {Goldina}, Tatiana and {Goldstein}, Daniel A. and {Goodenow}, Iain and {Graham}, Melissa L. and {Gressler}, William J. and {Gris}, Philippe and {Guy}, Leanne P. and {Guyonnet}, Augustin and {Haller}, Gunther and {Harris}, Ron and {Hascall}, Patrick A. and {Haupt}, Justine and {Hernandez}, Fabio and {Herrmann}, Sven and {Hileman}, Edward and {Hoblitt}, Joshua and {Hodgson}, John A. and {Hogan}, Craig and {Howard}, James D. and {Huang}, Dajun and {Huffer}, Michael E. and {Ingraham}, Patrick and {Innes}, Walter R. and {Jacoby}, Suzanne H. and {Jain}, Bhuvnesh and {Jammes}, Fabrice and {Jee}, M. James and {Jenness}, Tim and {Jernigan}, Garrett and {Jevremovi{\'c}}, Darko and {Johns}, Kenneth and {Johnson}, Anthony S. and {Johnson}, Margaret W.~G. and {Jones}, R. Lynne and {Juramy-Gilles}, Claire and {Juri{\'c}}, Mario and {Kalirai}, Jason S. and {Kallivayalil}, Nitya J. and {Kalmbach}, Bryce and {Kantor}, Jeffrey P. and {Karst}, Pierre and {Kasliwal}, Mansi M. and {Kelly}, Heather and {Kessler}, Richard and {Kinnison}, Veronica and {Kirkby}, David and {Knox}, Lloyd and {Kotov}, Ivan V. and {Krabbendam}, Victor L. and {Krughoff}, K. Simon and {Kub{\'a}nek}, Petr and {Kuczewski}, John and {Kulkarni}, Shri and {Ku}, John and {Kurita}, Nadine R. and {Lage}, Craig S. and {Lambert}, Ron and {Lange}, Travis and {Langton}, J. Brian and {Le Guillou}, Laurent and {Levine}, Deborah and {Liang}, Ming and {Lim}, Kian-Tat and {Lintott}, Chris J. and {Long}, Kevin E. and {Lopez}, Margaux and {Lotz}, Paul J. and {Lupton}, Robert H. and {Lust}, Nate B. and {MacArthur}, Lauren A. and {Mahabal}, Ashish and {Mandelbaum}, Rachel and {Markiewicz}, Thomas W. and {Marsh}, Darren S. and {Marshall}, Philip J. and {Marshall}, Stuart and {May}, Morgan and {McKercher}, Robert and {McQueen}, Michelle and {Meyers}, Joshua and {Migliore}, Myriam and {Miller}, Michelle and {Mills}, David J.},
        title = "{LSST: From Science Drivers to Reference Design and Anticipated Data Products}",
      journal = {\apj},
         year = 2019,
        month = mar,
       volume = {873},
       number = {2},
          eid = {111},
        pages = {111},
          doi = {10.3847/1538-4357/ab042c},
archivePrefix = {arXiv},
       eprint = {0805.2366},
 primaryClass = {astro-ph},
       adsurl = {https://ui.adsabs.harvard.edu/abs/2019ApJ...873..111I}
}
\newpage

\begin{appendix}
\onecolumn
\section{Extended image modeling of ESO0286}
In this appendix, we present the detailed results of the extended source reconstruction using an Elliptical Power Law (EPL) mass model for the lens ESO0286. The following figures illustrate the data, the reconstructed models, the source plane reconstructions, and the corresponding residuals for the F814W and F336W bands.

\begin{figure}[H]
    \centering
    \includegraphics[width=1.\textwidth]{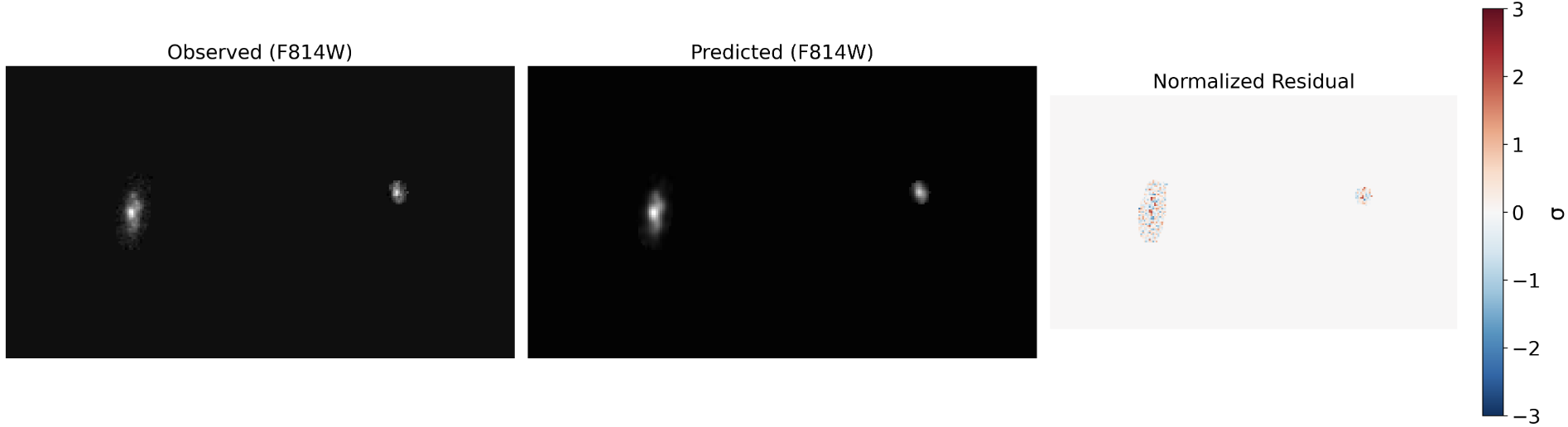}
    \caption{The best-fit EPL-ESR model in HST-imaging F814W band. }
    \label{fig:F814W}
\end{figure}

\begin{figure}[H]
    \centering
    \includegraphics[width=1.\textwidth]{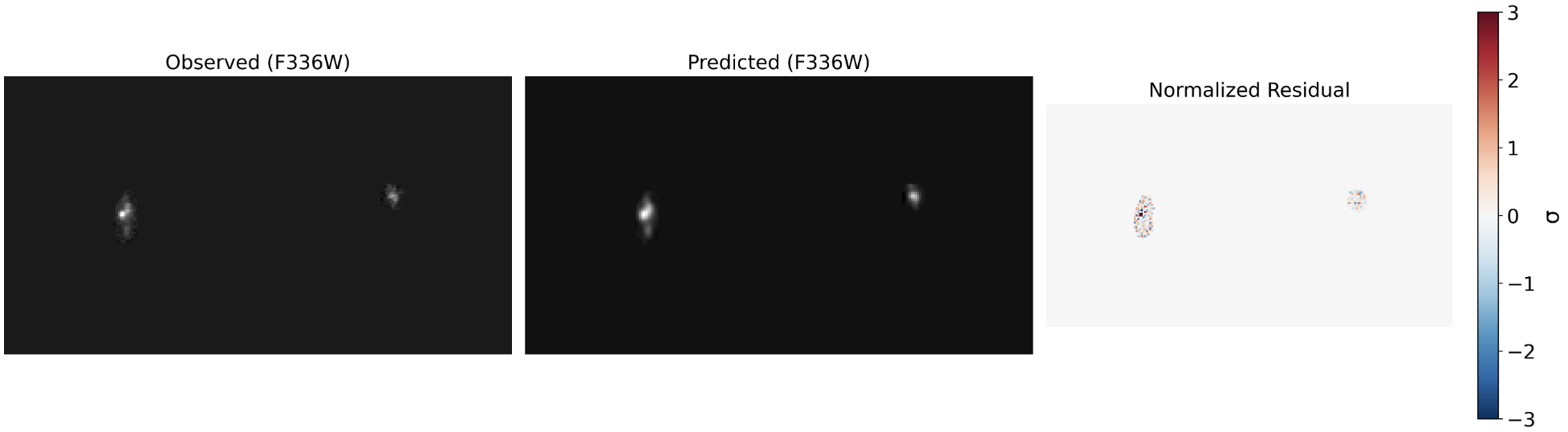}
    \caption{The best-fit model EPL-ESR in HST-imaging F336W band.}
    \label{fig:F336W}
\end{figure}

\begin{figure}[H]
    \centering
    \includegraphics[width=0.59\textwidth]{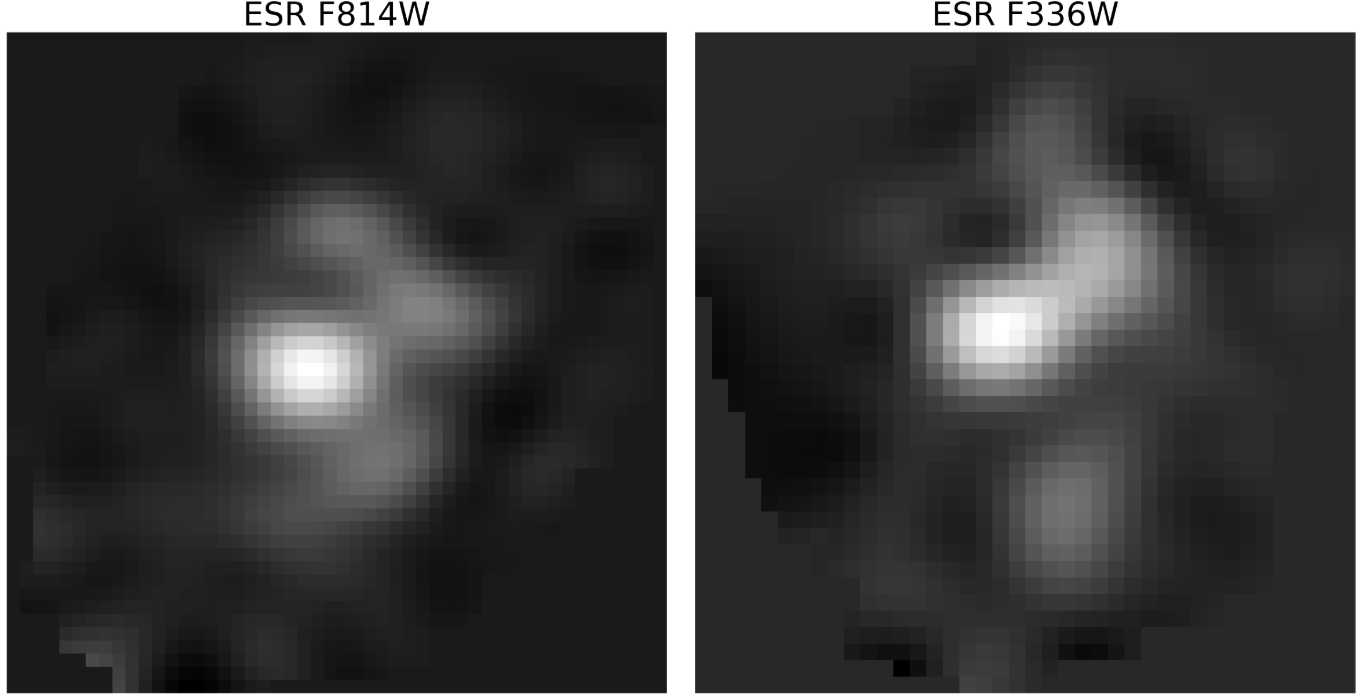}
    \caption{Extended source reconstruction on F814W and F336W bands based on the best-fit EPL-ESR model. }
    \label{fig:F336W}
\end{figure}

\section{Kinematics fitting}
\label{app:Kinematics fitting}
We demonstrate the quality of the kinematic extraction by detailing the best-fit model and resulting residuals for a representative Voronoi bin of the ESO0286 MUSE spectrum in Fig.~\ref{fig:spectrum_example}.

\begin{figure}[H]
    \centering
    \includegraphics[width=1.\textwidth]{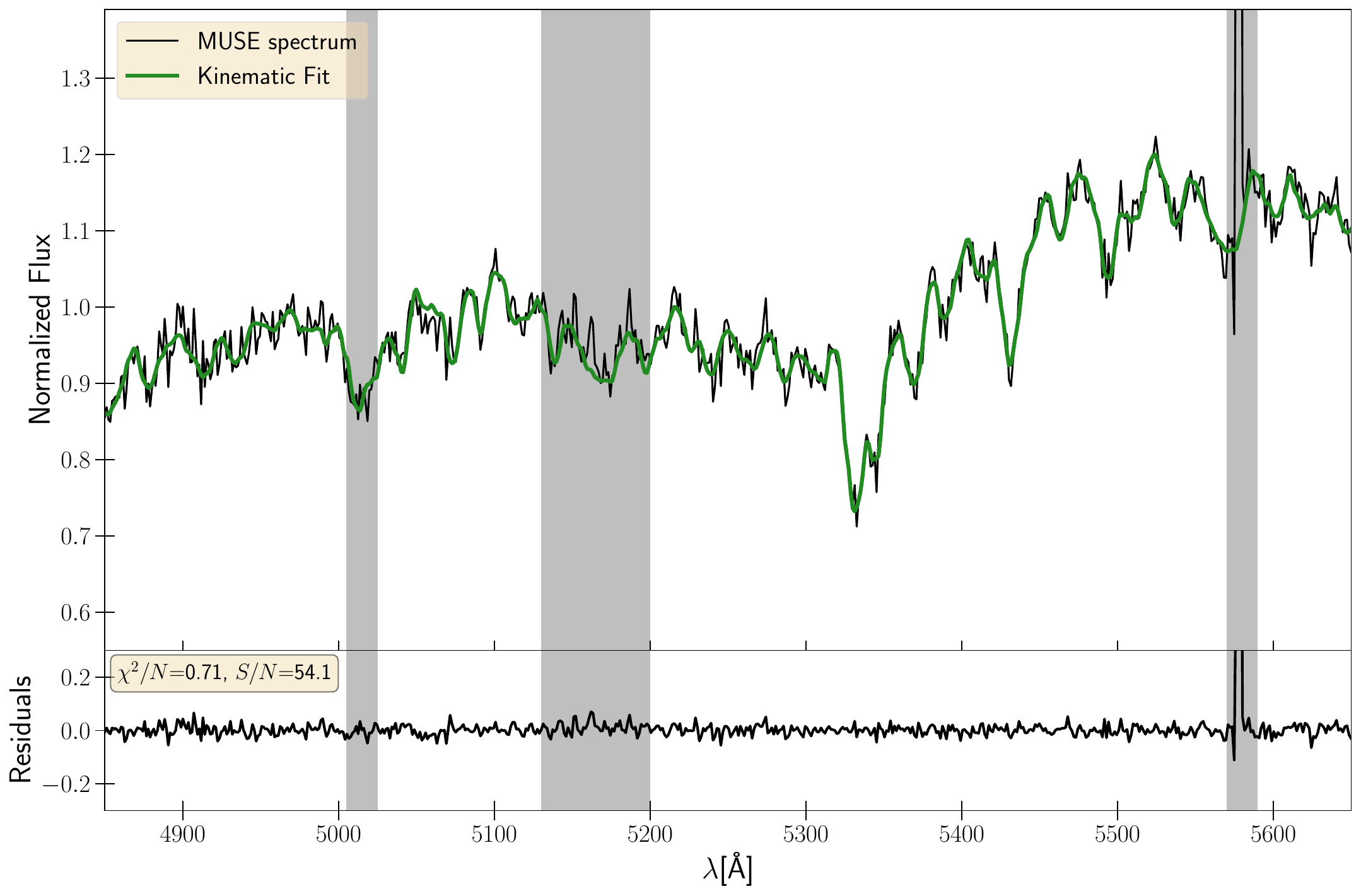}
    \caption{Top panel: WFM MUSE spectrum (black) of a typical Voronoi Bin of ESO0286 together with its kinematic-fit model (green). The 3 grey regions were masked globally for all Voronoi bins, because some of the bins had visible $\mathrm{H}\beta$ and \ion{O}{iii} emission, or were affected by telluric lines (\ion{O}{i}). The bottom panel shows the model residuals and fit statistics that was achieved for this example Voronoi bin.}
    \label{fig:spectrum_example}
\end{figure}

\section{Best-fit kinematics}
\label{app:Best-fit kinematics}
\subsection{Axisymmetric mass model}
To establish a baseline for our dynamical analysis, we first evaluate the kinematic fits and residuals produced by an axisymmetric mass model of ESO0286 (see Fig.~\ref{fig:SB_fitting_2D}).
\begin{figure*}
    \centering
    \includegraphics[width=1.\textwidth]{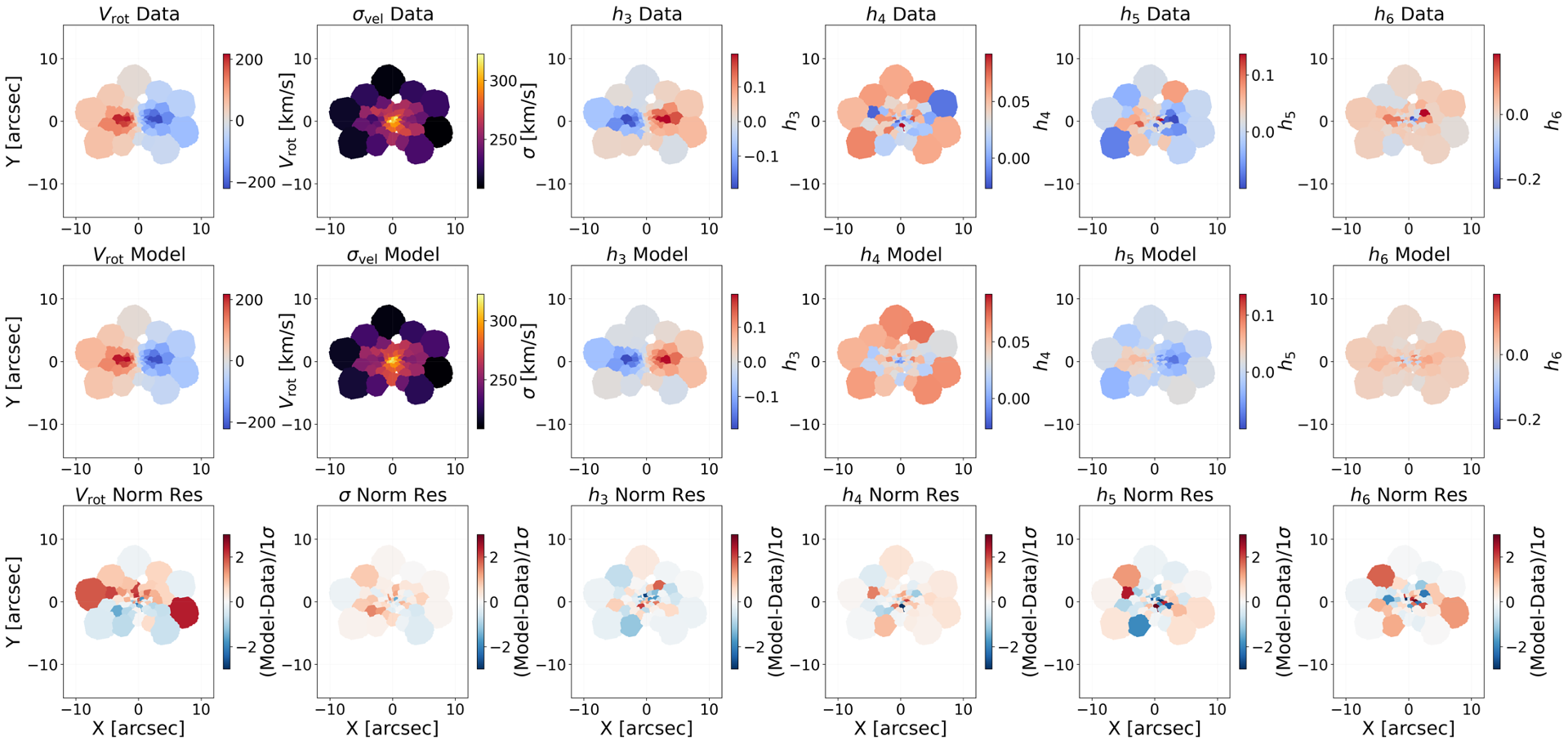}
    \caption{Comparison of the observed two-dimensional stellar kinematics of ESO0286 with the best-fit dynamical model. From left to right, the columns display the mean line-of-sight velocity ($V_{\rm rot}$), velocity dispersion ($\sigma_{\rm vel}$), and the higher-order Gauss-Hermite moments $h_3$, $h_4$, $h_5$, and $h_6$. The top row shows the extracted MUSE data, the middle row presents the corresponding predictions from the axisymmetric dynamical model, and the bottom row displays the normalized residuals, calculated as (Model $-$ Data) / $1\sigma$.}
    \label{fig:SB_fitting_2D}
\end{figure*}
\subsection{Triaxial mass model}
We show the predicted kinematic maps from the best-fit triaxial mass model.
\begin{figure*}
    \centering
    \includegraphics[width=1.\textwidth]{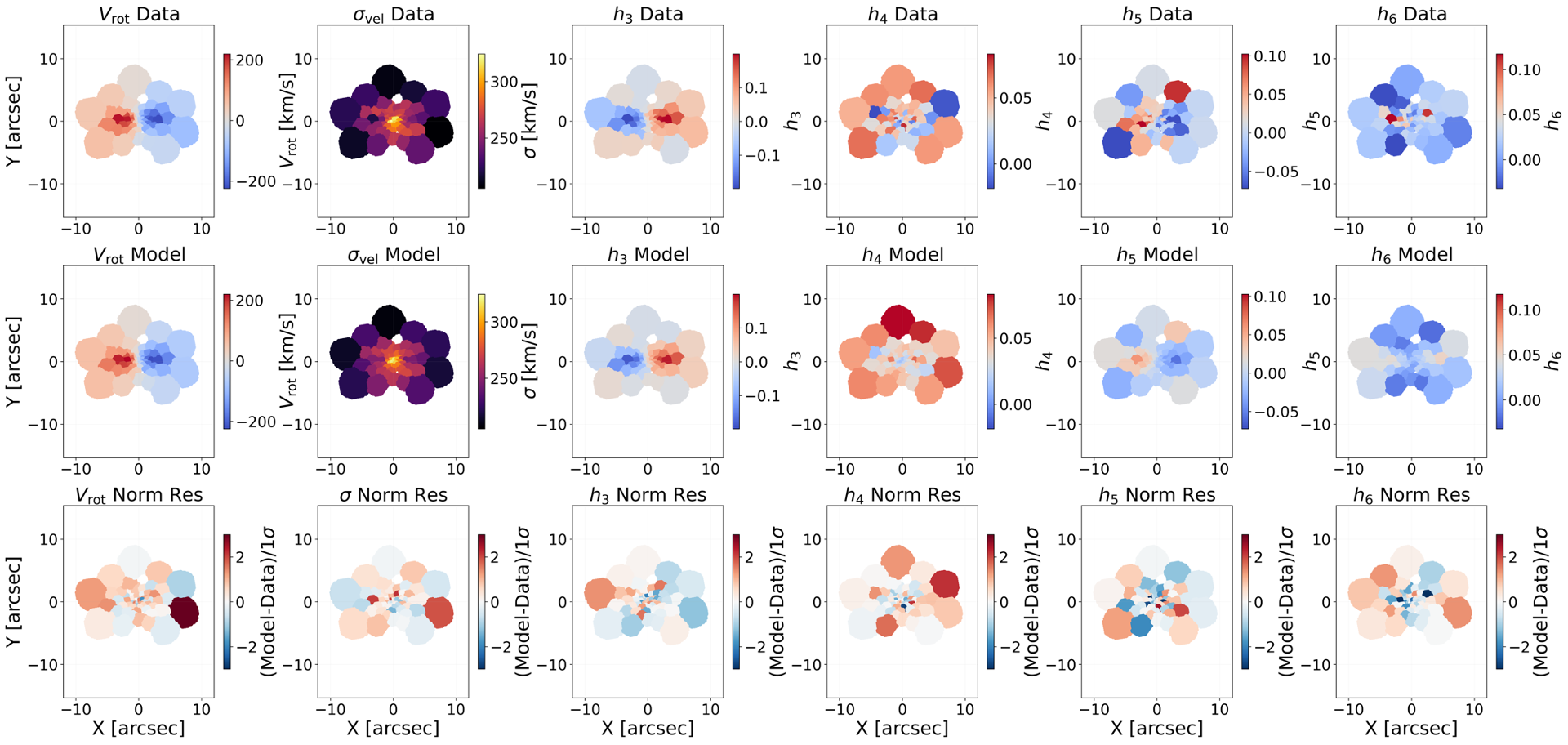}
    \caption{Comparison of the observed two-dimensional stellar kinematics of ESO0286 with the best-fit triaxial dynamical model. }
    \label{fig:SB_fitting_2D}
\end{figure*}

\section{The AIC envelope for all parameters across the triaxial models}
\label{app:The AIC envelope for all parameters across the models}
We present the complete set of AIC envelopes and $2\sigma$ confidence intervals for all sampled parameters across K and K\&L models.
\begin{figure}
    \centering
     \includegraphics[width=1.\textwidth]{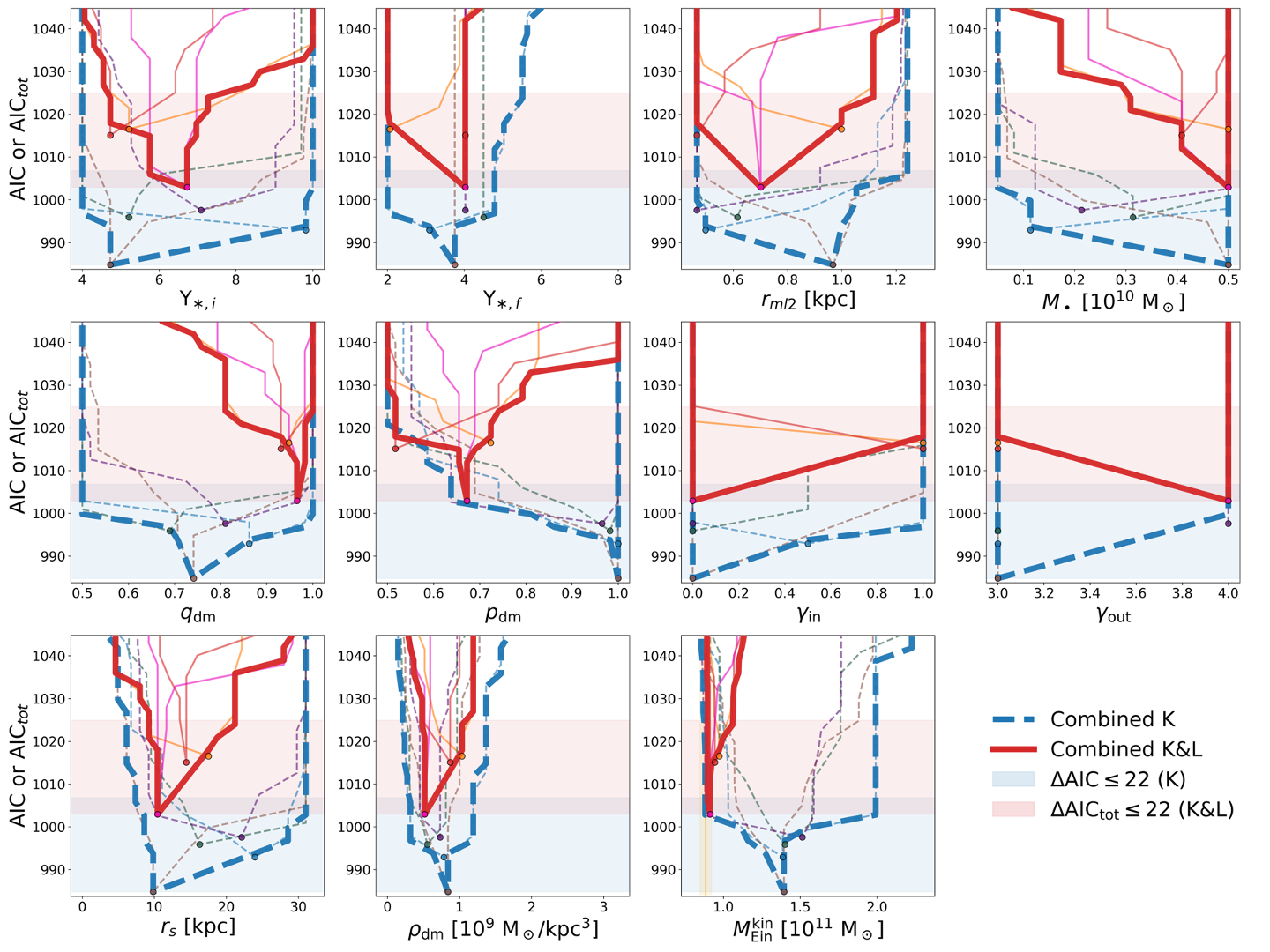}
    \caption{AIC envelopes for all sampled parameters across the different mass models. The individual colored lines represent the AIC lower envelopes for specific model families (Models II through VII), while the thick dashed blue and thick solid red lines represent the overall combined envelopes for the kinematics-only (K) and kinematics+lensing (K$\&$L) categories, respectively. Solid points mark the global best-fit value for each specific model family. Shaded horizontal regions represent the $2\sigma$ statistical uncertainty threshold derived from bootstrapping for the kinematics-only (blue) and kinematics+lensing (red) models. The vertical orange line and its corresponding shaded region in the $M_{\rm Ein}^{\rm kin}$ panel represent the independent mass measurement and uncertainty from the lensing-only model.}
    \label{fig:AIC_curves}
\end{figure}

\end{appendix}

\end{document}